\documentclass[twocolumn]{aastex63}

\usepackage{appendix}
\usepackage{natbib}
\usepackage{comment}
\usepackage{amsmath}
\usepackage{float}
\usepackage{xcolor}
\usepackage{import}
\usepackage{graphics,graphicx}
\graphicspath{ {plots/} }

\shorttitle{Optical-NUV survey of nearby M dwarfs}
\shortauthors{Paudel et al.}

\begin{document}

\title{A Multiwavelength Survey of Nearby M dwarfs: Optical and Near-Ultraviolet Flares and Activity with Contemporaneous TESS, Kepler/K2, \textit{Swift}, and HST Observations}

\correspondingauthor{Rishi R. Paudel}
\email{rpaudel@umbc.edu}

\author[0000-0002-8090-3570]{Rishi R. Paudel}
\affiliation{CRESST II and Exoplanets and Stellar Astrophysics Laboratory, NASA/GSFC, Greenbelt, MD 20771, USA}
\affiliation{Southeastern Universities Research Association, 1201 New York Ave NW, Suite 430, Washington. DC 20005}
\affiliation{University of Maryland, College Park, MD 20742, USA}
\affiliation{University of Maryland, Baltimore County, Baltimore, MD 21250, USA}

\author[0000-0001-7139-2724]{Thomas Barclay}
\author[0000-0002-1176-3391]{Allison Youngblood}
\author[0000-0003-1309-2904]{Elisa V. Quintana}
\author[0000-0001-5347-7062]{Joshua E. Schlieder}
\affiliation{NASA Goddard Space Flight Center, Greenbelt, MD 20771, USA}

\author[0000-0002-5928-2685]{Laura D. Vega}
\altaffiliation{Heising-Simons Astrophysics Postdoctoral Launch Program Fellow}
\affiliation{NASA Goddard Space Flight Center, Greenbelt, MD 20771, USA}
\affiliation{University of Maryland, College Park, MD 20742, USA}
\affiliation{CRESST II and Exoplanets and Stellar Astrophysics Laboratory, NASA/GSFC, Greenbelt, MD 20771, USA}

\author[0000-0002-0388-8004]{Emily A. Gilbert}
\affiliation{NASA Jet Propulsion Laboratory, 4800 Oak Grove Drive, Pasadena, CA 91109, USA}

\author[0000-0001-5643-8421]{Rachel A. Osten}
\affiliation{Space Telescope Science Institute, 3700 San Martin Drive, Baltimore, MD 21218, USA}

\author[0000-0002-1046-025X]{Sarah Peacock}
\affiliation{University of Maryland, Baltimore County,  Baltimore, MD, 21250, USA}
\affiliation{NASA Goddard Space Flight Center, Greenbelt, MD 20771, USA}

\author[0000-0001-5974-4758]{Isaiah I. Tristan}
\affiliation{Department of Astrophysical and Planetary Sciences, University of Colorado Boulder, 2000 Colorado Avenue, CO 80305, USA}
\affiliation{Laboratory for Atmospheric and Space Physics, University of Colorado Boulder, 3665 Discovery Drive, Boulder, CO 80303, USA}
\affiliation{National Solar Observatory, University of Colorado Boulder, 3665 Discovery Drive, Boulder, CO 80303, USA}

\author[0000-0002-2457-7889]{Dax L. Feliz}
\affiliation{American Museum of Natural History, 200 Central Park West, New York, NY 10024}

\author[0000-0003-0442-4284]{Patricia T. Boyd}
\affiliation{NASA Goddard Space Flight Center, Greenbelt, MD 20771, USA}

\author[0000-0002-0637-835X]{James R. A. Davenport}
\affiliation{Department of Astronomy, University of Washington, Box 351580, Seattle, WA 98195, USA}

\author[0000-0001-8832-4488]{Daniel Huber}
\affiliation{Institute for Astronomy, University of Hawai'i, 2680 Woodlawn Drive, Honolulu, HI 96822, USA}

\author[0000-0001-7458-1176]{Adam F. Kowalski}
\affiliation{Department of Astrophysical and Planetary Sciences, University of Colorado Boulder, 2000 Colorado Avenue, CO 80305, USA}
\affiliation{Laboratory for Atmospheric and Space Physics, University of Colorado Boulder, 3665 Discovery Drive, Boulder, CO 80303, USA}
\affiliation{National Solar Observatory, University of Colorado Boulder, 3665 Discovery Drive, Boulder, CO 80303, USA}

\author[0000-0003-3896-3059]{Teresa Monsue}
\affiliation{NASA Goddard Space Flight Center, Greenbelt, MD 20771, USA}
\affiliation{Institute of Astrophysics \& Computational Sciences, The Catholic University of America, 620 Michigan Ave NE, Washington, DC 20064, USA}

\author[0000-0003-2565-7909]{Michele L. Silverstein}
\altaffiliation{Resident guest investigator as a National Research Council (NRC) Research Associate}
\affiliation{U.S. Naval Research Laboratory, 4555 Overlook Avenue SW, Washington, DC 20375, USA}

\begin{abstract}
We present a comprehensive multiwavelength investigation into flares and activity in nearby M~dwarf stars. We leverage the most extensive contemporaneous dataset obtained through the Transiting Exoplanet Sky Survey (TESS), Kepler/K2, the Neil Gehrels Swift Observatory (\textit{Swift}), and the Hubble Space Telescope (HST), spanning the optical and near-ultraviolet (NUV) regimes. In total, we observed 213 NUV flares on 24 nearby M dwarfs, with $\sim$27\% of them having detected optical counterparts, and found that all optical flares 
had NUV counterparts. 
We explore NUV/optical energy fractionation in M dwarf flares. 
Our findings reveal a slight decrease in the ratio of optical to NUV energies with increasing NUV energies, a trend in agreement with prior investigations on G-K stars' flares at higher energies. Our analysis yields an average NUV fraction of flaring time for M0-M3 dwarfs of 2.1\%, while for M4-M6 dwarfs, it is 5\%. We present an empirical relationship between NUV and optical flare energies and compare to predictions from radiative-hydrodynamic and blackbody models. We conducted a comparison of the flare frequency distribution (FFDs) of NUV and optical flares, revealing the FFDs of both NUV and optical flares exhibit comparable slopes across all spectral subtypes. 
NUV flares on stars affect the atmospheric chemistry, the radiation environment, and the overall potential to sustain life on any exoplanets they host. We find that early and mid-M dwarfs (M0-M5) have the potential to generate NUV flares capable of initiating abiogenesis. 
\end{abstract}

\keywords{Stellar activity (1580); Stellar physics (1621); M dwarf stars (982); Red dwarf flare stars (1367); Stellar flares (1603); Stellar x-ray flares (1637); X-ray astronomy (1810); Near ultraviolet astronomy (1094); Optical astronomy (1776)}

\section{Introduction}
M dwarfs, characterized by their low mass (M $\lesssim$ 0.6 M$_{\odot}$) and cool effective temperatures (T$_{\rm eff}$ $\lesssim$ 4000 K), represent the most abundant stellar objects, accounting for approximately 75\% of the total stellar population within our galaxy \citep{2006AJ....132.2360H}. The small radii of M dwarfs make them favorable hosts to several exceptional planetary systems that enable deep follow-up investigations via mass measurements and transmission and emission spectroscopy. 
These stars are the primary targets of ongoing planet-search missions like the Transiting Exoplanet Survey Satellite (TESS; \citealt{2015JATIS...1a4003R}) and the CHaracterising ExOPlanets Satellite (CHEOPS; \citealt{2021ExA....51..109B}) and M dwarfs are more likely to host more Earth-sized planets, compared to higher mass stars \citep{2015ApJ...807...45D,2019AJ....158...75H}.

Stellar flares are transient events observed as photon emission across the electromagnetic spectrum, including gamma-rays, X-rays, UV, optical, infrared, and radio waves \citep{2005ApJ...621..398O,2008A&A...487..293F}. These events occur when the conversion of magnetic energy, previously stored in the magnetic field, to kinetic energy through magnetic reconnection leads to the acceleration of particles and the deposition of the particles' energy into the atmosphere through collisions with atmospheric molecules (e.g., heating).
The spectral energy distributions of solar and stellar flares consist of emission lines and continuum.
Typically, the continuum emission observed across ultraviolet and optical wavelengths is characterized as a high-temperature blackbody with a temperature around 10$^{4}$ K \citep{1992ApJS...78..565H}. However, this model has been shown to not accurately match all observations \citep{2023ApJ...944....5B,2019ApJ...871..167K,2023arXiv231212511B}.

Though M dwarfs are cooler and less luminous than the Sun, they have relatively strong magnetic fields for their sizes.  For example, WX UMa (an M5.5 dwarf) is reported to have average surface magnetic field $\langle B \rangle$ $\approx$ 7.0 kG \citep{2017NatAs...1E.184S}, 70x that of the Sun \citep{2009ASPC..405..135S}. M dwarfs are capable of producing very strong flares with energies larger than that of the strongest flare observed on the Sun, irrespective of their ages \citep{2020AJ....159...60G,2014ApJ...781L..24S,1991ApJ...378..725H,2018ApJ...861...76P,2018ApJ...858...55P,2020AJ....160..237F}. In order to understand the potential influence of M dwarf flares on exoplanet atmospheres, it is important to characterize flare activity. This will in turn be helpful to estimate the energy received by planets in the form of X-ray and UV photons and energetic particles at different stages of stellar evolution and the cumulative effect of flares over the lifetime of a planet.

The near-ultraviolet (NUV) radiation (1700--3200 Å) of cool stars plays a crucial role in determining the habitability of terrestrial exoplanets. The NUV flux received by these planets can have both positive and negative impacts. On one hand, an excessive dose of NUV flux can be detrimental as it has the potential to harm the DNA of surface organisms and dissociate key biosignature molecules like O$_{3}$ \citep{2015ApJ...809...57R,2015ApJ...806..137R}. On the other hand, NUV photons within the 2000-2800 Å range are essential for prebiotic photochemistry, which contributes to the emergence of life (abiogenesis; \citealt{2020ApJ...896..148R, 2018SciA....4.3302R}). Flares significantly enhance the NUV line and continuum emission flux by orders of magnitude \citep{2013ApJS..207...15K,2019ApJ...883...88B,2022RNAAS...6..201C}. 
In some cases, flares can provide the necessary NUV flux for abiogenesis when the quiescent NUV flux from the host star is insufficient. Therefore, flares play a vital role in determining the NUV conditions that impact exoplanetary atmospheres, biosignatures, and overall habitability.

Although numerous stellar flares have been observed by missions such as TESS and Kepler, only a limited number of flares have been simultaneously detected in both the optical and NUV wavelength bands. Consequently, the distribution of energy within flares across the optical and NUV wavelengths remains inadequately understood \citep{2023ApJ...944....5B, 2021ApJ...922...31P}. Most studies aiming to assess the impact of flares on exoplanets extrapolate findings from white light flare investigations to the UV range (see for e.g., \citealt{2020AJ....160..219F}, \citealt{2020ApJ...902..115H}).

Various models are utilized to estimate the flare energies, and their complexity can vary. One approach assumes a single-temperature blackbody to estimate the bolometric energies of flares detected through single bandpass photometry. For example, \cite{2013ApJS..209....5S} used a 9000 K blackbody model to estimate bolometric energies of flares detected with $Kepler$. On the other hand, alternative models combine these blackbody models with data derived from archived UV spectra, as demonstrated by \cite{2018ApJ...867...71L}. Nevertheless, the precision of UV predictions generated by these models has not been comprehensively tested. A study conducted by \cite{2019ApJ...871..167K} with the Hubble Space Telescope Cosmic Origins Spectrograph (HST COS) revealed that the 9000 K blackbody model underestimated both the near-ultraviolet (NUV) continuum and the overall NUV emission observed during two flares by a factor of 2-3. 
Furthermore, the peak flare temperatures can surpass temperatures of 9000 K (\citealt{2013ApJS..207...15K}; \citealt{2020ApJ...902..115H}). This introduces the need for greater complexity when constructing accurate flare models.

\cite{2023ApJ...944....5B} recently investigated the distribution of optical/NUV energy in stellar flares using overlapping datasets from the Galaxy Evolution Explorer (GALEX; \citealt{1999ASPC..164..182M}) and \textit{Kepler} \citep{2010Sci...327..977B}. Their analysis primarily focused on G and K stars and identified 1557 NUV flares in GALEX light curves for which contemporaneous \textit{Kepler} long-cadence (30 min) data is available, as well as two NUV flares in GALEX light curves for which contemporaneous \textit{Kepler} short-cadence (1 min) data is available. Interestingly, none of the flares detected by GALEX were identified in the contemporaneous \textit{Kepler} light curves. As a result, the authors established upper limits for flare energies in the \textit{Kepler} band and for optical to NUV flare energy ratios. 

The \cite{2023ApJ...944....5B} study revealed a correlation spanning approximately three orders of magnitude between the energy ratio of flares in the two bands and the energy in the GALEX NUV band (1693--3008 \AA). 
They demonstrate that at lower flare energies, the energy fractionation aligns with the predictions of some radiative-hydrodynamic (RHD) models \citep{2017ApJ...837..125K,2022FrASS...934458K}. Such models were updated to incorporate both continuum and emission lines, thereby providing the flare's spectral energy distribution from NUV wavelengths to the optical range. However, for larger flares, mostly exceeding energies of 10$^{33}$ erg, a significant deviation of up to approximately three orders of magnitude from the models was observed. The deviation appears as a disconnect between observed data and model predictions but the underlying cause remains unclear.

\cite{2023MNRAS.519.3564J} examined six empirical flare models to predict NUV and FUV flare rates in M dwarf stars based on white-light observations. Those models mainly consisted of a 9000 K blackbody (BB) spectrum, modified BB spectrum with flux contributions from emission lines, blend of 9000 K BB spectrum and UV spectra derived from the 1985 Great Flare of AD Leo \citep{1991ApJ...378..725H}, as well as the blend of 9000 K BB with NUV emission lines and an FUV flare model \citep{2018ApJ...867...71L}. \cite{2023MNRAS.519.3564J} used TESS data to determine average white-light flare rates for partially and fully convective M stars, supplementing this with GALEX data. They found that the 9000 K BB model underestimates the observed GALEX NUV flare energies of partially (M0-M2) and fully convective M dwarfs by up to a factor of 2.7 $\pm$ 0.6 and 6.5 $\pm$ 0.7 respectively. Furthermore, they found that the  \citep{2018ApJ...867...71L} flare model performs better than the other flare models considered, but still underestimates the observed GALEX NUV flare energies of M dwarfs. 
It should be noted that the \cite{2018ApJ...867...71L} flare model applies to inactive M dwarf flares and not to active stars like AD Leo. 

In this paper, we study a sample of NUV flares observed on nearby M dwarfs of varying spectral types and ages by \textit{Swift} and \textit{HST}. We compare various NUV flare properties such as flare durations, amplitudes, energies and rates with those of the optical flares using contemporaneous TESS and K2 data. We empirically determine the NUV/optical energy fractionation in M dwarf flares and compare with the predictive power of various RHD and blackbody flares models.
This paper is organized as follows. Section \ref{section:observations} outlines the observation details, while Section \ref{section:flare analysis} describes the analysis of both NUV and optical flares, as well as identification of NUV/optical flare counterparts. In Section \ref{section:results} we interpret and discuss the results from this study, and in Section \ref{section:summary} we give a summary of the main results from this project. 
\section{Observations Overview}\label{section:observations}
\subsection{The Stellar Sample}\label{subsec:sample}
Our sample of 34 M dwarfs was selected to cover a range of spectral sub-type and ages in order to gather a more complete picture of M dwarf flaring activity. We selected from the brightest targets in three spectral subtype categories: early (M0-2.5), mid (M3-5), and late (M5.5-7). Whenever possible, we included multiple stars of each category. Notably, our sample features some renowned flare stars and exoplanet host stars, including EV Lac, Proxima Cen, AU Mic, and GJ 876 (e.g., \citealt{2016Natur.536..437A,2018ApJ...867...71L,2020Natur.582..497P,2021ApJ...922...31P, 2022AJ....163..147G}).

We present the details of the stellar sample in Table \ref{table:targets info}, including each star's TESS Input Catalog (TIC) number, effective temperature ($T_{\rm eff}$), distance, spectral type (SpT), and age. Furthermore, we indicate whether the star is a planet host and/or part of a multiple system (MS), either binary or tertiary. Among the stars in our sample, nine are known to host one or more exoplanets, and an additional nine stars are known to have stellar companions.

\begin{figure}[htbp]
   \centering
   \includegraphics[width=0.45\textwidth]{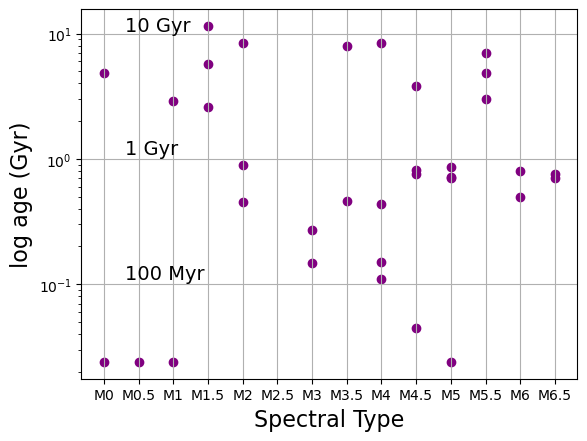}
   \caption{Distribution of M dwarfs in our sample by spectral type and age. The total number of M dwarfs is 34. We do not show error bars on the stellar ages, but do note that in many cases they are imprecise.}
   \label{fig:Mdwarf_distribution}
\end{figure}

Figure \ref{fig:Mdwarf_distribution} shows the distribution of our sample stars in spectral type and age. For the stars that are assigned with a range of ages (see Table \ref{table:targets info}), the average values are used. Most of these ages were compiled from the literature. For some stars whose ages were not available in the literature, we estimated them using the period-age relation from \cite{2023arXiv230701136E}. 
Stellar ages are notoriously difficult to estimate, and so caution should be exercised when using our age estimates. 
It is evident from Figure \ref{fig:Mdwarf_distribution} that our M dwarf sample encompasses stars of various spectral types, ranging from M0 to M6.5, and spans a wide range of ages, from 22 Myr (AU Mic) to 11.5 Gyr (Kapteyn's Star). We do note that there are some gaps in age/spectral type space to the unavailability of bright stars in those parameter spaces. 

\subsection{TESS data}
The Transiting Exoplanet Survey Satellite (TESS; \citealt{2015JATIS...1a4003R}) mission, launched in April 2018, was designed for a comprehensive photometric survey of the entire sky. It is equipped with four identical, red-sensitive, wide-field cameras, each of which enables surveillance of large portions of the sky in sectors spanning $24^{\circ} \times 96^{\circ}$. Each sector is observed continuously for $\sim$27 days. 
While its primary objective is to detect small planets orbiting the nearest and brightest stars, it also exhibits sensitivity to low-amplitude, short-duration phenomena such as stellar flares. TESS's photometric bandpass, spanning 600–1000 nm, is more sensitive to red objects like M dwarfs than the shorter-wavelength \textit{Kepler} mission. TESS's all-sky observational approach and its capacity to provide long-baseline and high-precision continuous time series optical data makes it particularly well-suited for studying stellar flares (e.g., \citealt{2020AJ....159...60G, 2022AJ....163..147G, 2022ApJ...926..204H}). 

We acquired TESS data for our targets during Cycles 1 and 2 in its 2-minute cadence mode simultaneously with \textit{Swift} as part of general investigator programs 1266 and 2252 (PI: J. Schlieder). The introduction of TESS 20-second cadence mode in Cycle 3 significantly advanced our ability to study the morphologies of white-light stellar flares. This mode allows for the detection of small flares which last for few tens of seconds and may remain unnoticed in longer cadence data or become entangled with other complex flares. We obtained simultaneous \textit{Swift} and 20-second TESS data for six stars (AP Col, AU Mic, GJ 3631, Wolf 359, YZ Ceti, YZ CMi) during TESS Cycles 3 and 4 as part of general investigator programs 3273 (PI: L. Vega), 4247 (PI: L. Vega) and 4212 (PI: R. Paudel). Table \ref{table:TESS and UVOT info} lists the TESS sectors, cadence utilized and total observation time for each star. We used 20-second TESS data whenever it was available. Nine stars were either undetected in \textit{Swift} UVOT data or their UV data was not obtained in \texttt{EVENT} mode (which is necessary for extracting high cadence light curves). Such stars are indicated by a dagger (\dag) sign next to their names in Table \ref{table:targets info}. In addition, TESS data is unavailable for one of the targets (WX UMa) in either 2-min and 20-second cadence mode simultaneous with \textit{Swift} mission. We only have Full Frame Image (FFI) data obtained at 30-min cadence for this target. Therefore, we only analyzed the TESS data of the remaining 24 stars.

We retrieved the light curves of our targets from the Mikulski Archive for Space Telescopes (MAST) using the \texttt{lightkurve} \citep*{lightkurve} package. We used the Pre-Search Data Conditioning (PDCSAP) light curve products produced and made available by the TESS Science Processing Operations Center (SPOC; \citealt{2016SPIE.9913E..3EJ}). 
We specifically selected data points with quality flags 0 and 512. The latter flag, known as ``impulsive outlier removed before cotrending," often eliminates the peak of flares. However, to ensure no flare data points were excluded, we included data with this quality flag as well. This approach helped minimize potential biases when estimating flare energies. 

\subsection{K2 mission data}
One of our targets, Wolf 359, was observed simultaneously by the Kepler Space Telescope as part of the extended K2 mission \citep{2014PASP..126..398H} and \textit{Swift} during Campaign 14 (2017 May 31 - 2017 Aug 19) of the K2 mission as a part of general investigator program 14082 (PI: E. Quintana). The K2 data was obtained in both long ($\sim$ 30 min; \citealt{2010ApJ...713L.120J}) and short cadence ($\sim$1 min; \citealt{2010ApJ...713L.160G}) mode for a total of $\sim$80 days and is publicly available in MAST. We extracted the short cadence light curve from the target pixel files (TPFs) using the methods explained in Feliz et al. in prep.

\subsection{\textit{Swift}/{\normalfont UVOT} Data}
Thirty-three targets in our original sample were observed by \textit{Swift}'s  Ultra-Violet/Optical Telescope (UVOT; \citealt{2005SSRv..120...95R}) via \textit{Swift} general investigator Cycle 14 program 1417229 (PI T. Barclay, Key Project ``A Comprehensive, Multiwavelength Survey of Cool Star Activity") as well as multiple additional programs through the mission's Target of Opportunity (ToO) program. The remaining target, GJ 3631, was observed through joint TESS-\textit{Swift} observation program 4212 (PI: R. Paudel) during TESS Cycle 4. Almost all observations were performed with the UVM2 filter centered at $\lambda$ = 2259.8 \AA\ ($\lambda_{min}$ = 1699.1 \AA, $\lambda_{max}$ = 2964.3 \AA, FWHM = 527.1 \AA). Few observations were performed with the UVW1 filter centered at $\lambda$ = 2629.4 \AA ($\lambda_{min}$ = 1614.3 \AA, $\lambda_{max}$ = 4730.0 \AA, FWHM = 656.6 \AA). We did not observe any flares in the UVW1 filter except on Proxima Cen. 
To enable the construction of high cadence UV light curves and flexible binning, the observations were obtained in UVOT's EVENT mode. A summary of total observation times with \textit{Swift}/UVOT for each target is listed in Table \ref{table:TESS and UVOT info}. We note that we did not acquire the \textit{Swift}/UVOT data of six targets (AX Mic, EZ AQr, GJ 205, HIP 23309, UV Ceti and NLTT 31625) in the \texttt{EVENT} mode. 

\textit{Swift} obtains simultaneous observations of each target using its X-ray Telescope (XRT; \citealt{2005SSRv..120..165B}) alongside UVOT. Hence, all the targets in our sample were also observed by \textit{Swift}/XRT via the same general investigator and ToO programs mentioned above. However, the primary goal of this paper is to compare optical and NUV flare properties and the activity of the targets. Therefore, we focus primarily on \textit{Swift} NUV  and data here. A detailed analysis of the X-ray flares observed on M dwarfs via this program and additional programs using the NICER X-ray telescope, and their comparison with contemporaneous TESS data, will be presented in a future publication. We present \textit{Swift}/XRT observation time and the quiescent luminosity of each star in the XRT band in Table \ref{table:TESS and UVOT info} and Table \ref{table:quiescent flux} respectively.

\subsubsection{\textit{Swift}/UVOT data reduction}
The raw data were obtained from the UK \textit{Swift} Science Data Centre (UKSSDC), and then processed in two steps to obtain a cleaned event list. First, we used the \texttt{COORDINATOR}\footnote{\url{https://heasarc.nasa.gov/ftools/caldb/help/coordinator.html}} FTOOL to convert raw coordinates to detector and sky coordinates. Second, we used \texttt{UVOTSCREEN} to filter the hot pixels and obtain a cleaned event list. 

We utilized the FTOOL \texttt{UVOTEVTLC} task to extract calibrated light curves from the cleaned event lists. For the sources, we followed the recommended approach of employing a circular extraction region with a radius of 5\arcsec centered around each source position except for some sources for which we used a radius of 7\arcsec. Such sources did not have a bright center and were somewhat elongated. Additionally, for a smooth background, we used a circular extraction region with a radius of 30\arcsec located away from the each source. We used \texttt{timebinalg = u} (uniform time binning), to create UVOT light curves with bins of 11.033 seconds. It is essential that the bin size is a multiple of the minimum time resolution of UVOT data, which is 11.033 milliseconds. When photon pile-up occurs on detectors, \texttt{UVOTEVTLC} incorporates a coincidence loss correction by utilizing the necessary parameters from CALDB. Subsequently, the light curves were corrected for barycentering using \texttt{BARYCORR}. To synchronize all the telescopes on a unified time system, the barycentric times were converted to the Modified Julian Date (MJD) system.

We converted the count rate in each filter (UVM2 and UVW1) to flux density in units of erg cm$^{-2}$ s$^{-1}$ \AA$^{-1}$ by using an average count rate to flux density conversion ratio of 8.446 $\times$ 10$^{-16}$ for UVM2 filter and 4.209 $\times$ 10$^{-16}$ for UVW1 filter. The conversion ratio is part of the {\it Swift}/UVOT CALDB and was obtained using GRB models\footnote{\url{https://heasarc.gsfc.nasa.gov/docs/heasarc/caldb/swift/docs/uvot/uvot_caldb_counttofluxratio_10wa.pdf}}. We estimated the quiescent NUV fluxes of each star by using these conversion ratios and the FWHM ($\Delta_{uv}$ = 530 \AA\ (UVM2) and $\Delta_{uv}$ = 660 \AA\ (UVW1)) of each filter. We list the quiescent NUV fluxes of the stars in Table~\ref{table:quiescent flux}. We note that three stars (GJ 1061, GJ 4353 and LHS 292) were undetected in \textit{Swift}/UVOT data. 
\subsection{HST/STIS Data}
We obtained time-series \textit{HST} NUV spectra simultaneously with TESS for two targets, GJ 832 and GJ 1061, through general observer program 15463 (PI: A. Youngblood). All data were obtained with the STIS instrument's photon-counting NUV-MAMA detector with the G230L grating and 52\arcsec $\times$ 2\arcsec\ aperture. The time-series spectral data products cover wavelengths 1570-3180 \AA\ at resolving power R$\sim$500. GJ 832 was observed simultaneously with TESS Sector 1 for five consecutive orbits on August 15, 2018 from 15:52:07-22:47:13 for a total exposure time of 12,246 seconds. GJ 1061 was observed simultaneously with TESS Sector 3 for one orbit (one orbit lost due to guide star acquisition error) on September 25, 2018 23:39:13 to September 26, 2018 00:23:55. Three consecutive orbits (one orbit lost due to guide star acquisition error) were also obtained on February 21, 2019 10:43:44-14:34:06 for a total exposure time of 11,054 seconds; these three orbits were not performed simultaneously with TESS. The data were calibrated with the default CALSTIS v3.4.2 pipeline. We used the \texttt{stistools} Python package's \texttt{inttag} function to extract time-series spectra at a cadence of 10 seconds.

\section{Flare Analysis} \label{section:flare analysis}
\subsection{White Light Flares} \label{subsection:WLFs}
\subsubsection{Flare Identification} \label{subsubsection:flare identification}
We used the \texttt{stella} package \citep{2020JOSS....5.2347F} to identify flares in the TESS light curve. It employs convolutional neural networks (CNNs) and utilizes the flare catalog presented in \cite{2020AJ....159...60G} to generate the required training, validation, and test sets for the CNNs.
By utilizing \texttt{stella}, we obtained an initial estimate of the flare's location in the light curve and a corresponding ``probability" value indicating the likelihood that each data point belongs to a flare. We considered data points with a probability greater than 0.5 and at least one directly adjacent flaring point as potential flare candidates. We further applied an additional criterion to each identified flare candidate. Specifically, the amplitude of the flare must exceed 2.5 times the standard deviation ($\sigma$) of the quiescent level. This criterion is often used in other flare studies such as \cite{2018ApJ...858...55P} and \cite{2019ApJ...883...88B}. We list the number of optical flares identified in TESS light curves for each target in Table \ref{table:TESS and UVOT info} in column $N_{\rm T}$. We identified hundreds of flares in the K2 light curve of Wolf 359, the details of which will be provided in Feliz et al. (in prep.). Additionally, see \cite{2021AJ....162...11L} for an analysis of the K2 lightcurve of Wolf 359. 

Subsequently, we determined the stellar rotation periods by excluding all potential flare events, focusing solely on the quiescent light curve, and employing the Lomb-Scargle Periodogram. As we utilized single TESS sector data for most of our targets, this approach is applicable only to stars that exhibit photometric spot variability in the TESS light curve and stars with rotation periods less than 15 days. For all other stars, we gathered the rotation period values from existing literature. The rotation periods, which range from 0.2073 to 200 days, are provided in Table \ref{table:rotation periods}. For the periods we estimated, we also calculated uncertainties based on the half width at half maximum (HWHM) of the periodogram peaks, following the methodology suggested by \cite{2013AJ....145..148M}.

The light curves were then detrended using the rotation period. To accomplish this, we 
used a periodic Gaussian Process (GP) to model the starspot-modulated
stellar rotation on the quiescent light curve using a ``Rotation term" from \texttt{celerite} \citep{2017zndo....806847F}. The flux predicted by the model was subtracted from the total flux, resulting in the detrended light curve. Figure \ref{fig:V1005 Ori light curve, TESS} demonstrates this procedure for one of our targets, V1005 Ori.

%
\begin{figure*}
   \centering
   \includegraphics[width=\textwidth]{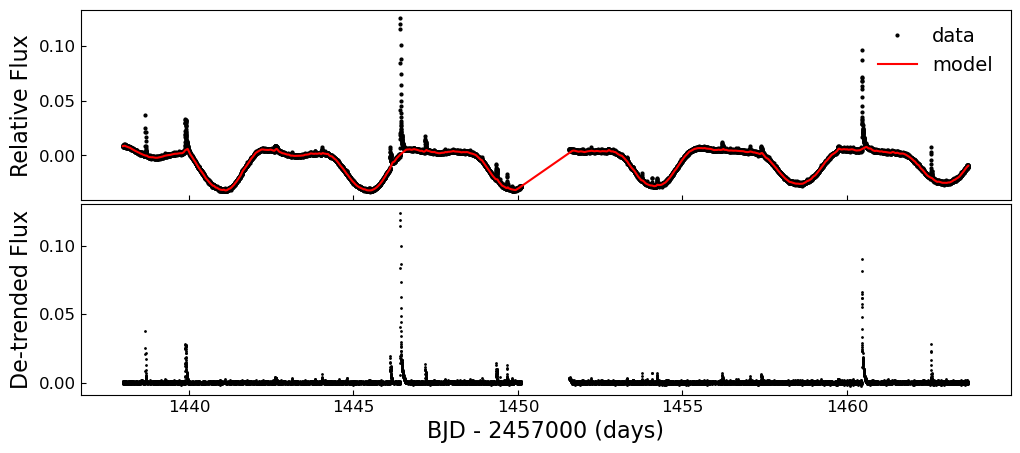}
   \caption{TESS light curve of V1005 Ori, obtained during Sector 5 (2018 November 15 to 2018 December 11). The top panel shows the original light curve (black) and GP fit of the spot modulation (red). The rotation period of this star is 4.379$\pm$0.337 d. The lower panel displays the detrended light curve after subtracting the original quiescent light curve with rotational modulation.}
    \label{fig:V1005 Ori light curve, TESS}
\end{figure*}
%
\begin{figure}
    \centering
    \includegraphics[width=0.45\textwidth]{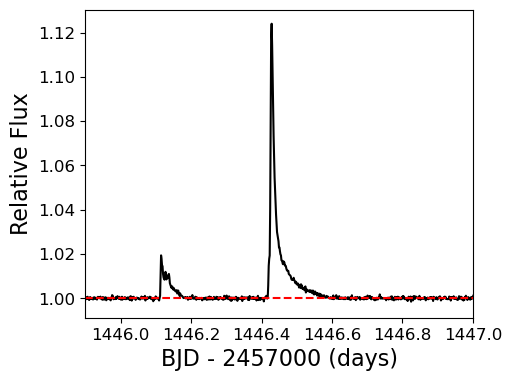}
    \caption{Example of flares identified in the TESS light curve of V1005 Ori shown in Figure \ref{fig:V1005 Ori light curve, TESS}. The red dashed line represents the quiescent level of the detrended light curve. The flux is normalized by the quiescent level flux.}
    \label{fig:tess example flare}
\end{figure}
\subsubsection{Determining Flare Properties}
\label{subsubsec:determining flare properties}
To determine the energies of the white light flares, we first estimated the equivalent duration (ED) for each individual flare. The ED represents the duration over which a flare emits the same amount of energy as the star does during its quiescent state \citep{1972Ap&SS..19...75G}. It is measured in units of time and depends on the filter used but is independent of the distance to the
flaring object and is therefore widely used for determining flare energies. The ED of a flare is expressed as:
\begin{equation}
    ED =  \int \frac{\Delta F}{F_{q}} dt =  \int \frac{F_{f} - F_{q}}{F_{q}} dt
\end{equation}
where $F_{\rm f}$ is the flaring flux and $F_{\rm q}$ is the quiescent flux of the star. We utilized the detrended light curves of the targets to estimate the equivalent durations (ED) of the flares identified. 

We can convert ED to absolute flare energy by using each star's spectral information whenever available (e.g., \citealt{2021ApJ...922...31P}, \citealt{2022AJ....163..147G}). But, we do not have homogeneous spectra for all the targets. 
Instead, we adopted the approach of \cite{2013ApJS..209....5S} which assumes a blackbody (BB) for the spectral energy distribution of each flare and estimates its energy using: 

\begin{equation}
E_{\rm flare,bol} = L_{\rm flare} \times ED
\label{equation:flare energy}
\end{equation} 
where, 
\begin{equation} \label{eq:Lflare}
\begin{split}
L_{\rm flare} = \sigma T^4_{\rm flare} A_{\rm flare} \\
= \sigma T^4_{\rm flare} \times \pi R^{2}_{\star} \times \frac{\int B(T_{\rm flare}, \lambda)S(\lambda) d\lambda} {\int B(T_{\rm eff}, \lambda)S(\lambda)  d\lambda}
\end{split}
\end{equation} 
In Equation \ref{eq:Lflare}, $A_{\rm flare}$ is the area of flare emitting region, $L_{\rm flare}$ is the bolometric luminosity of the flare and gives the energy emitted by a flare corresponding to ED of 1 s on a given star. Likewise, $\sigma$ is the Stefan–Boltzmann constant, $T_{\rm flare}$ is flare BB temperature, $R_{\star}$ and $T_{\rm eff}$ are radius and effective temperature of a given star, B($T$, $\lambda$) is the Plank function, and S($\lambda$) is the spectral response function of the TESS instrument. We list the values of $L_{\rm flare}$ for a 10,000 K BB and $R_{\star}$ for each star in Table \ref{table:rotation periods}. Those values are listed for only those stars on which we observed flares with TESS. 

Note that the value of $L_{\rm flare}$ depends on the assumed BB temperature. A 9,000 K BB flare has 0.84$\times$ as much energy as a 10,000 K BB flare. 21\% of a 9,000 K BB flare's bolometric energy is emitted in the TESS band, and likewise 18\% for a 10,000 K BB. Similar estimates were obtained by previous studies, such as those by \cite{2021PASJ...73...44M} and \cite{2019A&A...628A..79S}.   

\subsection{{\normalfont NUV} Flares}
\subsubsection{\textit{Swift}/{\normalfont UVOT}}
The \textit{Swift}/UVOT observations of any given target are broken into segments of 20-30 minutes due to observing constraints of the mission. Because of this complexity and short duration observation of each target, we examined each light curve by eye to identify the flare candidates. Any event with two consecutive data points above the local noise level was identified as a flare. This noise level is different for each \textit{Swift}/UVOT observing window. From here onwards, any flare in the \textit{Swift}/UVOT lightcurve will be referred to as a ``NUV flare". We identified a total of 213 NUV flares, among which 199 (93.4\%) were observed for their full duration. 
Furthermore, 15 NUV flares were observed on Wolf 359 by \textit{Swift}/UVOT simultaneously with K2. 

To compute the flare energies, we first estimated the ED of each flare in a similar manner as done for the TESS flares. The flare energies $E_{NUV}$ were computed by using the equation:
\begin{equation}
    E_{NUV} =  ED \times 4 \pi d_{\star}^{2} \times F_{q}\label{eq:NUV flare energy}
\end{equation}
\noindent where $d_{\star}$ is the distance to the star, and $F_{q}$ is the quiescent flux in units of erg cm$^{-2}$ s$^{-1}$. We list the properties of all the NUV flares in  Appendix \ref{appendix:NUV flares}. 

Taking into account the small sample of NUV flares, we divide our targets in only two spectral bins: M0-M3 and M4-M6, which also approximately corresponds to partially convective and fully convective M dwarfs. In addition, we also divided the observed flare energies in four energy bins: (10$^{27}$ - 10$^{29}$) erg, (10$^{29}$ - 10$^{31}$) erg, (10$^{31}$ - 10$^{33}$) erg and $>$10$^{33}$ erg. We list the number of optical and NUV flares observed in each spectral and energy bin in Table \ref{table:flares_by_SpT_and_E-bin}. We find that the highest number of NUV flares are observed in the energy bin (10$^{29}$ - 10$^{31}$) erg for both spectral bins. Likewise, we find spectral bin M0-M3 has highest number of optical flares in the energy bin (10$^{31}$ - 10$^{33}$) and M4-M6 has highest number of optical flares in the energy bin (10$^{29}$ - 10$^{31}$) erg. We did not observe any NUV flares with energies $>$10$^{33}$ erg.

\subsubsection{{\normalfont HST/STIS}} \label{subsubsec:HSTSTIS_flare_ids}
\begin{figure*}
   \centering
   \includegraphics[width=\textwidth]{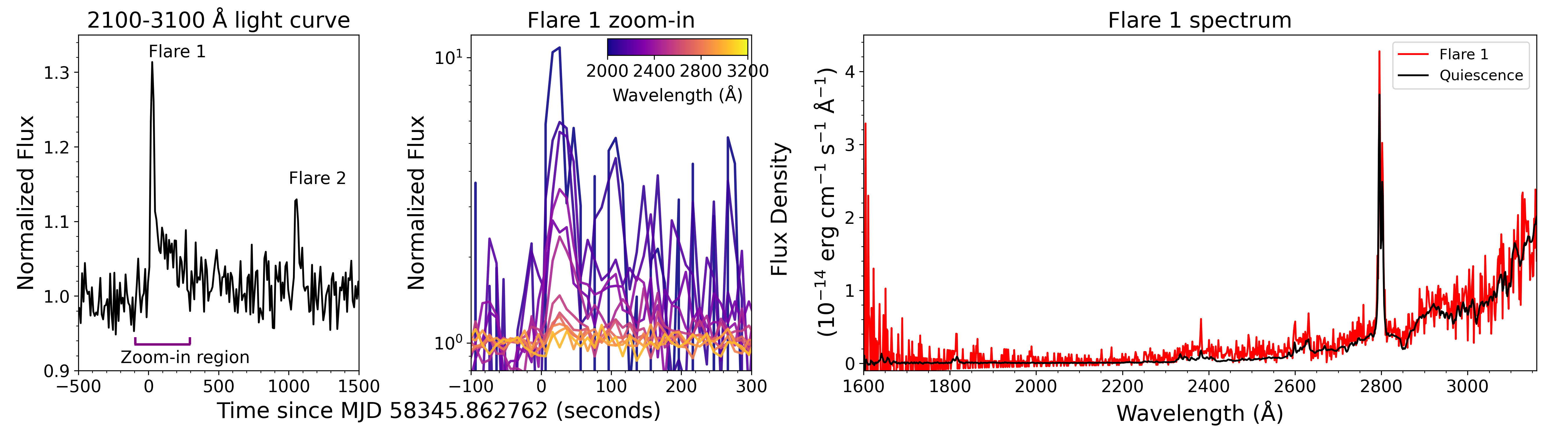}
    \caption{HST/STIS observations of GJ 832. Left: the 2100-3100 \AA\ light curve at 10 s cadence normalized to the star's quiescent flux shown covers the 2000 s period when two flares, labeled Flare 1 and Flare 2, occurred. The purple horizontal bar shows the time period corresponding to the middle panel. Middle: A zoom-in of the 400 s period around Flare 1. 12 light curves are shown, each corresponding a 100 \AA\ wide spectral bin from 2000-3200 \AA. Note that the vertical axis is logarithmic to better visualize the wide range of flux enhancements seen over the STIS bandpass. Right: The average quiescent stellar spectrum (black) is compared to the stellar spectrum during a 27-second time period encompassing the brightest part of Flare 1 (red).}
    \label{fig:GJ832_STIS_flare_fig1}
\end{figure*}

We observed GJ 832 and GJ 1061 with HST for 5 and 4 orbits, respectively. Like the \textit{Swift} observations, the HST/STIS observations are also broken into segments of 30-45 minutes. Examining the light curves by eye, we identified two flares from GJ 832 (both occurred in the 4th orbit, obsid: odsx01020) and no flares in the GJ 1061 data. Figure~\ref{fig:GJ832_STIS_flare_fig1} shows the two GJ 832 flare light curves and the spectrum of GJ 832 during quiescence and at the peak of the first flare. We computed a mean quiescent spectrum by averaging all five orbits. Note that the extremely short duration of the flares means that they have a negligible impact on the cumulative spectrum. We extracted flare spectra from obsid odsx01020 using the \texttt{inttag} and \texttt{x1d} functions with a start time of 988 s and a duration of 27 s for the first flare and a start time of 2019 s and duration of 23 s for the second flare. Focusing on the 2100-3100 \AA\ bandpass (i.e., ignoring the noisiest regions of the STIS spectrum), we determine that the first flare reached a peak relative amplitude 1.31$\times$ and the second flare reached 1.12$\times$. The sharp rise and decay of each flare lasted approximately 60 seconds and 40 seconds, respectively. The first flare exhibits an extended decay phase of over 10 minutes and then the second flare occurred soon after. Simultaneous 2-min TESS data show no evidence of either flare. We measure the two flares' equivalent durations in the 2100-3100 \AA\ bandpass to be 22.4 s and 5.6 s. We calculate the flare energies following Equation~\ref{eq:NUV flare energy} and find log$_{10}$ E$_f$ = 29.3 erg and 28.7 erg. Note that the 2100-3100 \AA\ wavelength range of this calculation is slightly different than the nominal UVM2 bandpass (1700-2960 \AA).
\subsection{Identifying White-Light/{\normalfont NUV} Flare Counterparts}\label{subsection:optical/nuv counterparts}
To identify the optical counterparts to NUV flares, we visually examined the TESS/K2 and \textit{Swift}/UVOT light curves after synchronizing the observation times of TESS/K2 and \textit{Swift}/UVOT on a unified time system. Among the detected 213 NUV flares, we identified a sample of 51 flares which have optical counterparts. Out of those 51 NUV flares, 45 were observed for the full duration by \textit{Swift}/UVOT. Furthermore, among the 213 NUV flares, 23 did not have simultaneous TESS data. When these flares occurred, either the corresponding stars were outside of TESS's field of view or TESS was downlinking data and not observing. Therefore, we found that $\sim$27\% of NUV flares have optical counterparts.

The details of all NUV flares which have optical counterparts can be found in Appendix \ref{appendix:NUV flares} 
and those of corresponding optical counterparts can be found in Appendix \ref{appendix:TESS flares}.  
In Appendix \ref{appendix:NUV flares}, 
the NUV flares observed for their full duration and having detected optical counterparts are indicated by a dagger ($\dag$) sign in the third column labeled $N$, while those not observed for their full duration but still having optical counterparts are indicated by a double dagger ($\ddag$) sign. Likewise, an asterisk (*) signifies that simultaneous TESS/K2 data was unavailable for a given NUV flare.

In Figure \ref{fig:example,simultaneous flare}, we present two large flares which were observed simultaneously with both TESS and \textit{Swift}/UVOT on two of our targets: AP Col (left panel) and YZ CMi (right panel).
\begin{figure*}
   \centering
   \includegraphics[width=0.40\textwidth]{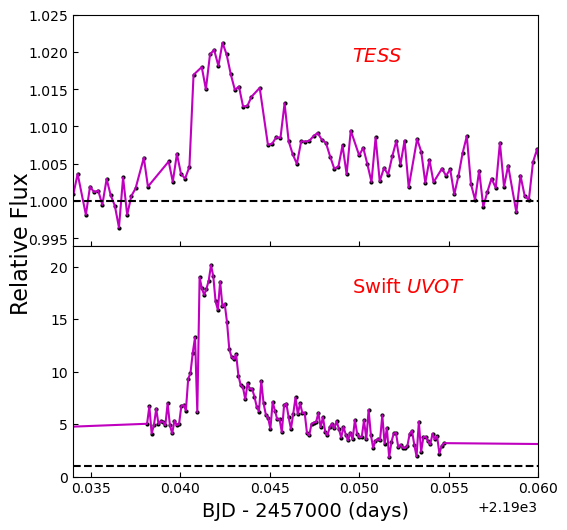}
   \vspace{0.00mm}
   \includegraphics[width=0.40\textwidth]{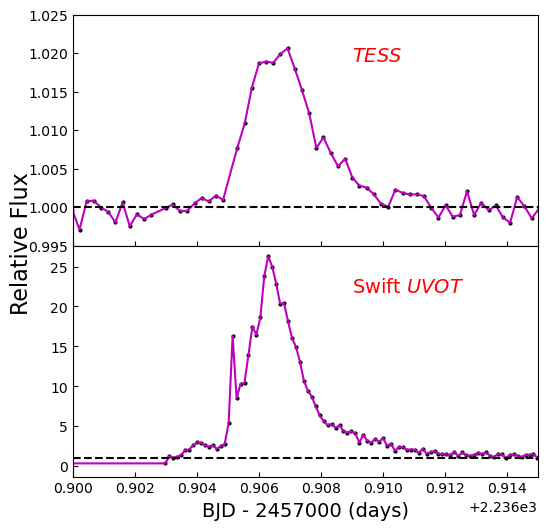}
    \caption{Flares observed simultaneously with both TESS and \textit{Swift}/UVOT on two of our targets: AP Col (left panel) and YZ CMi (right panel) during TESS Sector 32 and 34 respectively. Both stars were observed by TESS in 20-second cadence mode and the displayed cadence of the \textit{Swift}/UVOT data is $\sim$10 seconds. The black dashed line represents the quiescent level of each star in the corresponding band. The duration of optical flare observed on AP Col is comparable to that of its NUV counterpart. Based on the optical counterpart, we can confirm that \textit{Swift}/UVOT observed the NUV flare on AP Col for almost the entire duration, missing only a small portion of both the rise and decay phases. On the other hand, the duration of NUV flare observed on YZ CMi is greater than its optical counterpart. The optical flares have peak fluxes $\sim$2\% above the quiescent level and the NUV flares have peak fluxes a factor of $\sim$20-25 above quiescence.}
    \label{fig:example,simultaneous flare}
\end{figure*}

The total number of NUV flares that have simultaneous TESS data but do not have optical counterpart is 135 among which 8 were not observed for full duration by Swift/UVOT. So, we establish an upper limit on the optical flare energies for the remaining 127 NUV flares. We identified the TESS cadences closest to the NUV flare start times using the criterion that the maximum difference between the closest TESS cadence and start time of a given NUV flare is less than size of single TESS cadence (i.e., 20 s or 120 s). We then calculated the maximum equivalent duration for the flare. The equivalent duration is the product of the maximum (flare) flux above quiescence at those TESS cadences (oftentimes a single TESS cadence for short duration NUV flares) and the duration of a TESS cadence (20 s or 120 s). The maximum (flare) flux above the quiescent level is the difference of the quiescent flux and the sum of the observed flux and its corresponding uncertainty.
The equivalent durations were then calibrated to flare energies using bolometric flare luminosities listed in Table \ref{table:rotation periods}. There are 11 cases for which the maximum flux in TESS light curve is slightly below the quiescent level, which makes the equivalent duration is negative. We did not take into account such cases for further analysis of energy fractionation.

\begin{figure*}
   \centering
   \includegraphics[width=0.30\textwidth]{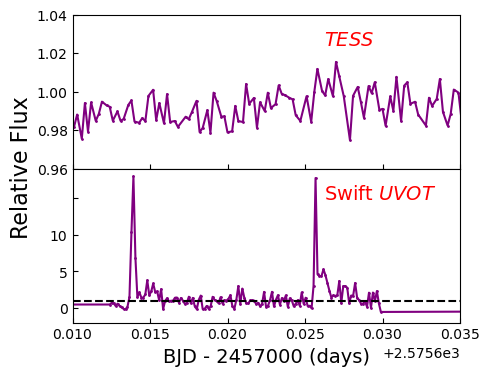}
   \vspace{0.00mm}
   \includegraphics[width=0.30\textwidth]{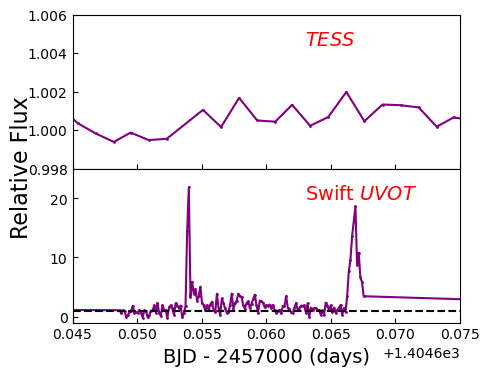}
   \vspace{0.00mm}
   \includegraphics[width=0.30\textwidth]{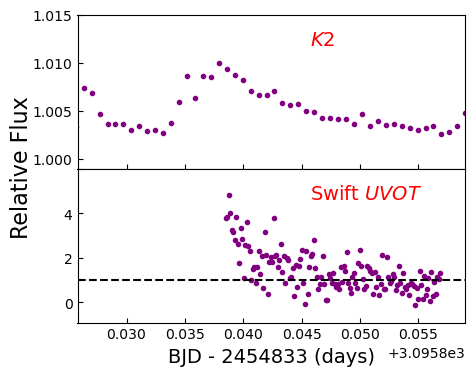}
    \caption{\textit{Left Panel}: NUV flares (lower plot) undetected in the TESS 20-second light curve of GJ 3631 (upper plot).  \textit{Middle Panel}: NUV flares (lower plot) undetected in the TESS 2-min light curve of YZ Ceti (upper plot). \textit{Right Panel}: Elevation in the NUV flux of Wolf 359 (lower plot) during a flare in the $K2$ 1-min light curve (upper plot). The horizontal dashed line in each lower plot represents the NUV quiescent level of a given star.}
    \label{fig:NUV flare undetected}
\end{figure*}

We demonstrate four NUV flares for which no optical counterparts were identified in the 20-second and 2-minute TESS light curves in the left and middle panels of Figure \ref{fig:NUV flare undetected}. From the two plots, we see that the flares are undetected because they are either diluted by the noise in the high cadence light curve or suppressed due to a longer cadence size.

We list the number of NUV flares for which optical counter parts were identified in TESS/K2 light curves for each spectral bin and energy bin mentioned above in Table \ref{table:flares_by_SpT_and_E-bin}. We find that 0\%, 22.2\% and 50\% of NUV flares observed on M0-M3 had optical counterparts in the energy bins (10$^{27}$ - 10$^{29}$) erg, (10$^{29}$ - 10$^{31}$) erg and (10$^{31}$ - 10$^{33}$) erg respectively. Likewise, 18\%, 24.5\%  and 100\% of the NUV flares observed on M4-M6 dwarfs had optical counter parts in the energy bins (10$^{27}$ - 10$^{29}$) erg, (10$^{29}$ - 10$^{31}$) erg and (10$^{31}$ - 10$^{33}$) erg respectively. Overall, 28\% and 23.4\% of NUV flares observed on M0-M3 and M4-M6 dwarfs had optical counter parts respectively. 

We also estimated the number of optical flares observed on our targets during the \textit{Swift}/UVOT observation windows, which had NUV counterparts. We found that a total of 65 optical flares were observed by TESS/K2 simultaneously with \textit{Swift}/UVOT. We confirmed this by visually inspecting the light curves. It is interesting to note that all except two of them had NUV counterparts. In addition to the 51 NUV flares discussed above, Swift/UVOT observed a very small fraction of the rise/decay phase of an additional 14 flares during the start/end of a given observing window. This is confirmed by an elevation in the NUV flux in the light curves of the stars and is demonstrated in the right panel of Figure \ref{fig:NUV flare undetected}. Such flares are not used for further analysis in this paper since we have very limited information about their properties. However, taking into account those 14 additional events, the total number of NUV flares we observed on our stars is 227, among which 65 ($\sim$ 29\%) had optical counterparts. 

 
Surprisingly, the likelihood of an NUV flare occurring alongside an optical flare is very high. Conversely, the probability of an optical flare accompanying an NUV flare is notably lower, around 28\%. Examples of NUV flares lacking optical counterparts, as depicted in the left and middle panels of Figure \ref{fig:NUV flare undetected}, indicate that the combination of the cadence of optical sampling and the relatively low flare amplitude may be the primary factors limiting the detection of optical flare counterparts. In the provided examples, there are observable signs of flux increase in the TESS light curves during NUV flares. However, these signs do not manifest as distinct flares in the light curves, as they are diluted within the noise of the light curves.


\begin{table*}
\scriptsize
    \caption{Target List}
 	\centering
     \begin{tabular}{cccccccccccccccc}
     \hline
     Name & Other name(s) & TIC & $T_{\rm mag}$ & Dist. & Ref. & $T_{\rm eff}$ & SpT & Ref. & Planet Host & MS & Age & Ref. \\
     \hline
     & & & mag & pc & & K & & & & & & \\ 
     \hline
     AP Col & LP 949-15 & 160329609 & 9.66 & 8.66 & 1 &  2998 & M4.5Ve & 1,2 & & & 40-50 Myr (Argus) & 2,47  \\
     AU Mic & HD 197481 & 441420236 & 6.76 & 9.72 & 1 & 3700 & M1Ve & 3 & Yes (2) & & 24$\pm$3 Myr ($\beta$Pic) & 49 \\
     AX Mic$^{\dag}$ & GJ 825 & 159746875 & 5.12 & 3.97 & 1 & 3729 & M0V & 4 & & & 4.8 Gyr & 39 \\
     DG CVn & G 165-8 & 368129164 & 9.29 & 18.29 & 5 & 3175 & M4Ve & 6 &  & Bin. & $\sim$20-200 Myr & 40 \\
     DX Cnc & LHS 248 & 3664898 & 10.50 & 3.58 & 1 & 2814 & M6.5V & 1,7 & & & 0.71 Gyr & 42 \\
     EV Lac & GJ 873,LHS 3853 & 154101678 & 7.73 & 5.05 & 1 & 3270 & M3.5e & 8 & & &125-800 Myr & 8 \\
     EZ AQr$^{\dag}$ & GJ 866 & 402313808 & 8.30 & 3.40 & 9 & 3030 & M5e & 10 &  & Tert. & 0.87 Gyr & 42 \\
     Fomalhaut C & LP 876-10 & 47423224 & 9.78 & 7.67 & 1 & 3132 & M4V & 11 & & Tert. & 440$\pm$40 Myr & 43  \\
     GJ 1061$^{\dag}$ & LHS 1565& 79611981 & 9.47 & 3.67 & 1 & 2905 & M5.5V & 1,12 &Yes (3) & & $>$7.0$\pm$0.5 Gyr & 44\\
     GJ 1284 & HIP 116003 & 9210746 & 8.73 & 15.95 & 1 & 3406 & M2Ve & 1,13 & & Bin. & 110-800 Myr & 45 \\
     GJ 205$^{\dag}$ & HD 36395 & 50726077 & 5.97 & 5.70 & 1 & 3700 & M1.5 & 14 & & & 2.57 Gyr & 22 \\
     GJ 3631 & LHS 2320 & 281744438 & 11.7 & 13.9 & 1 & 3145 & M5V & 37,34 & & & 0.72 Gyr & 42\\
     GJ 4353$^{\dag}$ & 	HIP 116645 & 224270730 & 9.80 & 20.71 & 1 & 3542 & M2 & 15 & & & 0.89 Gyr & 42 \\
     GJ 832 & 	HIP 106440 & 139754153 & 6.68 & 4.96 & 1 & 3590 & M2 & 16 & Yes (2) & & 8.4 Gyr & 46\\
     GJ 1 & HD 225213 & 120461526 & 6.67 & 4.35 & 1 & 3575 & M1.5 & 17 & & & 5.7 Gyr & 42\\
     GJ 876 & 	HIP 113020 & 188580272 & 7.58 & 4.68 & 1 & 3272 & M4.0 & 18 & Yes (4) & &8.4$^{2.2}_{-2.0}$ Gyr &  48\\
     HIP 17695 & G 80-21 & 333680372 & 9.32 & 16.79 & 1 & 3393 & M4 & 19 & & & 149$^{+51}_{-19}$ Myr (ABDor) & 49 \\
     HIP 23309$^{\dag}$ & CD-57 1054 & 220473309 & 8.46 & 26.86 & 1 & 3886 & M0 & 19 & & & 24$\pm$3 Myr ($\beta$Pic) & 49\\
     Kapteyn's Star & HD 33793 & 200385493 & 7.05 & 3.93 & 1 & 3570 & M1.5V  &  20,21 & Yes (2) & & 11.5$^{+0.5}_{-1.5}$ Gyr & 50 \\
     Lacaille 9352 & GJ 887, HD 217987 & 155315739 & 5.57 & 3.29 & 1 & 3676 & M1V & 1,22 &  Yes (2) & & 2.9 Gyr & 22 \\
     LHS 292$^{\dag}$ & GJ 3622 &55099399 &  11.34 & 4.56 & 1 & 2784 & M6.5 & 17 & & & 0.75 Gyr & 42 \\
     LP 776-25 & NLTT 14116 & 246897668 &  9.24 & 15.83 & 1 & 3414 & M3.0V & 1,23 & & & 149$^{+51}_{-19}$ Myr (ABDor) & 51,49 \\
     Luyten's Star & GJ 273 & 318686860 & 7.31 & 5.92 & 1 & 3382 & M3.5V & 24,25 & Yes(4) & & $>$8 Gyr & 52 \\
     NLTT 31625$^{\dag}$ & L 471-3 & 144490040 &  9.75 & 28.35 & 1,26 & 3594 & M3 & 27 & & & 0.27 Gyr & 42 \\
     Proxima Cen. & GJ 551 & 1019422535 & 7.58 & 1.30 & 1 & 3050 & M5.5V & 28 &  Yes (2) & & 4.85 Gyr & 41 \\
     Ross 614 & GJ 234 A & 711366839 & 8.35 & 4.12 & 1 & 3154 & M4.5Ve & 5,29 &  & Bin. & 0.76 Gyr & 42 \\
     Twa 22 & 	SSSPM J1017-5354 & 272349442 & 10.55 & 19.81 & 1,26 & 3000 & M5 & 30 &   & Bin. & 24$\pm$3 Myr ($\beta$Pic) & 54,49 \\
     UV Ceti$^{\dag}$ & GJ 65B & 632499595 & 9.40 & 2.69 & 1 & 2728 & M6.0 & 31,32 & & Bin. & $>$0.5 Gyr & 53 \\
     V1005 Ori & HIP 23200 & 452763353 & 8.43 & 24.38 & 1 & 3866 & M0.5 & 1,33  & & & 24$\pm$3 Myr ($\beta$Pic) & 55,49\\
     Wolf 359 & CN Leo, GJ 406 & 365006789 & 9.28 & 2.41 & 1,26 & 2800 & M6V & 34 & & & 0.1-1.5 Gyr & 56 \\
     Wolf 424 & GJ 473 & 399087412 & 9.00 & 4.48 & 1 & 3013 & M5.0 & 17 &  & Bin. & 0.70 Gyr & 42 \\
     WX UMA$^{\dag}$ & LHS 39, GJ 412 B & 252803606 & 10.71 & 4.90 & 1 & 2900 & M5.5V & 34 & & Bin. & 3.0 Gyr & 22 \\
     YZ Ceti & HIP 5643 & 610210976	 &  12.30 & 3.72 & 1 & 3056 & M4.5 & 35 & Yes (3) & & 3.8$\pm$0.5 Gyr & 42\\
     YZ CMi & GJ 285, HIP 37766 & 266744225 & 8.34 & 5.99 & 1 & 3100 & M4.5e & 1,36 & & & 0.81 Gyr & 42 \\
    \hline
     \end{tabular}
     \\
     \textbf{Note:} A $\dag$ sign added with the name of star indicates that either it was undetected in \textit{Swift}/UVOT data or its \textit{Swift}/UVOT data was not obtained in \texttt{EVENT} mode. The column `MS' indicates whether the target is part of a multiple-star system.\\
     \textbf{References:}\\
        1) \cite{2019AJ....158..138S}; 2) \cite{2011AJ....142..104R}; 3) \cite{2020Natur.582..497P}; 4) \cite{2013ApJ...771...45D}; 5) \cite{2019AJ....158...56H};
        6) \cite{2003ApJ...583..451M}; 7) \cite{2013MNRAS.431.2063S}; 8) \cite{2021ApJ...922...31P}; 9) \cite{1997A&A...320..865M}; 10) \cite{2019ApJ...879..105L};
        11) \cite{2013AJ....146..154M}; 12) \cite{2002AJ....123.2002H}; 13) \cite{2006A&A...460..695T}; 14) \cite{2018A&A...620A.180R}; 15) \cite{2014MNRAS.443.2561G};
        16) \cite{2016ApJ...822...97H}; 17) \cite{2013MNRAS.431.2063S}; 18) \cite{2016MNRAS.463.1844S}; 19) \cite{Pineda_2021}; 20) \cite{2014MNRAS.443L..89A}; 21) \cite{1997AJ....113..806G}; 22) \cite{2015ApJ...804...64M}; 23) \cite{2018A&A...614A..76J}; 24) \cite{2012ApJ...757..112B}; 25) \cite{1996AJ....112.2799H}; 
        26) \cite{2021AJ....161..147B}; 27) \cite{2014MNRAS.443.2561G}; 28) \cite{2016Natur.536..437A}; 29) \cite{1991ApJS...77..417K}; 30) \cite{2014A&A...562A.127B}; 31) \cite{2018ApJ...860...15M}; 32) \cite{1974ApJS...28....1J}; 33) \cite{2008MNRAS.390..545D}; 34) \cite{2020A&A...642A.115C}; 35) \cite{2017A&A...605L..11A};
         36) \cite{2020A&A...641A..69B}; 37) \cite{2021A&A...645A.100S}; 38) \cite{2014MNRAS.445.2169M}; 39) \cite{2013ApJ...768...25G}; 40) \cite{2014AJ....147...85R}; 41) \cite{2003A&A...404.1087K};
         42) \cite{2023ApJ...954L..50E}; 43) \cite{2012ApJ...754L..20M}; 44) \cite{2020MNRAS.493..536D};
         45) \cite{2021arXiv210901624C}; 46) \cite{2009ApJ...705.1226B};
         47) \cite{2019ApJ...870...27Z}; 48) \cite{2018ApJ...863..166V}; 49) \cite{2015MNRAS.454..593B};
         50) \cite{2016ApJ...821...81G}; 51) \cite{2017AJ....154..151B};  52) \cite{2020A&A...641A..23P};
         53) \cite{2016A&A...593A.127K}; 54) \cite{2014ApJ...783..121G}; 55) \cite{2012AJ....143...80S}; 
         56) \cite{2023AJ....166..260B}\\
         \label{table:targets info}
\end{table*}
\begin{table*}
    \caption{Summary of TESS, \textit{Swift}/XRT and \textit{Swift}/UVOT observations}
    \begin{tabular}{cccccccc}
    Name & TESS Sector & TESS cadence & TESS obs & $N_{T}$ & XRT obs & UVOT obs & $N_{\rm NUV}$ \\
     \hline
     & & & [d] & & [ks] & [ks] & \\
     \hline
    AP Col & 6 & 2 min & 20.6 &  65 & 36.0 & 36.6 & 10 \\
            & 32 & 20 s  & 23.5 & 102 & 24.5 & 26.6 & 6 \\
      AU Mic & 1 &2 min & 25.1 &44 & 48.1 & 48.0 & 5  \\
                & 27 & 20 s & 22.6 & 71  & 20.8 & 17.2 & 1 \\
    DG CVn & 23 &2 min & 18.7 & 96 & 24.5 & 25.5 & 3 \\
    DX Cnc & 21 & 2 min& 23.1 &27 & 30.0 & 29.6 & 0 \\
      EV Lac & 16 &2 min & 23.2 &56 &  18.0 & 18.0 & 9 \\
     Fomalhaut C &2 &2 min & 25.0 & 62 & 42.1 & 32.8 & 8\\
     GJ 1284 & 2 &2 min & 25.4 & 60 & 35.7 & 9.9 & 1 \\
     GJ 3631 & 46 &20 sec & 22.2 & 82 & 85.4 & 86.1 & 16\\
     GJ 832 & 1 & 2 min& 25.4 & 0 & 39.0 & 38.7 & 0 \\
     GJ 1 & 2 & 2 min& 25.4 & 0  & 29.5 & 29.2 & 0 \\
     GJ 876 & 2 & 2 min & 25.4 & 0 & 38.1 & 37.9 & 0 \\
      HIP 17695 & 5 & 2 min& 24.1 & 72 & 33.7 & 33.5 & 2 \\
      Kapteyn's Star &4 &2 min &  20.5 & 0 & 35.1 & 34.9 & 0 \\
      Lacaille 9352 & 2 & 2 min& 25.4 & 2 & 37.9 & 37.9 & 1  \\
    LP 776-25 & 5 &2 min & 24.1 & 75 & 34.8 & 30.6 & 3 \\
    Luyten's star & 7 &2 min & 22.7 & 1 & 35.0 & 34.9 & 1\\
      Prox. Cen. & 11,12 & 2 min& 47.7 & 134  & 82.9 & 64.9 & 20  \\
     Ross 614 & 6 &2 min & 20.4  & 60  & 34.7 & 34.0 & 22 \\
      TWA 22 &  9& 2 min & 22.7 & 78 & 37.1 & 37.3 & 6  \\ 
      V1005 Ori & 5& 2 min & 24.0 & 35  & 32.6 & 34.2 & 4 \\
      Wolf 359 & 46& 20 sec & 22.2 & 151 & 123.9 & 121.1 & 33 \\
      Wolf 424 & 23  &2 min & 18.1 & 63 &  13.0 & 12.9 & 5\\
      YZ Ceti & 3 &2 min & 17.9 & 63 & 36.6 & 36.5 & 11 \\
        & 30 &20 sec & 21.6 & 91 & 15.0  & 15.0 & 0\\
      YZ CMi & 7& 2 min& 22.7 & 123 & 37.1 & 37.0 & 21 \\
             & 34 & 20 sec& 22.3 & 157 & 21.6 & 17.4 & 10\\ 
    \hline
     \end{tabular}
     \tablecomments{$N_{T}$ and $N_{\rm NUV}$ stands for the number of flares identified in the TESS and UVOT data respectively.}
     \label{table:TESS and UVOT info}
\end{table*}
\begin{table}
    \caption{Quiescent X-ray and NUV luminosities of \\
    targets in our sample. $L^{'}_{\rm NUV}$ is the excess NUV luminosity.}
     \begin{tabular}{cccc}
     \hline
     Name & log $L_{\rm X}$ & log $L_{\rm NUV}$ & log $L^{'}_{\rm NUV}$ \\
      &  [erg s$^{-1}$] & [erg s$^{-1}$] & [erg s$^{-1}$]\\
     \hline
     AP Col & 0.17 & 27.45 & 27.45 \\
      AU Mic  & 1.6 & 28.75  & 28.75 \\
      DG CVn  & 0.19 & 28.40 & 28.40 \\
      DX Cnc  & 0.035 & 27.10 & 27.10 \\
      EV Lac  & 1.0 & 27.90 & 27.90 \\
      Fomalhaut C  & 0.10 & 27.00 & 27.0 \\
      GJ 1284  & 0.29 & 28.30 & 28.30  \\
      GJ 3631 & 0.057 & 26.90 & 26.90 \\
      GJ 832 & 0.02 & 26.90 & 26.88 \\
      GJ 1 & 0.0098 & 26.70 & 26.68 \\
      GJ 876  & 0.0096 & 26.30 & 26.29 \\
      HIP 17695  & 0.21 & 28.30 & 28.30 \\
      Kapteyn's Star & 0.0077 & 26.50 & 26.48 \\
      Lacaille 9352 & 0.03 & 27.30 & 27.29 \\
      LP 776-25 & 0.25 & 28.20 & 28.20 \\
      Luyten's Star & 0.0051 & 26.50 & 26.49 \\
      Prox. Cen. & & 26.0  & 26.0 \\
      Ross 614 & 0.32 & 27.10 & 27.10 \\
      TWa 22 & 0.10 & 27.80 & 27.80 \\
      V1005 Ori & 0.38 & 29.0 & 29.0 \\
      Wolf 359 & 0.13 & 26.10 & 26.10 \\
      Wolf 424 & 0.22 & 27.10 & 27.10  \\
      YZ Ceti & 0.10 & 26.20 & 26.20 \\
      YZ CMi & 0.58 & 27.90 & 27.90 \\
     \hline
     \end{tabular}
     \label{table:quiescent flux}
\end{table}

\begin{table*}[h!]
    \caption{Rotation periods, bolometric flare luminosity ($L_{\rm bol,flare}$) assuming 10,000 K BB and radii of targets}
 	\centering
     \begin{tabular}{ccccccc}
     \hline
     Name & Period & $P_{err}$  & Reference & $L_{\rm bol,flare}$  & Radius & Reference \\
     & [d] & [d] &  &  [10$^{30}$ erg/s] & [$R_{\odot}$] &  \\
     \hline
      AP Col & 1.016 & 0.020  & this work & 7.57 &  0.29 & 14 \\
      AU Mic & 4.872 & 0.361  & this work &161 & 0.75 & 15 \\
      DG CVn & 0.268 & 0.001 & this work & 20.2 & 0.46 & 16 \\
      DX Cnc & 0.459 & 0.003 & this work & 0.88 & 0.12 & 14 \\
      EV Lac & 4.319 & 0.337 & this work & 18.6 & 0.35 & 17 \\
      Fomalhaut C & 0.318 & 0.002 & this work & 6.16 & 0.23  & 14 \\
      GJ 1284 & 7.335 & 0.946 & this work, 1 & 57.4 & 0.56 & 18  \\
      GJ 3631 & 0.6920 & 0.0077 & this work & 4.31 & 0.19 & 14  \\
      GJ 832 & 55 & - & 2 &  \\
      GJ 1 & 60.1 & 5.7 & 5 &  \\
      GJ 876 & 96.7 & - & 6 & \\
      HIP 17695 & 3.860 & 0.2452 & this work & 45.6 & 0.50 & 19 \\
      Kapteyn's Star & 124.71 & 0.19 & 7 &  \\
      Lacaille 9352 & $\sim$200  & - & 8 &  61.6 & 0.47 & 20 \\
      LP 776-25 & 1.3708 & 0.031 & this work & 38.2 & 0.45 & 14 \\
      Luyten's Star & 99 &  & 2 & 15.4 & 0.29 & 3 \\
      Prox. Cen. & 82.600 & 0.100 & 10 & 1.98 & 0.14 & 4 \\
      Ross 614 & 1.585 & 0.049 & this work & 9.23 & 0.26 & 14 \\
      TWa 22 & 0.732 & 0.009 & this work & 15.3 & 0.41 & 18 \\
      V1005 Ori & 4.379 & 0.337 & this work & 142 & 0.63 & 9 \\
      Wolf 359 & 2.72  & 0.04 & Lin et al. 2021 & 1.16 & 0.14 & 14  \\
      Wolf 424 & 0.2073 & 0.001 & this work & 2.09  & 0.15 & 9 \\
      YZ Ceti & 67 & 0.8 & 11 & 2.85 & 0.17 & 12 \\
      YZ CMi & 2.774 & 0.133 & this work & 15.0 & 0.37 & 13 \\
     \hline
     \end{tabular}
     \\
        \textbf{References:}\\
        1) \cite{2021arXiv210901624C}; 2)\cite{2017A&A...600A..13A};\\
        3) \cite{2015ApJ...800...85N}; 4) \cite{2012ApJ...757..112B};
            5) \cite{2015MNRAS.452.2745S}; 6) \cite{2010A&A...511A..21C}; 
            7)\cite{2021AJ....161..230B}; 8) \cite{2020Sci...368.1477J}; 
            9) \cite{2019AJ....158...56H}; \\ 10) \cite{2017A&A...602A..48C}; 
            11)\cite{2017ATel10678....1E}; 12) \cite{2015ApJ...804...64M}; 
            13) \cite{2020A&A...641A..69B}; 14) \cite{2021A&A...645A.100S};
            15) \cite{2020Natur.582..497P}; 16) \cite{2016ApJ...832..174O};
            17) \cite{2021ApJ...922...31P}; 18) \cite{2021arXiv210804778P};
            19) \cite{2021ApJ...911..111P}; 20) \cite{2020Sci...368.1477J}; 21) \cite{2015ApJ...800...85N};
            22) \cite{2012ApJ...757..112B}; 
            23) \cite{2019AJ....158...56H}; 24) \cite{2015ApJ...804...64M};
            25) \cite{2020A&A...641A..69B} \\
            \label{table:rotation periods}
\end{table*}

\begin{table*}
\centering
\caption{Number of NUV/optical flares in each spectral and energy bin. Fifth and sixth rows provide information regarding the number of NUV flares which had optical counterparts in a given energy bin.}
\begin{tabular}{|c|cc|cc|cc|cc|cc|}
\hline
Sp. Type & \multicolumn{2}{c|}{10$^{27}$-10$^{29}$ erg} &\multicolumn{2}{c|}{10$^{29}$-10$^{31}$ erg} & \multicolumn{2}{c|}{10$^{31}$-10$^{33}$ erg}&  \multicolumn{2}{c|}{$>$10$^{33}$ erg} & \multicolumn{2}{c|}{Overall} \\
\hline
 & $N_{\rm NUV}$ & $N_{\rm optical}$ & $N_{\rm NUV}$ & $N_{\rm optical}$& $N_{\rm NUV}$ & $N_{\rm optical}$& $N_{\rm NUV}$ & $N_{\rm optical}$ & $N_{\rm NUV}$ & $N_{\rm optical}$ \\
\hline
M0-M3 & 1&0 &18 & 85&6 &250 & 0 & 9 &25 &344 \\
M4-M6 &78 & 79 &106 &858 &4 & 287& 0& 3 & 188 &1227 \\
\hline
\multicolumn{11}{|c|}{Number of  NUV flares with optical counterparts}   \\
\hline
M0-M3 & \multicolumn{2}{c|}{0}  &\multicolumn{2}{c|}{4}  &\multicolumn{2}{c|}{3}  &\multicolumn{2}{c|}{0} & \multicolumn{2}{c|}{7}  \\
M4-M6 & \multicolumn{2}{c|}{14} &\multicolumn{2}{c|}{26}  &\multicolumn{2}{c|}{4}  &\multicolumn{2}{c|}{0} & \multicolumn{2}{c|}{44} \\
\hline
\end{tabular}
\label{table:flares_by_SpT_and_E-bin}
\end{table*}

%
%
\section{Results} \label{section:results}
\subsection{Flaring Rates and Durations}
\begin{figure*}
   \centering
\includegraphics[width=\textwidth]{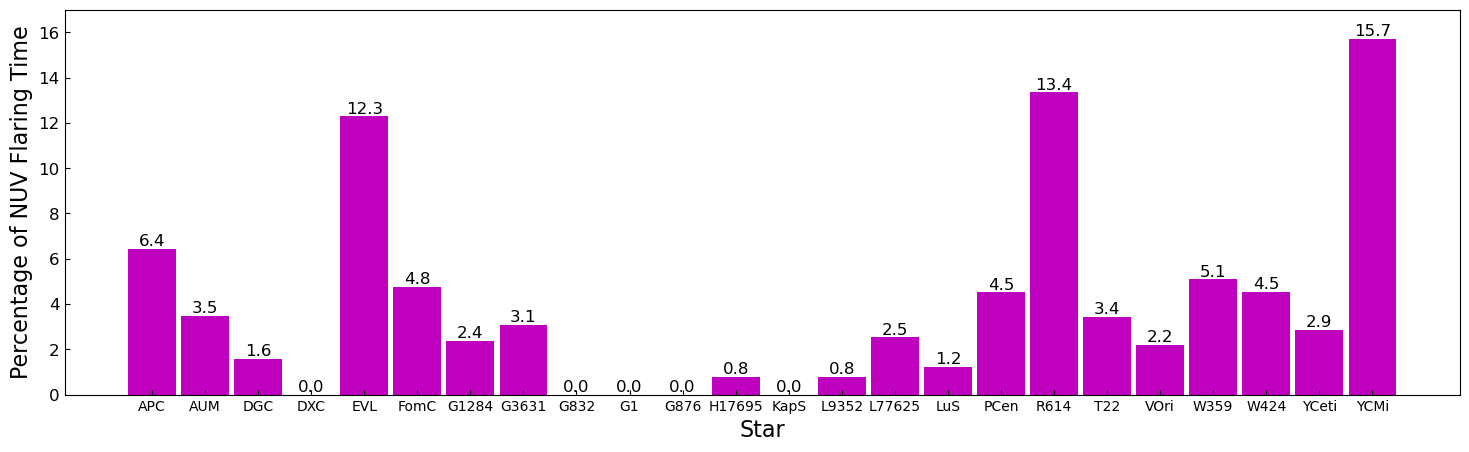}
    \caption{\textit{Swift} NUV flaring fraction of observing time for each star in the sample. The names of stars are given in abbreviated form and are in the same order as in Table~\ref{table:targets info}.}
    \label{fig:flaring fraction}
\end{figure*}

In Figure~\ref{fig:flaring fraction} we present the NUV flaring fraction time for each star in our sample based on the duration of all flares observed (listed in Table \ref{table:UV flare properties})  and total observation time (listed in Table \ref{table:TESS and UVOT info}) of that star. 
EV Lac, Ross 614 and YZ CMi were in a flaring state for $>$10\% of the observing time, a much greater percentage than the other stars. 
On the other hand, five stars (DX CNc,  GJ 1, GJ 832, GJ 876 and Kapetyn's star) did not flare. This result applies to \textit{Swift} observations only. On average, the NUV flaring time is 2.1\% for M0-M3 dwarfs and 5\% for M4-M6 dwarfs.  However, this fraction also depends on the age of the stars since they tend to show different flaring behaviours in various ages.

Almost all NUV flares in this study were observed on known active M dwarfs. However, the occurrence of two NUV flares on the optically-inactive M dwarf GJ 832 (approximately 8.4 Gyr old) in HST/STIS band 
indicates that even inactive M dwarfs can produce NUV flares. This is in agreement with the findings from X-ray and FUV wavelengths for GJ 832 and other optically-inactive M dwarfs \citep{2018ApJ...867...71L}. The detection of two short-duration flares in the HST light curve, with an exposure time of only $\sim$12 ks (shorter than most of the \textit{Swift}/UVOT observations), suggests that observing optically-inactive M dwarfs requires instruments with greater sensitivity than \textit{Swift}/UVOT. Such stars exhibit very low count rates in \textit{Swift}/UVOT light curves, which limits our sensitivity to the energies of very small flares. 
%

\begin{figure}[htbp]
   \centering
   \includegraphics[width=0.5\textwidth]{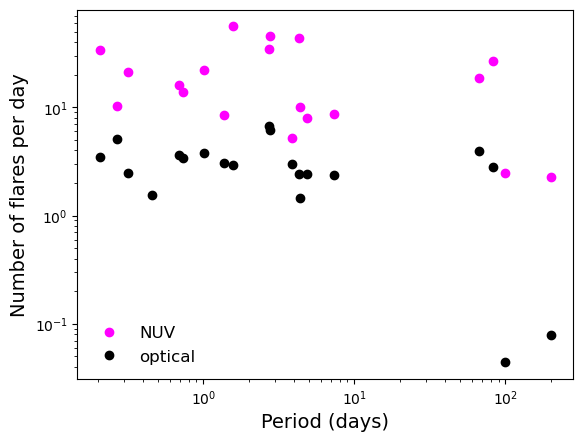}
    \caption{Flare rates as a function of rotation periods of stars. The magenta dots correspond to NUV flares and the black dots correspond to optical flares.}
    \label{fig:period vs flare rates}
\end{figure}

Next, we investigate whether there is any correlation between the rotation period and the NUV/optical flare rates of stars. Our analysis includes only those stars that exhibited at least one flare in both NUV and optical passbands. 
The flare rate of each star is calculated as the ratio of the total number of flares observed in a given passband to the total observation time in that same passband. Figure \ref{fig:period vs flare rates} shows no discernible relationship between the rotation period and flare rates for both NUV and optical passbands. Interestingly, the NUV flare rates consistently appear to be higher than the optical flare rates in all cases. However, we must consider potential biases that could affect these results, specifically related to the differences in the cadence size used for identifying flares in NUV and optical observations. Smaller cadences are crucial for detecting very small flares that last for only a few tens of seconds. Additionally, the contrast between flaring and quiescent states in much larger in the NUV than in the optical; this also aids in detection of NUV flares.

In Figure \ref{fig:nuv,tess flare duration comparison}, we explicitly compare the flare durations as measured in the NUV and optical. For this comparison, we used only the sample of flares that were observed by TESS in 20-second cadence mode and have detected NUV counterparts. The three longest-duration flares were observed on AP Col, YZ CMi and GJ 3631. The duration of NUV flares range (0.37 - 23.0) minutes, and that of TESS flares range (0.62-47.3) minutes. The black dashed line in Figure \ref{fig:nuv,tess flare duration comparison} represents the case when the durations of both optical and NUV flares are equal, and is not a fit. It is evident from the plot that many of the flares do not show similar durations between the bands; there are more flares with longer NUV flare durations than longer optical flare durations. We need a bigger sample of flares observed by using similar cadence size in both bands to verify if the NUV flares tend to have longer durations than the optical flares. We also note that \cite{2022PASJ...74..477K} found a linear duration between e-folding times of soft X-ray (SXR) and H$_{\alpha}$ light curves in the decay phase of flares for a time range of 1-10$^{4}$ seconds.
\begin{figure}[htbp!]
   \centering
   \includegraphics[width=0.40\textwidth]{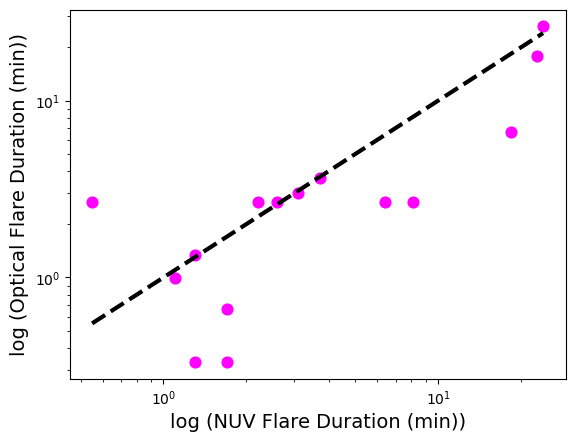}
    \caption{Comparison of NUV flare durations with those of the corresponding optical counterparts. For the most accurate comparison, we used only the sample of 15 flares which were observed by TESS in 20-second cadence mode. The black dashed line represents the case when NUV and optical flare durations are equal for each flare.}
    \label{fig:nuv,tess flare duration comparison}
\end{figure}
In Figure \ref{fig:flare duration vs nuv energy}, we plot the NUV flare durations against the corresponding NUV flare energies, limiting only to those flares that were observed for their full duration and with 20-second TESS data in order to minimize the disparity in cadence size between \textit{Swift} NUV data and TESS data. Next, we compare the trends in durations and integrated energies of all optical and NUV flares studied here with other stellar and solar flare studies in Figure \ref{fig:flare duration vs energy}. To convert NUV flare energies to bolometric energies, we utilized the ratio $E_{\rm NUV}$/$E_{\rm bol}$ = 0.217. This ratio is based on the ratio of bolometric flare energy to TESS band energy (for a 10,000 K BB) discussed in Section \ref{subsubsec:determining flare properties} and the ratio of energies in TESS and \textit{Swift} NUV (UVM2) band for best RHD model predicting NUV flare energies from TESS flare energies as discussed in Section \ref{subsubsec:best hybrid model}.\\ 
The solar and stellar flares used in Figure \ref{fig:flare duration vs energy} were observed with Solar Dynamics Observatory (SDO)/Helioseismic and Magnetic Imager (HMI) \citep{2017ApJ...851...91N}, GALEX \citep{2019ApJ...883...88B}, the Dark Energy Camera (DECam; \citealt{2021MNRAS.506.2089W}), and \textit{Kepler} \citep{2013ApJS..209....5S,2015EP&S...67...59M}. The DECam flares were observed on stars at distances up to 500 pc as part of DECam's Deeper, Wider, Faster program \citep{2019IAUS..339..135A}. However, we do not have information regarding the types of stars in the DECam sample. The $Kepler$ flares, on the other hand, were all observed on solar-type stars.\\ 
Figure \ref{fig:flare duration vs energy} also shows the expected relationship between flare duration and energy for emitting regions with a uniform magnetic field strength ($B$) but varying length scale ($L$). This is represented by forward slanted dotted lines. Likewise, the backward slanted dashed lines represent the expected relationship in the case of uniform $L$ but varying $B$ \citep{2017ApJ...851...91N}. 
These laws were formulated by equating the flare duration ($\tau$) to the reconnection timescale ($\tau_{\rm rec}$) and establishing a relationship between flare energy ($E$) and the magnetic energy ($E_{\rm mag}$) released, while eliminating either the length scale or the magnetic field strength. \\
We can see that the majority of the M dwarf NUV flares have energies and durations comparable to the solar white light flares. Some of them even have energies and durations smaller than those of solar flares, the smallest being 2.6$\times$10$^{28}$ erg and $\sim$0.3 minutes. Likewise, $\sim$50\% of M dwarf optical flares have energies greater than that of the solar white light flares. The grouping of the NUV and optical M dwarf flares around the solar flares indicate that both solar and the M dwarf flares follow a common scaling law between the flare duration and energy. This suggests a common mechanism of magnetic reconnection produces both solar and M dwarf flares. Interestingly, our M dwarf NUV and optical flares have durations comparable to the GALEX, $Kepler$ short cadence, and DEC flares, yet far lesser energies. 
\begin{figure}
   \centering
   \includegraphics[width=0.45\textwidth]{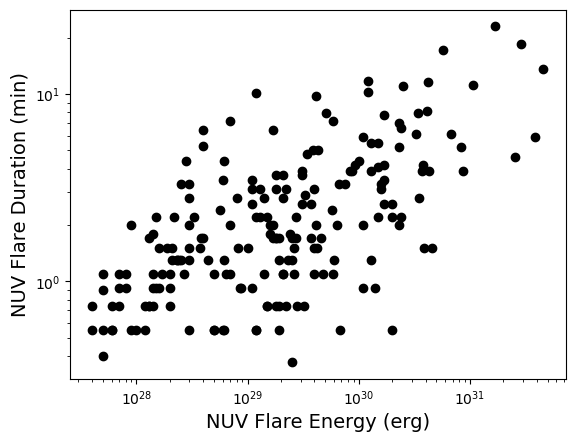}
    \caption{Duration of NUV flares versus the corresponding NUV flare energies. Only the sample of NUV flares observed for full duration is used for this plot.}
     \label{fig:flare duration vs nuv energy}
\end{figure}
\begin{figure*}[htbp!]
   \centering
   \includegraphics[width=\textwidth]{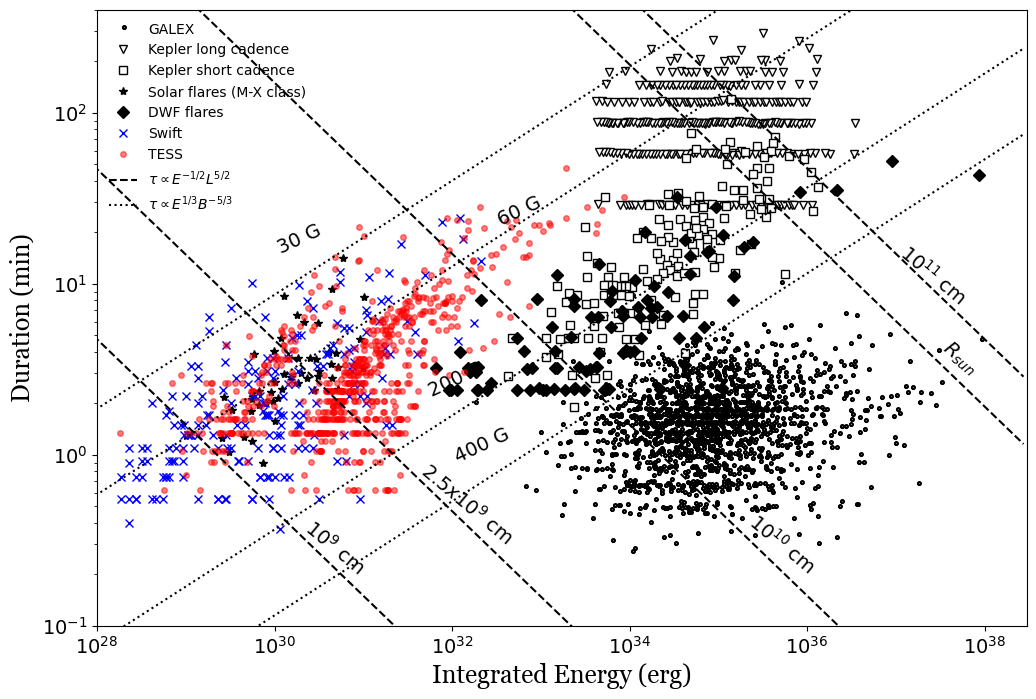}
    \caption{Comparison of bolometric energies and durations of NUV (blue crosses) and optical (red dots) M dwarf flares from this work with those of solar and stellar flares observed with SDO/HMI (black solid stars; \citealt{2017ApJ...851...91N}), GALEX (black astericks; \citealt{2019ApJ...883...88B}), DEC (solid diamonds; \citealt{2021MNRAS.506.2089W}) and \textit{Kepler} (open triangles and open squares; \citealt{2013ApJS..209....5S,2015EP&S...67...59M}). The cadence sizes are 45 s, 10 s, 20 s, and 1 or 30 min, respectively. The dashed and dotted lines correspond to scaling laws between flare durations and energies corresponding to either constant magnetic field ($B$) with varying emitting region length scale ($L$) or constant emitting region length scale with varying magnetic field.}
    \label{fig:flare duration vs energy}
\end{figure*}\\
\subsection{Flare Amplitudes}
\begin{figure*}[h!]
   \centering
   \includegraphics[width=0.45\textwidth]{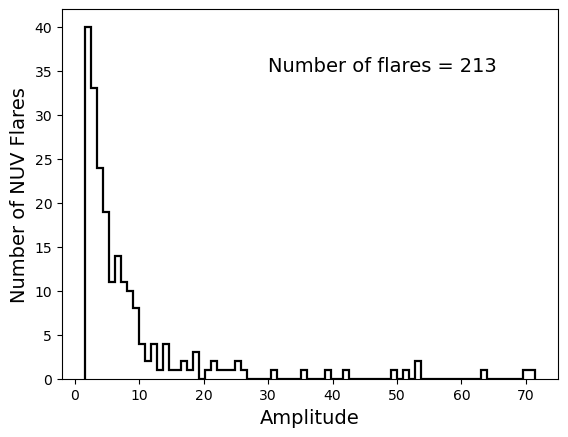}
   \vspace{0.00mm}
   \includegraphics[width=0.45\textwidth]{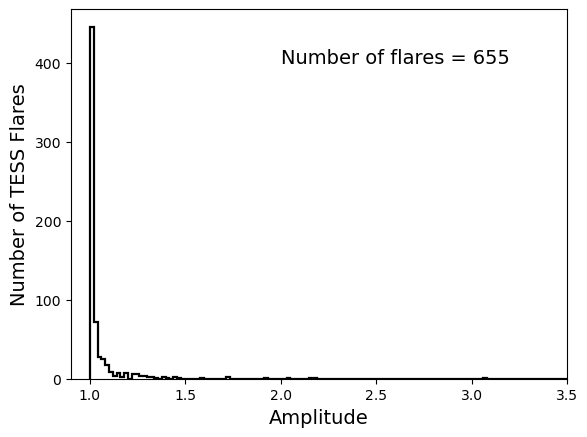}
    \caption{Histogram of flare amplitudes, i.e., the peak flare fluxes measured relative to the quiescent flux of the flaring star, measured in the NUV with \textit{Swift} (left panel) and optical with 20-second TESS data (right panel).} 
     \label{fig:flare amplitudes}
\end{figure*}

Figure \ref{fig:flare amplitudes} shows the distributions of amplitudes relative to quiescence for both NUV and TESS/optical flares. We used the sample of TESS flares observed in 20-second cadence mode to ensure greater accuracy, because flare amplitudes can be underestimated when using longer cadence data. Among the 213 NUV flares, 209 (98\%) have amplitudes greater than 1.5$\times$, and 198 (93\%) have amplitudes greater than 2$\times$ the quiescent flux. Among the 655 TESS flares, 280 (43\%) have amplitudes less than 1\%, 445 (68\%) have less than 2\%, and 539 (82\%) have less than 5\% of the quiescent flux. 
Furthermore, we estimate the mode of NUV flare amplitudes is 2.1, while the mode of optical flares is 1.004; both amplitudes are expressed relative to the corresponding quiescent levels.
This discrepancy between NUV and optical flare amplitudes arises because M dwarfs have low levels of quiescent UV fluxes, so the contrast between quiescent and flaring periods is far greater than in the optical.

\begin{figure}[htbp]
   \centering
   \includegraphics[width=0.45\textwidth]{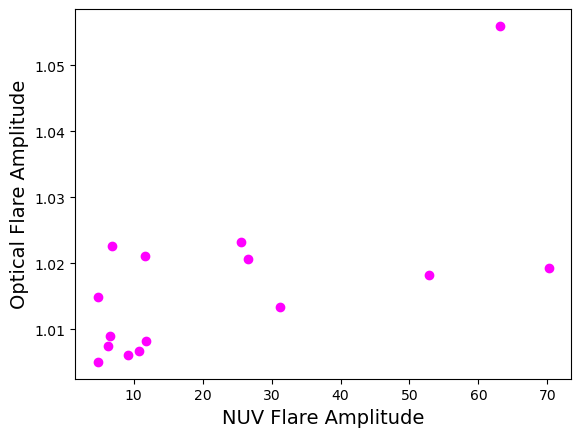}
    \caption{Comparison between amplitudes of NUV and optical flares observed simultaneously with \textit{Swift} and TESS (20-second cadence only). }
    \label{fig:flare amplitudes comparison}
\end{figure}

Next, we compare the amplitudes of flares that were observed simultaneously by both \textit{Swift} and TESS (20-second cadence only) in Figure \ref{fig:flare amplitudes comparison}. 
It is evident that all optical flares, except one, have peak fluxes that are $\lesssim$2.5\% higher than the quiescent flux. However, the NUV counterparts exhibit very large peak fluxes, reaching levels as high as $\sim$70$\times$ the quiescent flux. 

Since the mid-to-late M dwarfs and L dwarfs are redder than early M dwarfs, the optical flares exhibit higher amplitudes in the case of such stars with peak fluxes sometimes exceeding 100 times the quiescent flux (see for example, \citealt{2020MNRAS.494.5751P,2018ApJ...858...55P}). In addition to this, results in Section \ref{subsection:optical/nuv counterparts} suggest that almost all optical flares have NUV counterparts. This might be why we observed the highest number of optical counterparts (totaling 18) on Wolf 359 (an M6 dwarf) compared to the other stars in our sample. Wolf 359 also has high flare rate and it was observed by \textit{Swift} for longer duration than other stars.

Our results regarding the flare durations and amplitudes suggest that one of the reasons for the lack of optical counterparts to NUV flares in G-K  stars pointed by \cite{2023ApJ...944....5B} might be the cadence size mismatch between GALEX data and \textit{Kepler} data, rather than any physical process. The results in Section \ref{subsection:optical/nuv counterparts} strongly support this claim. \cite{2019ApJ...883...88B} reported that the majority (98\%) of the GALEX flares studied in \cite{2023ApJ...944....5B} had durations less than 9 minutes, and there were very few flares with durations on the order of 10 minutes. They used 10-second cadence GALEX data, and contemporaneous $Kepler$ long cadence (30 minutes) data, except for two flares which had $Kepler$ short cadence (1 minute) data, to search for optical counterparts to 1559 GALEX flares.   

The next possible reason might be very low amplitudes of optical flares in G-K stars. This is also evident from the M dwarf flares we observed simultaneously with \textit{Swift} and TESS. Even though they have cadence sizes closer than any existing data, we found that the optical flares have significantly lower amplitudes compared to their NUV counterparts (see Figure \ref{fig:flare amplitudes comparison}). Since G-K dwarfs are optically brighter than M dwarfs, the amplitudes of the optical flares produced by G-K stars would be smaller than that of M dwarfs and hence might be diluted by the noise in the data. However, the results of \cite{2023ApJ...944....5B} requires confirmation through a larger sample of G-K flares observed simultaneously in optical and NUV bands using similar cadence sizes, as well as more sensitive optical telescopes capable of capturing flare amplitudes in G-K stars to less than 0.1\%.
%
%
%
\subsection{NUV/optical energy fractionation}
\label{sec:energy fractionation}

\subsubsection{Expectations from flare models and previous observations}
\label{subsubsection:flares models}
A single temperature blackbody model is insufficient in explaining certain observed features in flare spectral energy distributions, such as the Balmer jump and emission/absorption lines. In order to address this limitation and provide a more realistic representation of stellar flare spectral energy distributions, \cite{2023ApJ...944....5B} employed a range of radiative-hydrodynamic (RHD) models using the RADYN code \citep{Carlsson_1992, Carlsson_1995, Carlsson_1997, Allred_2015} to investigate the energy fractionation between TESS/Kepler and GALEX bandpass energies. Specifically, they utilized different combinations of two RHD models, namely F13 and 5F11 (where the last two digits denote the logarithm of the electron beam flux), to more accurately reproduce the optical and NUV spectra of M dwarf flares. The M dwarf flares exhibit continuum enhancements spanning from ultraviolet to optical wavelengths, along with an additional continuum component beyond the H$\beta$ wavelength ($\geq$ 4900 \AA; \citealt{2013ApJS..207...15K}). The two flare models use injected electron beams with differing properties, like maximum fluxes and low-energy cutoffs, to reproduce both features. The F13 flare model demonstrates increased blue continuum emission characterized by a color temperature exceeding 9000 K, while the 5F11 model displays a more pronounced Balmer jump ratio. The two models are combined using a relative filling factor, $X_{\rm 5F11}$ = 0 - 12.5. The combined model incorporates the increase in both Balmer jump ratio and Balmer line flux without disrupting the flare color temperature, thereby aligning with spectral observations of M dwarf flares within the blue-optical wavelength range. For more comprehensive information on these models, we refer the reader to \cite{2023ApJ...944....5B}.
We compare these models to our observed flux ratios between the TESS optical and \textit{Swift} NUV bandpasses. Table \ref{table:energy ratios} provides a summary of the model names, along with the corresponding expected energy ratios $E_{\rm TESS}$/$E_{\rm Swift}$ and $E_{\rm Kepler}$/$E_{\rm Swift}$. 
These energy ratios were calculated using the total luminosity within each band
\citep[see Equation 2 of][]{Tristan_2023} and assuming identical flaring durations between bands. 
The TESS and $Kepler$ fluxes were calculated by integrating the model spectra with the filter response functions, following the method outlined in \cite{2005PASP..117.1049S}.
\begin{table*}
\scriptsize
\centering
\caption{Flare Models and Energy Ratios between Bandpasses}
\begin{tabular}{ccc}
\hline
Model &  TESS/\textit{Swift} Ratio & Kepler/\textit{Swift} Ratio  \\
\hline
F13 & 0.741 & 1.460 \\
F13+2.5$^{*}$5F11 & 0.760 & 1.486  \\
F13+5.0$^{*}$5F11 & 0.778 & 1.510 \\
F13+7.5$^{*}$5F11 & 0.795 & 1.534 \\
F13+10.0$^{*}$5F11 & 0.812 & 1.557 \\
F13+12.5$^{*}$5F11 & 0.828 & 1.579   \\
5F11 & 1.886 & 3.029 \\
9000 K BB & 2.648 & 4.704 \\
18,000 K BB & 0.296 & 0.661  \\
36,000 K BB & 0.114 & 0.283 \\
\hline
\end{tabular} \label{table:energy ratios}
\end{table*}
\\
\subsubsection{Empirical relation between \texorpdfstring{$E_{\rm NUV}$}{E\_NUV} and \texorpdfstring{$E_{\rm TESS}$}{E\_TESS}} \label{subsubsection: PL fit}

\begin{figure}
   \centering
   \includegraphics[width=0.45\textwidth,height=0.3\textwidth]{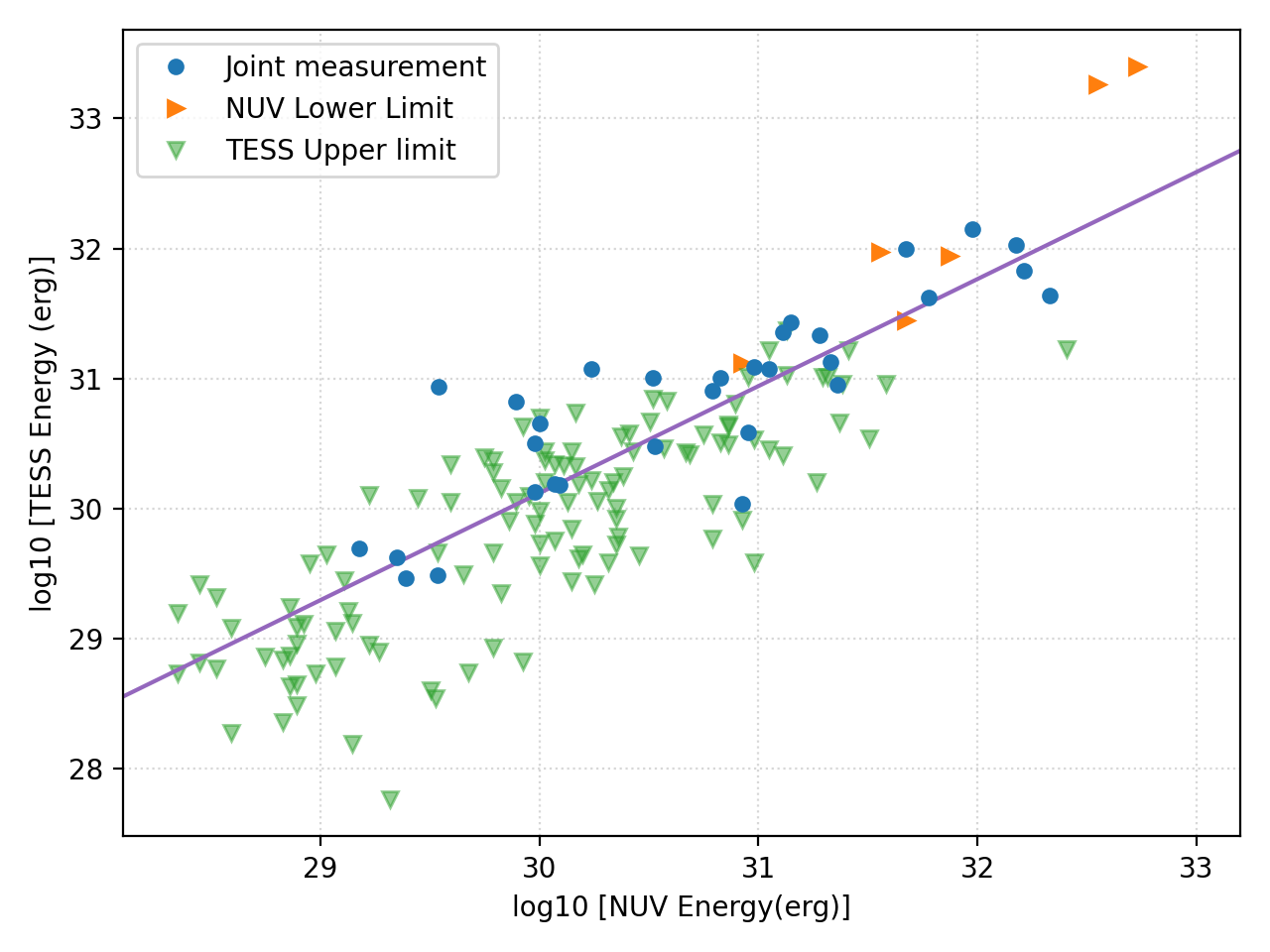}
   \caption{Observed NUV flare energies versus observed TESS flare energies. The blue dots represent the flares which were observed simultaneously by both TESS and $Swift$/UVOT for their full duration. The orange triangles facing right represent the flares which were not observed for full duration by $Swift$/UVOT but have optical counter parts. The green downward-facing triangles represent the NUV flares with upper limits on the TESS flare energies. The blue line represents the linear fit with the form log $E_{\rm TESS}$ = 0.823$\pm$0.059 log $E_{\rm NUV}$ + 5.630$\pm$1.8.}
    \label{fig:observed NUV vs TESS energies}
\end{figure}

We calculated an empirical relationship between our observed NUV and TESS flare energies (Figure \ref{fig:observed NUV vs TESS energies}). We included in the model the upper limits on the TESS energies for the NUV flares without optical counterparts in the TESS data, as well as lower limits on the Swift flare energies for incomplete flares. We note that the TESS flare energies ($E_{\rm TESS}$) are estimated by converting the bolometric flare energies, obtained for a 10,000 K BB from Equation \ref{equation:flare energy}, to TESS band energies using the energy ratio $E_{\rm TESS}$/$E_{\rm bol}$ = 0.18, as given in Section \ref{subsubsec:determining flare properties}. We performed a linear regression analysis on the logarithmic data and obtained a relationship of the form log $E_{\rm TESS}$ = 0.823$\pm$0.059 $E_{\rm NUV}$ + 5.630$\pm$1.8. We refer to this fit as PL fit in the following sections. We considered other functional forms such as a quadratic relation, but found the linear relationship in log-log space (e.g. a power law) better fits the data. Using the power law fit results, we predicted the NUV energies based on the observed TESS energies. We compare those predicted NUV energies with the observed NUV energies in panel A of Figure \ref{fig:linear, 9000K, mixed models comparison}. The residual plot in the bottom of panel A of Figure \ref{fig:linear, 9000K, mixed models comparison} indicates that the NUV energies estimated from TESS energies, utilizing the PL fit results, align moderately well with the observed values. It is important to note that some scatter may result from a potential mismatch in the cadence size of NUV and TESS data.

\subsubsection{Choice of best model for predicting \texorpdfstring{$E_{\rm NUV}$}{E\_NUV} using \texorpdfstring{$E_{\rm TESS}$}{E\_TESS}} \label{subsubsec:best hybrid model}

Due to the distinct spectral characteristics of M dwarf flares, our primary interest lies in determining the most suitable fit among the five available hybrid models. This is in contrast to opting for a single-temperature BB model or models exhibiting significant enhancements solely in the continuum (F13 model) or solely in the Balmer jump ratio (5F11 model). Consequently, our focus remains exclusively on hybrid models. 

To assess the optimal hybrid model among the five presented in Table \ref{table:energy ratios}, we calculate NUV energies from the observed TESS energies using the energy ratios associated with all five hybrid models listed in the same table. For this, we used the sample of flares which were observed simultaneously by both TESS and Swift/UVOT for full duration. Subsequently, we carried out four statistical tests: the Kolmogorov-Smirnov test, the Anderson-Darling test, Bayesian Information Criterion (BIC), and the Akaike Information Criterion (AIC) to ascertain the model that fits the data best. A summary of the results from these statistical tests is provided in Table \ref{table:statistical test, mixed models}. We also include the statistical tests for the results of PL fit for comparison with those of hybrid models. 
We can see that the KS test and AD tests are inconclusive for all models. However, the BIC and AIC values are somewhat different for each model and the lowest value determines the best model. 
Here, the model F13+12.5$^{*}$5F11 has the lowest value of both BIC and AIC, so we consider it to be the best among the five hybrid models.

%
\begin{table*}
\centering
\caption{Results of statistical tests. I. }
\begin{tabular}{c|cccc}
\hline
 Model & KS test & AD test & BIC & AIC \\
 \hline
F13+2.5$^{*}$5F11 & 0.25,0.27 & 0.70,0.17 & 73.2 & 71.7 \\
F13+5.0$^{*}$5F11 & 0.25,0.27 & 0.70,0.17  & 71.7 & 70.3\\
F13+7.5$^{*}$5F11 & 0.25,0.27 & 0.60,0.19 & 70.3 & 68.9  \\
F13+10.0$^{*}$5F11 & 0.25,0.27 & 0.59, 0.19 & 69.0 & 67.5\\
F13+12.5$^{*}$5F11 & 0.25,0.27 & 0.51, 0.21 & 67.7 & 66.3 \\ 
PL fit & 0.09,0.999 & -0.93, 0.25 & 63.2 & 61.7 \\
\hline
\end{tabular}
\label{table:statistical test, mixed models}
\\
\textbf{Note:} Among the two numbers listed in KS test and AD test, \\
the first number is the statistic and the second number is the $p$-value.
\end{table*}

\subsubsection{Comparison between PL fit, 9000 K BB and F13+12.5*5F11 hybrid model }
\label{subsubsec:comparison of linear fit, 9000k and hybrid model}
Next, we compared the observed NUV flare energies with those predicted from TESS/K2 flares using the PL fit, 9000 K BB\footnote{It is chosen instead of 10,000 K BB here because we have used the same RHD models as those in \cite{2023ApJ...944....5B}.} and F13+12.5*5F11 hybrid model for the sample of flares that were simultaneously observed by TESS/K2 and \textit{Swift}/UVOT throughout their complete durations. The comparisons are shown in panels A, B and C of Figure \ref{fig:linear, 9000K, mixed models comparison}. We included a comparison with the 9000 K blackbody (BB) model, aiming to provide readers with insight into its distinctions from the hybrid model and the PL fit. This particular model holds significant usage in the literature for predicting NUV flare energies based on optical flare energies. Some scattering of the data points in the upper panels of the two plots might be due to a mismatch in the cadence size between the TESS and NUV data, leading to differences in the accuracy of the estimated energy values in the two bands. 

The flare energies predicted by the two models (9000 K BB and F13+12.5*5F11) differ by a factor of 3.2 (see Table \ref{table:energy ratios}). So, we do not see distinct differences in the upper panel of two plots. However, we see some difference in the corresponding residual plots. The expected NUV flare energy versus observed NUV energy plot and residual plot in panel B of Figure \ref{fig:linear, 9000K, mixed models comparison} suggests that the 9000 K BB model is able to predict the NUV flare energies with values of less than $10^{30}$ ergs  with more accuracy than those with energies greater than $10^{30}$ ergs. We estimate that it underestimates the NUV flare energies with values greater than $10^{30}$ ergs by a factor of $\sim$2.5 which is based on the median of differences in the observed and predicted NUV energies with values greater than $10^{30}$ ergs. Likewise, the expected NUV flare energy versus observed NUV energy plot and the residual plot in  panel C of Figure \ref{fig:linear, 9000K, mixed models comparison} suggests that the F13+12.5*5F11 model is able to predict energy of flares with values greater than $10^{30}$ ergs with more accuracy than those with energies less than $10^{30}$ ergs. We estimate that it overestimates the NUV flare energies with values less than $10^{30}$ ergs by a factor of $\sim$2.9, which is based on the median of differences in the observed and predicted NUV energies with values less than $10^{30}$ ergs.  
%
\begin{figure*}[h!]
\includegraphics[width=0.33\textwidth]{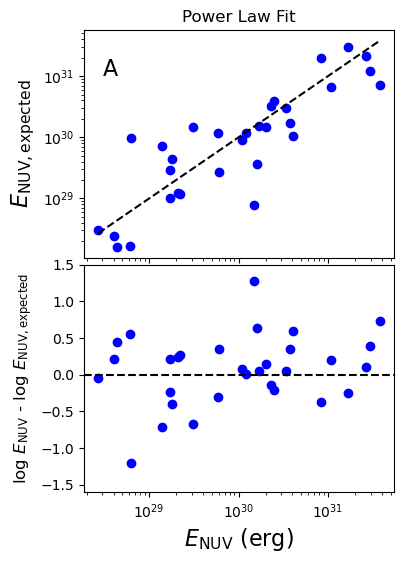}
\vspace{0.00mm}
\includegraphics[width=0.33\textwidth]{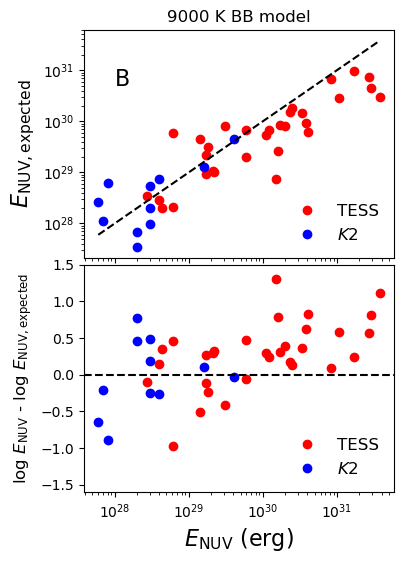}
\vspace{0.00mm}
\includegraphics[width=0.33\textwidth]{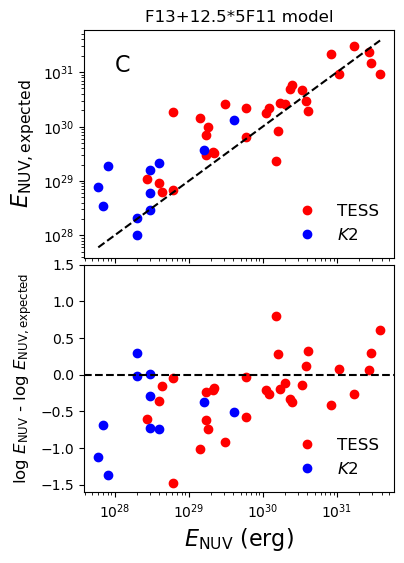}
\caption{A: \textit{Upper panel}: Comparison of observed NUV flare energies with those predicted from fitting of the observed TESS flare energies i.e PL fit. \textit{Lower panel}: Differences in logarithm of observed and expected NUV flare energies versus the observed NUV flare energies. B: Same as panel A but for 9000 K BB model. C: Same as panel A but for F13+12.5*5F11 hybrid model. The slanted dashed line in the upper plot and the horizontal dashed line in the lower plot of all panels represents the case when the expected NUV energy is equal to the observed NUV energy.} 
\label{fig:linear, 9000K, mixed models comparison}
\end{figure*}
%
\subsubsection{$E_{\rm TESS}$/$E_{\rm NUV}$ versus $E_{\rm NUV}$}
Similar to the study conducted by \cite{2023ApJ...944....5B}, which utilized overlapping GALEX and $Kepler$ data, we investigate the relationship between NUV and optical flare energies using simultaneous TESS and \textit{Swift}/UVOT data. One advantage we possess over their investigation is a larger sample of flares that were identified in both TESS and \textit{Swift}/UVOT light curves. We have a total of 31 NUV flares that were observed for the entire duration by both telescopes and were identified in both light curves. Additionally, six more NUV flares were identified in both light curves, but they were not observed for the complete duration by \textit{Swift}/UVOT. The remaining NUV flares were not identified in TESS light curves, and we derived upper limits on their TESS flare energies. 

\begin{figure*}[h!]
\centering
\includegraphics[width=0.45\textwidth]{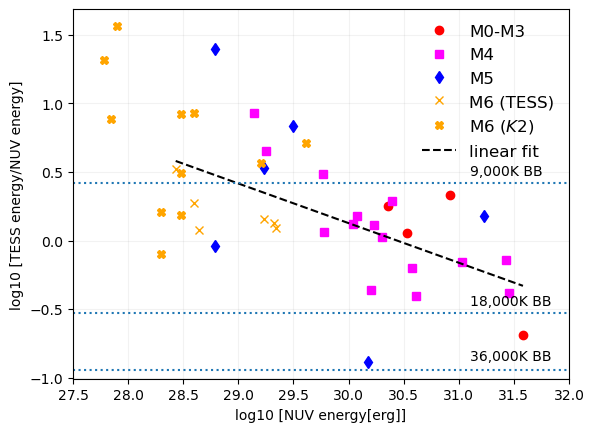}
\vspace{0.00mm}
\includegraphics[width=0.45\textwidth]{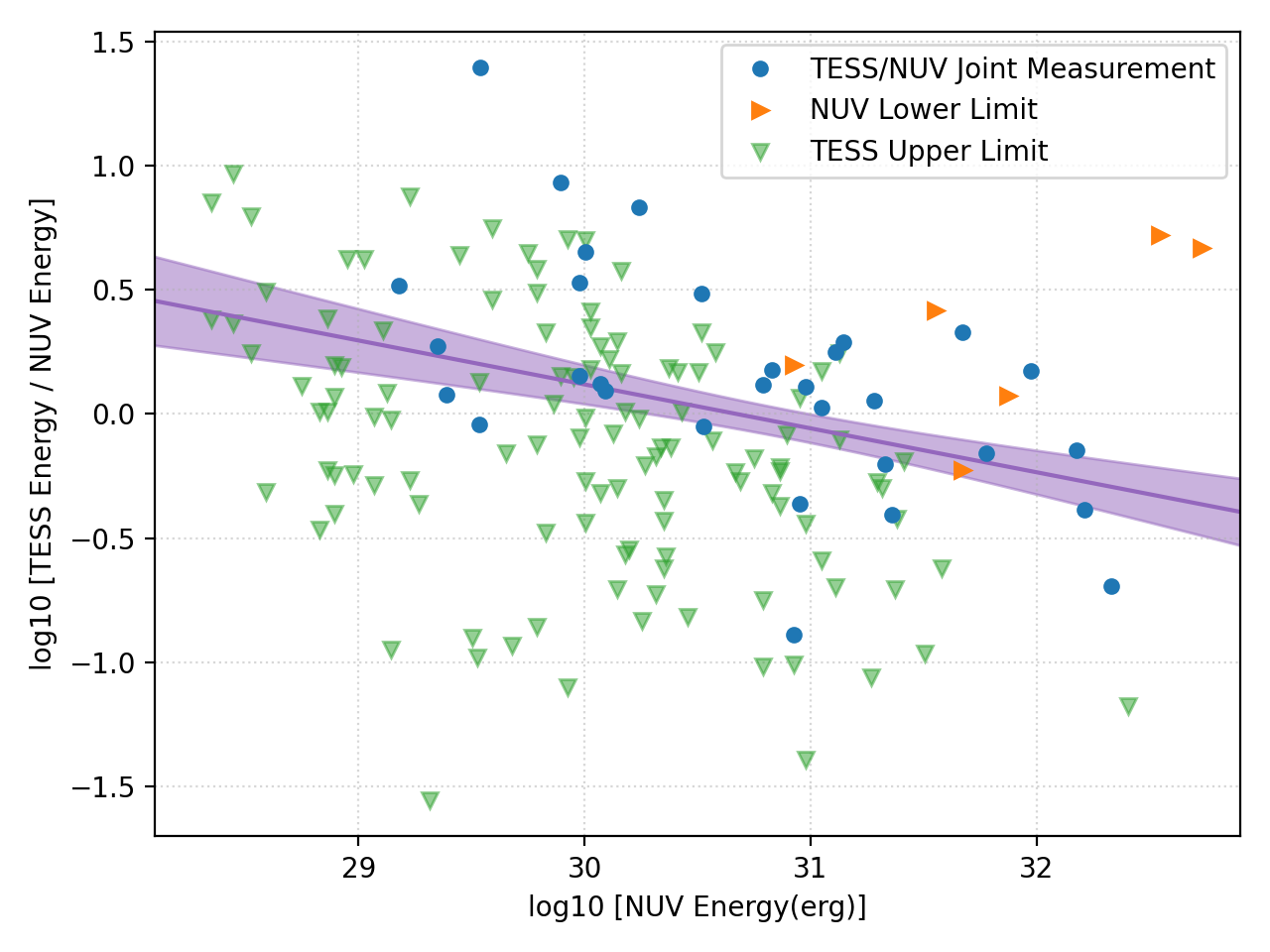}
\includegraphics[width=0.45\textwidth]{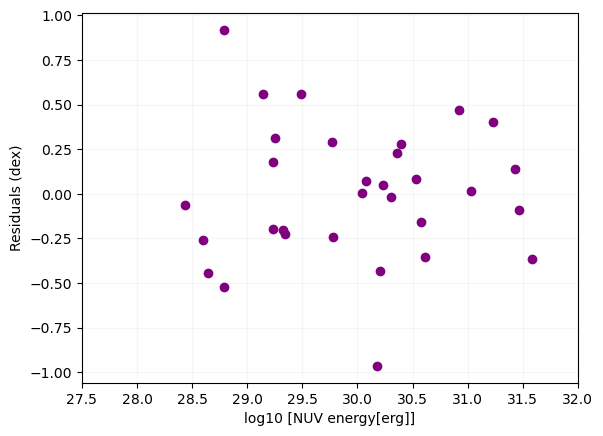}
\includegraphics[width=0.45\textwidth]{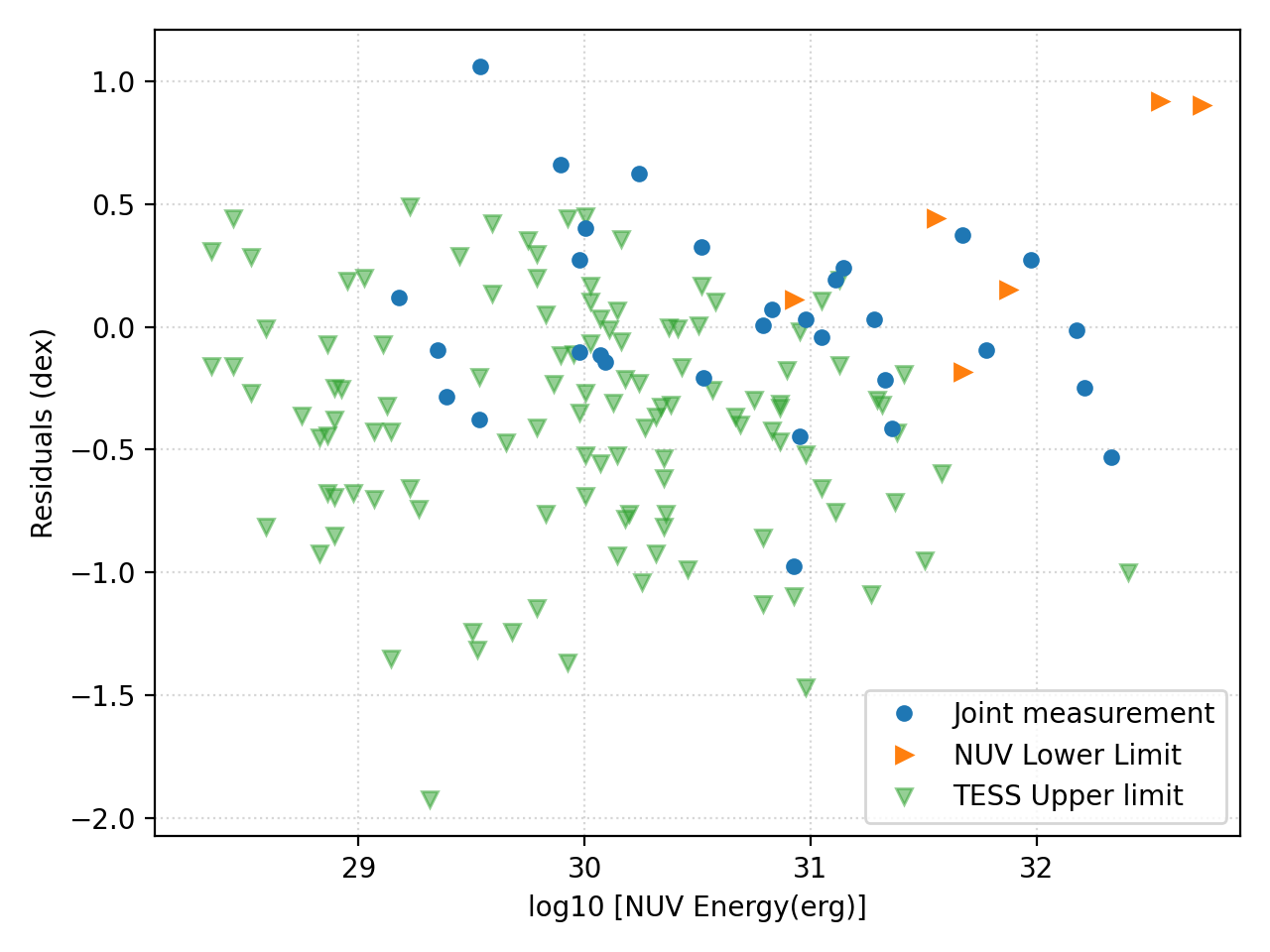}
\caption{\textit{Upper left}: Energy fractionation between TESS and \textit{Swift} NUV flare energies for a sample of 31 NUV flares which were observed for the entire duration by \textit{Swift} and had optical counterparts identified in TESS light curves. The data points are classified by spectral type of the stars on which the flares were observed. The black solid line represents the power law fit to the data. This line shows that there is a slight decreasing trend in the energy fractionation as the NUV flare energy increases. Additionally, the K2 flare energies and their corresponding ratios with NUV energies are overplotted for comparison. The dotted horizontal lines correspond to the anticipated energy fractionation for blackbodies at temperatures of 9,000~K, 18,000~K and 36,000~K. \textit{Lower left}: Corresponding residuals of the upper left plot. \textit{Upper right}: Energy fractionation between TESS and \textit{Swift} NUV flare energies for the entire sample of NUV flares in this study. The purple dots represent the energies of NUV flares that were observed for the entire duration by \textit{Swift} and had optical counterparts identified in TESS light curves. The triangles represent the energies of NUV flares that were either not observed for the complete duration by \textit{Swift} or lacked optical counterparts in TESS light curves. The purple solid line represents the power law fit to the data, and the shaded region represents the 1-$\sigma$ uncertainty on the fit \textit{Lower right}: Corresponding residuals of the upper plot.}
\label{fig:NUV/optical energy fractionation}
\end{figure*}

In left panel of Figure~\ref{fig:NUV/optical energy fractionation}, we present a plot of the ratio of TESS to \textit{Swift} NUV energy ($E_{\rm TESS}$/$E_{\rm NUV}$) as a function of \textit{Swift} energy ($E_{\rm NUV}$), utilizing only the energies of NUV flares that were observed for the entire duration by \textit{Swift} and had optical counterparts identified in TESS light curves. For comparison purposes, we also overlay the K2 flare energies and their corresponding ratios with NUV energies.  
To assess any potential correlation between $E_{\rm TESS}$/$E_{\rm NUV}$ and $E_{\rm NUV}$, we fit a line log $E_{\rm TESS}$/$E_{\rm NUV}$ = -0.289$\pm$0.080 log $E_{NUV}$ + 8.797$\pm$2.383 by excluding K2 data. The fit is represented by the black solid line. The fit to the data was obtained using \texttt{pymc3}, revealing a slight decreasing trend in the energy fractionation as the NUV flare energy increases. The corresponding residuals are shown in the lower plot of left panel. A more pronounced decreasing trend was observed in the higher GALEX flare energies studied by \cite{2023ApJ...944....5B}. The main distinction between the two studies is that the flare energies in \cite{2023ApJ...944....5B} ranged from log $E$ (erg) $\sim$(32-36), whereas in this study, the flare energies have values in the range log $E$ (erg) $\sim$(28-32). We want to remind the readers about the different stellar populations being probed in this paper versus the \cite{2023ApJ...944....5B} paper. While the sample of stars is all M dwarfs in this study, it is mostly G-K dwarfs in \cite{2023ApJ...944....5B} except one. They included additional literature values of M dwarf flares which spanned a wider range of energies than solar flares, and suggested that M dwarf flares might have energy fractionations closer to what the Sun exhibits at lower energies. 

Next, we examined the energy fractionation using the entire sample of NUV flares. The right panel of Figure~\ref{fig:NUV/optical energy fractionation} illustrates the relationship between the two energies for this case. The fitted line has the form log $E_{\rm TESS}$/$E_{NUV}$ = -0.177$\pm$0.060 log $E_{NUV}$ + 5.40$\pm$1.80 and hence we find that the correlation is even weaker compared to the aforementioned case. The corresponding residuals are shown in the lower plot of right panel. This fit includes the upper and lower limits in the observed TESS and Swift data sets.

Observations in \cite{2013ApJS..207...15K} demonstrated 
that there was a systematic dependence of the amount of excess Balmer continuum emission above a linear fit to the blackbody 
component; the ratio of the two continuum fluxes decreases as the flare luminosity increases. This suggests that for small to moderate luminosity flares the larger component of NUV emission, on top of any blackbody emission, would show up as a larger energy fractionation between the NUV and optical bandpasses. 
\citet{2016ApJ...820...95K} also demonstrated that moderate to large flares have smaller Balmer jump ratios and hotter color temperatures, which would also suggest a systematic behavior
in energy fractionation between the NUV and optical bandpasses.

%
\subsubsection{Comparison of models with HST/STIS NUV spectra}
\begin{figure}[htbp]
   \centering
   \includegraphics[width=0.5\textwidth,height=0.35\textwidth]{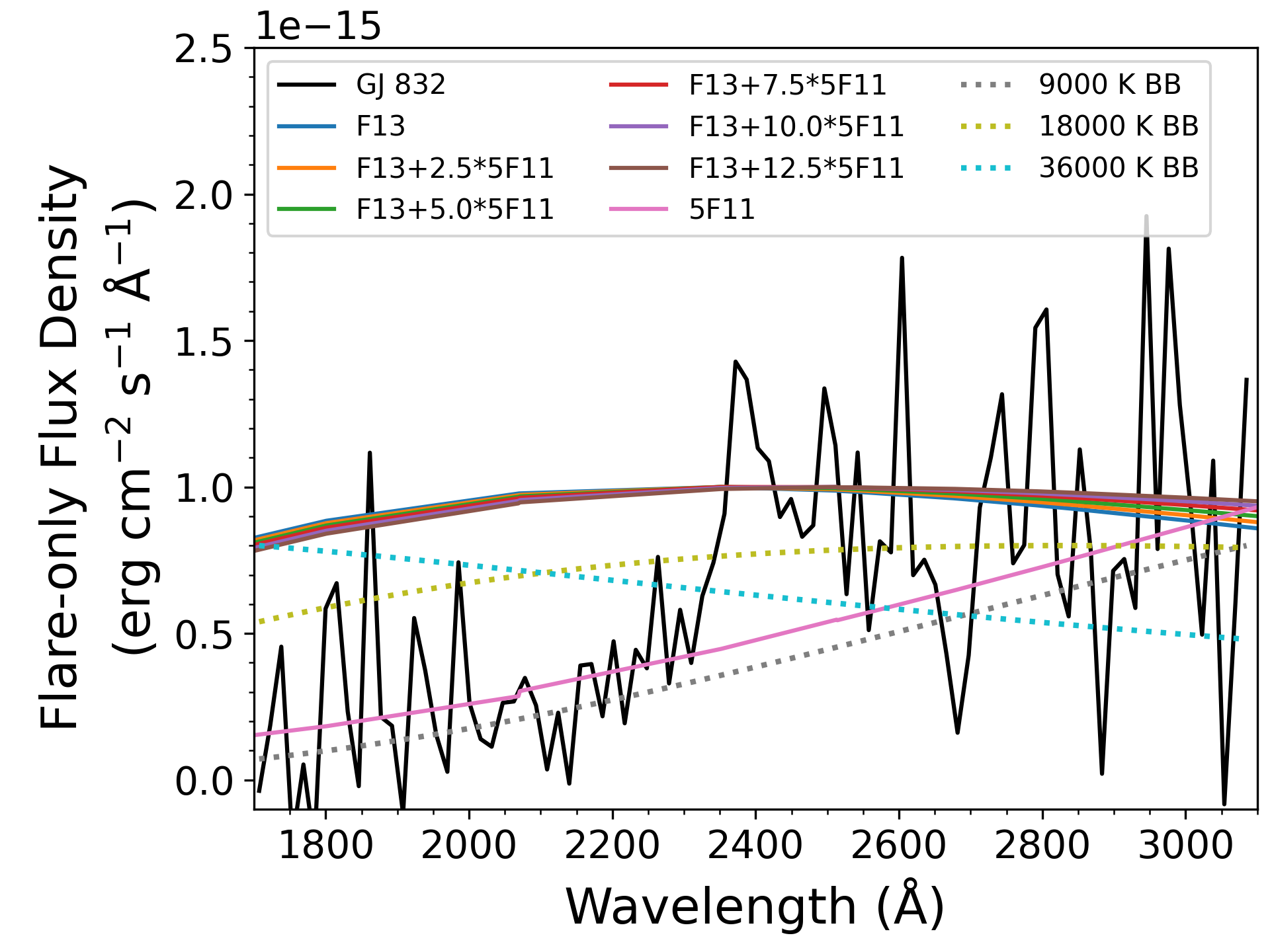}
    \caption{Comparison of the GJ 832 NUV flare-only spectrum (black line; binned to 15 \AA\ for improved signal-to-noise) with the model spectra discussed in Section \ref{subsubsection:flares models}. The model spectra have been scaled for visual comparison of the spectral shapes.}
    \label{fig:HST flare_spectrum GJ 832}
\end{figure}

We analyzed the NUV spectral energy distribution of the first GJ 832 flare by subtracting the mean quiescent spectrum from the flare spectrum (Sec~\ref{subsubsec:HSTSTIS_flare_ids}) to generate a ``flare-only" spectrum, shown in Figure~\ref{fig:HST flare_spectrum GJ 832}. The flare-only spectrum is dominated by emission lines with no clear evidence of strong continuum emission. The majority of the flare emission occurs between 2200-3000 \AA\ from a forest of \ion{Fe}{2} emission lines and the \ion{Mg}{2} h\&k emission lines. 

We also compared our flare-only spectra to the hybrid models discussed in Section \ref{subsubsection:flares models} and find significant disagreement. This can be seen in Figure \ref{fig:HST flare_spectrum GJ 832}. The non-hybrid 5F11 model and the cooler 9000 K BB model best match the data, but appear to be missing the \ion{Fe}{2} and \ion{Mg}{2} emission between 2300-2800 Å. This indicates that emission lines are an important component of some flare emissions, and these are not fully taken into account by the available RADYN models. 
%
%
\subsection{Flare frequency distributions (FFDs)}
\label{sec:FFDs}
\subsubsection{FFD generation and fitting}
\label{subsubsection:ffd generation and fitting}
A power law relationship has been observed in the cumulative distribution of flares for a flaring star. This relationship can be expressed as a linear equation: log $\tilde{\nu}$ = C + $\beta$ log $E$, where $\tilde{\nu}$ represents the cumulative frequency of flares occurring per unit time with energies higher than $E$. The values of the coefficient $C$ and the spectral index $\beta$ may vary for each individual star depending on its age and spectral type (\citealt{2005stam.book.....G}; \citealt{1976ApJS...30...85L}). 

We used the maximum-likelihood method described in \cite{2010arXiv1008.4686H} and implemented in the routine \texttt{emcee} \citep{2013PASP..125..306F} to fit a straight line to our data (in log scale) and hence obtain the optimal values of parameters $\beta$. To avoid bias, we neglected some low energies which show deviation from a power law distribution for fitting in the case of both observed as well as predicted energies. The deviation is most likely due to instrumental sensitivity in detecting very weak flares. We also neglected the highest observed energies to reduce any bias in fitting. In some cases, particularly when the number of flares is small, we observe that the largest flare energy deviates from the straight line, leading to a change in the true value of the slope of the line, especially towards lower flare rates. We list the values of fitted parameters of FFDs in Table \ref{table:FFD_fitting} for NUV flares and optical flares. 
\cite{2023ApJ...955...24R} also studied NUV M dwarf flares using archival GALEX data. However, they estimated that the slopes of the NUV FFDs were slightly less steep than the values we estimated in this study. One of the factors that might lead to different slope values could be the different energy ranges considered in the two studies. The range of NUV flare energies is approximately (10$^{29}$ - 10$^{35}$ erg) in \cite{2023ApJ...955...24R}.

\begin{figure}[htbp]
   \centering
   \includegraphics[width=0.5\textwidth]{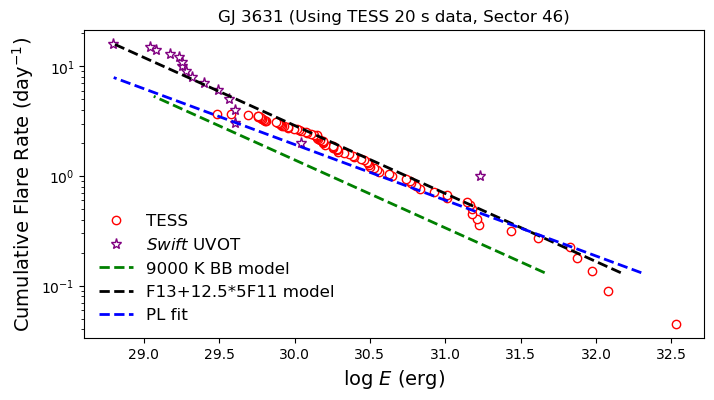}
    \caption{Frequency distribution of flares on GJ 3631 observed with TESS (red circles) and \textit{Swift}/UVOT (purple stars) using the corresponding bandpass energies. The green, black and blue dashed lines represent the linear fit the frequency distribution of flares predicted using 9000 K BB, F13+2.5*5F11 model and the power-law fit. All the fits are obtained by using the same range of TESS energies used for fitting. The blue and black dashed lines are extrapolated to the observed minimum NUV energy.}
    \label{fig:FFD_GJ3631}
\end{figure}

We aim to compare the flare frequency distributions (FFDs) of M dwarf flares observed by TESS and \textit{Swift}/UVOT as well as those predicted from the TESS flare energies using the PL fit and the best hybrid model. This comparison will enable us to test the outcomes of the PL fit and the best hybrid model presented in Sections \ref{subsubsection: PL fit} and \ref{subsubsec:best hybrid model} respectively. To start, we focus on GJ 3631 as one of our targets, which has a high flare rate in both TESS and NUV passbands. During TESS Sector 46, we recorded 16 NUV flares with \textit{Swift}/UVOT and 81 optical flares with TESS. \textit{Swift}/UVOT and TESS observed this star for 86.1 ks and 22.2 days respectively. The TESS data was collected in 20-second cadence mode. The FFD of GJ 3631 is depicted in Figure \ref{fig:FFD_GJ3631}. 

In the same figure, we compare the observed NUV FFD with those projected from optical flare rates using the 9000 K BB and F13+12.5*5F11 models, and the PL fit. For our projections with the models and the PL fit, we first predicted the NUV FFDs from TESS energies using the energy ratios outlined in Table \ref{table:energy ratios} and the results of PL fit from Section \ref{subsubsection: PL fit}. We then used power law to fit those predicted NUV FFDs. Subsequently, we extrapolated these fitting outcomes to the minimum observed NUV energies. Thus the dashed lines in Figure \ref{fig:FFD_GJ3631} represent the fits to the predicted NUV FFDs, not the direct prediction of the models and the PL fit. From the figure, we see that the F13+12.5*5F11 model predicts the NUV FFD with greater proximity to the observed NUV FFD compared to the prediction by the 9000 K BB model. We find that the fit of NUV FFD predicted by PL fit deviates from the observed distribution at lower energies for the cases when the sample of flares with energies $>$10$^{31}$ erg is small, and we show that it can predict the NUV FFD more accurately when the sample of flares with energies $>$10$^{31}$ erg is large as discussed in the following paragraph.

\begin{figure}[htbp]
   \centering
   \includegraphics[width=0.5\textwidth]{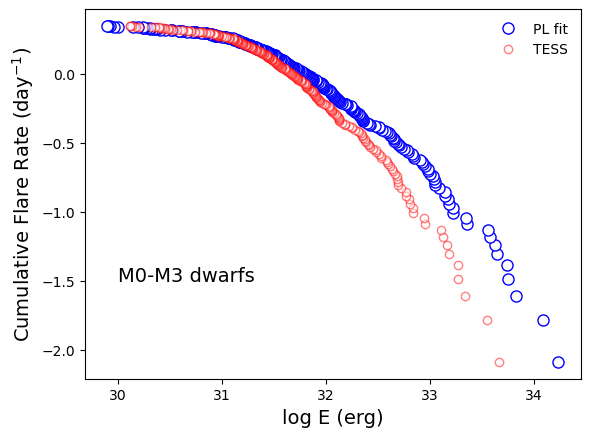}
    \caption{Comparison between observed TESS FFD of M0-M3 dwarfs and the FFD estimated by using the PL fit.}
    \label{fig:TESS-PL fit comparison}
\end{figure}

\begin{figure}[htbp]
   \centering
   \includegraphics[width=0.4\textwidth,height=0.25\textwidth]{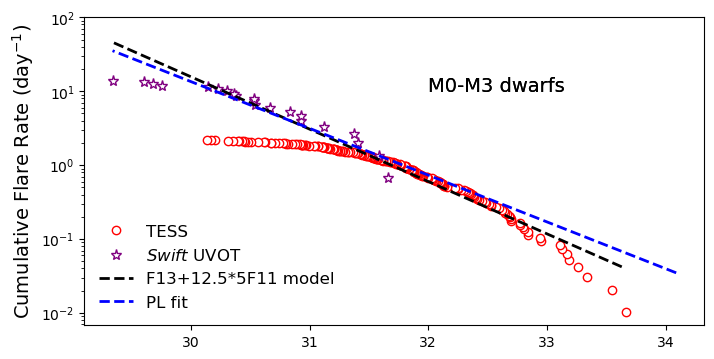}
   \includegraphics[width=0.4\textwidth,height=0.25\textwidth]{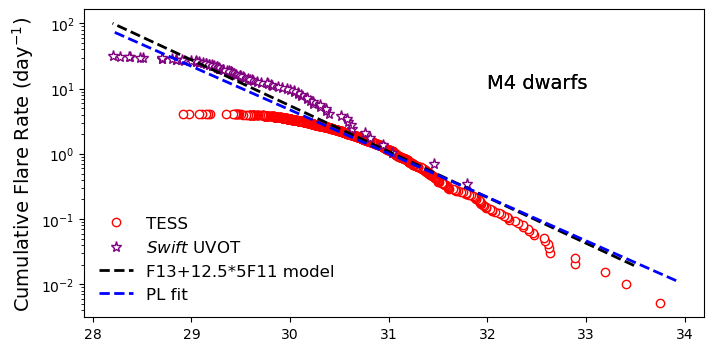}
   \includegraphics[width=0.4\textwidth,height=0.25\textwidth]{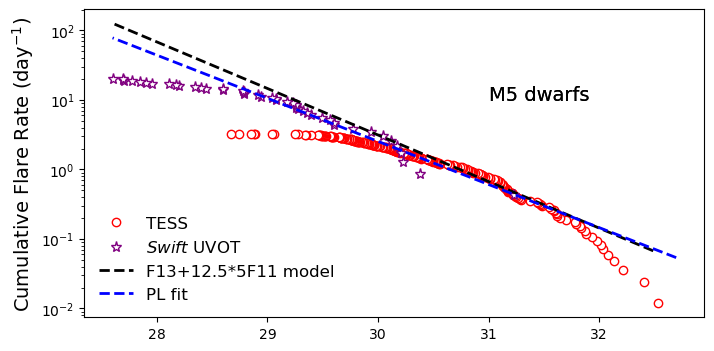}
   \includegraphics[width=0.4\textwidth,height=0.25\textwidth]{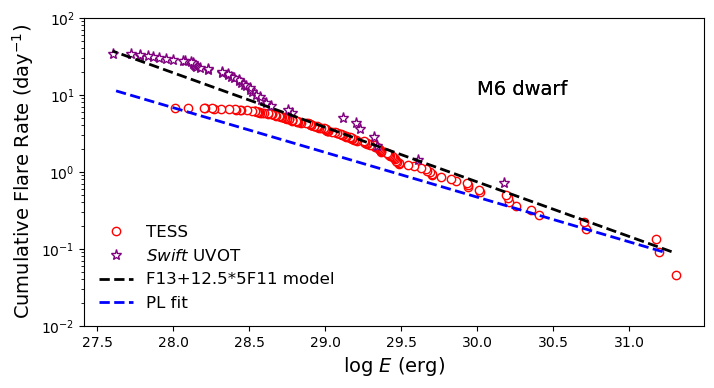}
    \caption{Cumulative frequency distribution of flares on M0-M3, M4, M5 and M6 dwarfs observed with TESS data (red circles) and \textit{Swift}/UVOT data (purple stars), utilizing the respective bandpass energies. Additionally, the blue and black dashed lines corresponds to the fit to the NUV FFD predicted by the PL fit and the F13+12.5$^{*}$5F11 model from TESS FFD respectively.} 
     \label{fig:FFD by spt with models and power-law}
\end{figure}

The minimum energy ($E'_{\mathrm{min}}$) used for fitting the predicted NUV FFDs is determined by the minimum energy $E_{\mathrm{T,min}}$ applied in fitting the observed TESS FFDs. If the predicted NUV FFD does not exhibit significant flattening at lower energies, similar to the observed TESS FFD, we use all predicted energies except the largest for fitting. However, if there is flattening, $E'_{\mathrm{min}}$ for the NUV FFD predicted by the blackbody and hybrid models is the energy corresponding to $E_{\mathrm{T,min}}$ using the energy ratio in Table \ref{table:energy ratios}. For PL fit, $E'_{\mathrm{min}}$ is not solely the energy predicted from $E_{\mathrm{T,min}}$. The predicted value needs adjustment in cases where there are many flares with energies greater than $10^{31}$ ergs. This is because the PL fit predicts NUV energies with values lower than TESS energies for $\lesssim10^{31}$ ergs and higher than TESS energies for $>10^{31}$ ergs. This is demonstrated in Figure  \ref{fig:TESS-PL fit comparison}. If the value of $E_{\mathrm{T,min}}$ is less than $10^{31}$ ergs, the new value of the minimum energy $E'_{\mathrm{min}}$ will be lesser than $E_{\mathrm{T,min}}$. Consequently, the slope of the FFD becomes less steep, given that the slope of FFD depends on the range of energies chosen for fitting. The fitted NUV FFD based on this minimum may not accurately predict the observed high flare rates of lower NUV energies. While we are uncertain about adjusting the minimum value of the predicted NUV energy for FFD fitting correctly, we use a new value $E''_{\mathrm{min}}$, which is the average of $E'_{\mathrm{min}}$ and the NUV energy predicted by the PL fit corresponding to the largest TESS energy. This way the fitted FFD predicted by PL fit is able to match the high flare rates of observed NUV flares with lower energies to some extent. This is the case when the sample of flares with energies $>$10$^{31}$ erg is large, and is demonstrated in the FFDs shown in Figure \ref{fig:FFD by spt with models and power-law}.
%
%
%
\begin{table*}
\centering
\caption{Summary of power-law fits to FFDs. In this table, $\beta_{\rm NUV}$ and $\beta_{\rm optical}$ are the fitted slopes of FFDs, $N_{\rm NUV}$ and $N_{\rm optical}$ are the number of flares used for fitting, and $E_{\rm min}$ and $E_{\rm max}$ are the minimum and maximum energies used for fitting. The errors in the slopes are obtained by dividing the values of $\beta$ by $\sqrt{N}$, where $N$ is the number of flares used for fitting.}
\begin{tabular}{|c|cccc|cccc|}
\hline
Sp. Type &  NUV  & & & & Optical & & & \\
\hline
& $\beta_{\rm NUV}$ & $N_{\rm NUV}$ & log $E_{\rm min}$ & log $E_{\rm max}$ & $\beta_{\rm optical}$ & $N_{\rm optical}$ & log $E_{\rm min}$ & log $E_{\rm max}$ \\
& & & erg & erg & & & erg & erg \\
\hline
M0-M3 & -0.62$\pm$0.15 & 18 & 30.2 & 32.0  & -0.71$\pm0.05$ & 225 & 31.0 & 33.6 \\
M4-M6 &  -0.55$\pm$0.08 & 44 & 28.5 & 31.5  &  -0.76$\pm$0.04 & 362 & 29.5 & 32.5 \\
(TESS 20 s) & & & & & & & & \\
M4-M6 & -0.59$\pm$0.06 & 105  & 28.5 & 31.8 & -0.80$\pm$0.04 & 534 & 30.0 & 32.5 \\
(TESS 2 min + 20 s) & & & & & & & & \\
\hline
M4 & -0.63$\pm$0.07 & 76 & 29.0 & 31.5 & -0.70$\pm$0.03 & 666 & 30.0 & 33.4 \\
M5 & -0.61$\pm$0.11 & 32 & 28.6 & 30.4 & -0.67$\pm$0.05 &181 & 30.0 & 32.4 \\
M6 & -0.82$\pm$0.13 & 39 & 28.0 & 29.6 & -0.71$\pm$0.06 & 137 & 28.5 & 31.2 \\
\hline
\end{tabular}
\label{table:FFD_fitting}
\end{table*}
%
\subsubsection{FFD as a function of spectral type}
In this section, we analyze the FFDs of the M dwarfs in our sample as a function of spectral type in both NUV and optical passbands. We used a sample of M dwarfs with multiple flares in each spectral type. The sample consists of five M0-M3 dwarfs (AU Mic, EV Lac, GJ 1284, LP 776-25, V1005 Ori), seven M4 dwarfs (AP Col, DG CVn, Fomalhaut C, HIP 17695, Ross 614, YZ Ceti, YZ CMi), four M5 dwarfs (Proxima Cen., GJ 3631, TWA 22, Wolf 424), and one M6 dwarf (Wolf 359) that exhibited multiple flares. We estimated the average FFD for each spectral bin. Due to the small number of flares in the early M dwarfs and the limited number of stars, we grouped M0-M3 dwarfs into a single bin. We note that we did not account for differing flare energy sensitivity and observation duration across various spectral types when grouping them into a single bin. The FFDs were fitted using the method described above. Additionally, we investigated the fits by combining M4-M6 dwarfs into a single bin to compare the FFDs between fully convective and partially convective M dwarfs, although the exact boundary between these two classes is hard to identify with spectral type alone. The results of FFD fittings for each spectral type and each energy bands are summarized in Table \ref{table:FFD_fitting}. We also list the number of flares and the minimum and maximum energies used for fitting.

In Figure \ref{fig:FFD by spt with models and power-law}, we present the FFDs for each spectral type: M0-M3, M4, M5, and M6. Within each subplot, a comparison is drawn between the observed NUV FFD and the observed TESS FFD, alongside the fit to NUV FFDs predicted from the TESS flare energies using the F13+12.5$^{*}$5F11 model and the PL fit.  The description of data points and lines is akin to that in Figure \ref{fig:FFD_GJ3631}. As we are no longer interested in comparing the FFD estimated using the energy ratio of 9000 K BB, the corresponding fit to FFD for the 9000 K BB model is not depicted. Like the case in Figure \ref{fig:FFD_GJ3631}, we find that the fit to NUV FFD predicted by the PL fit deviates away from the observed NUV FFD at lower flare energies in the case of M6 dwarf, the reason for which is explained in Section \ref{subsubsection:ffd generation and fitting}. 

Notably, this study represents the first time we can compare the FFDs of NUV and optical data using contemporaneous data. To visualize the results, we plotted the observed and the corresponding fitted FFDs of each spectral type for both NUV and TESS passbands in Figure \ref{fig:FFDs by spt,nuv & tess}, respectively. It is evident that the FFDs follow similar trends in both passbands. Specifically, the FFDs exhibit less steep slopes for NUV flares compared to TESS flares, and the slopes are consistent for all spectral types within the given errors. We also remind the readers that the values of slopes depend on the ranges of energies and the number of flares used for fitting. NUV flares are fitted for lower energy ranges, while TESS flares are fitted for higher energy ranges. Additionally, the number of flares used for fitting TESS flares is higher than that of NUV flares.
%
\begin{figure*}[htbp]
\centering
\includegraphics[width=0.45\textwidth,height=0.3\textwidth]{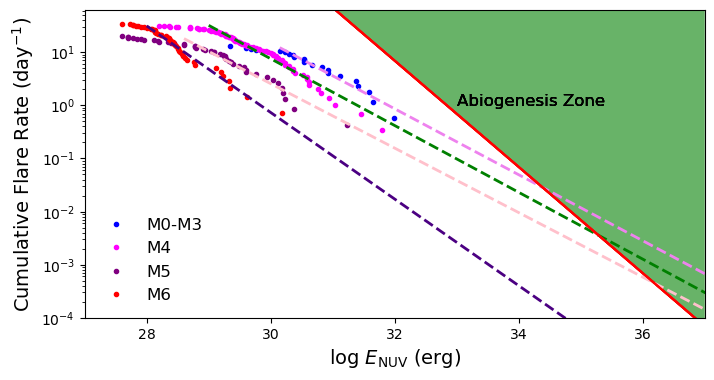}
\vspace{0.00mm}
\includegraphics[width=0.45\textwidth,height=0.3\textwidth]{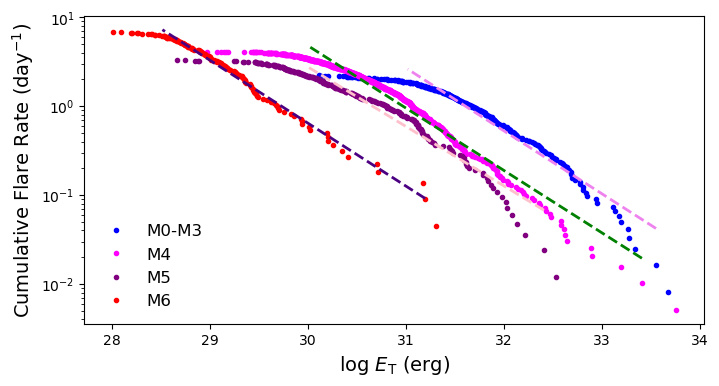}
\caption{\textit{Left Panel}: Frequency distribution of NUV flares as a function of spectral type. The dots represent the observed data, and the dashed lines are the linear fit to the FFDs. The FFDs of M0-M3, M4 and M5 have comparable slopes. The green area in the upper right corner corresponds to the abiogenesis zone estimated using the \cite{2018SciA....4.3302R} relation. This figure suggests that M0-M3, M4, and M5 dwarfs have NUV flare rates necessary for abiogenesis. However, the sample size of flares in each spectral type is small. \textit{Right Panel:} Frequency distribution of TESS flares as a function of spectral type. The dots represent the observed data, and the dashed lines are the linear fit to the FFDs. The FFDs of TESS flares have steeper slopes compared to those of NUV flares. However, it is important to note that the range of energies used for fitting is different. We observe that both TESS and NUV flares follow a similar distribution trend as a function of spectral type.}
\label{fig:FFDs by spt,nuv & tess}
\end{figure*}

\subsubsection{Abiogenesis} \label{subsubsectiion:agiogenesis}
\cite{2018SciA....4.3302R} performed a study to investigate the influence of NUV light on the initiation of photochemical reactions that contribute to the formation of fundamental components necessary for the origin of life, known as abiogenesis. Their results are useful for exploring the possibility of emergent life on exoplanets located within the liquid water habitable zones of various stars. They found that the application of NUV light within the wavelength range of 200-280 nm to a mixture of HCN and SO$^{2-}_{3}$ resulted in the production of RNA pyrimidine nucleotide precursors. Conversely, the absence of NUV light in the same mixture led to the formation of inert adducts that lacked any prebiotic potential. \cite{2018SciA....4.3302R} defined the minimum NUV flux required for abiogenesis as the level of flux that initiates a 50\% yield of the photochemical product. This minimum flux corresponds to a specific flux of 6.8 $\times$ 10$^{9}$ photons cm$^{-2}$ s$^{-1}$ \AA$^{-1}$ across the 200-280 nm wavelength band, which upon integrating across those wavelengths is equivalent to $F_{min}$ $\sim$45 erg cm$^{-2}$ s$^{-1}$. According to \cite{2018SciA....4.3302R}, the minimum cumulative rate ($\nu$) of flares at a given $U$-band energy ($E_{U}$) that satisfies the necessary yield is:
\begin{equation}
\nu = \frac{8 \times 10^{27} erg/s}{E_{\rm U}}
\end{equation}
However, this equation is based on the assumption of the Great AD Leo flare of 1985 and its spectrum to calculate the number of photons per cm$^{-2}$ \AA$^{-1}$, and hence, the NUV energy in the range of 200 nm to 280 nm. 
We no longer need to rely on this assumption, as we now have a good sample of NUV flares and our own estimation of NUV flare energies. Therefore, we use the NUV flare energies ($E_{\rm NUV}$) to constrain the abiogenesis zone and express the above equation as:
 \begin{equation}
\nu = \frac{6.9 \times 10^{32} erg/day}{E_{\rm NUV}}
\label{eq:abiogenesis zone}
\end{equation} 
In the left panel of Figure \ref{fig:FFDs by spt,nuv & tess}, the green zone corresponds to the abiogenesis zone defined by Equation \ref{eq:abiogenesis zone}. In this figure, the fits to NUV FFDs of each spectral type are extrapolated to higher energies to see where they intersect the boundary line defining the abiogenesis zone. We observe that the fits to FFDs of M0-M3, M4, and M5 dwarfs intersect the line at NUV energies log $E\sim$(34 - 36) erg. Since M dwarfs are capable of emitting large flares with such energies \citep{1976ApJS...30...85L,2016ApJ...829...23D,2016ApJ...832..174O}, left panel of Figure \ref{fig:FFDs by spt,nuv & tess} suggests that M0-M5 dwarfs can produce flares that could initiate abiogenesis while M6 might not be able to produce such flares. Since we have only one flaring M6 dwarf in our sample, we need a larger sample of stars with spectral type M6 to confirm this.
%
\subsection{The NUV activity-rotation relation}\label{subsection:nuv activity-rotation relation}
For the most rapid rotators, 
two crucial indicators of magnetic activity exhibit saturation. The first indicator, X-ray emission, is observed to saturate at values of $L_{\rm X}$/$L_{\rm bol}$ $\approx$ 10$^{-3}$ \citep{1984A&A...133..117V,1985ApJ...292..172M,2003A&A...397..147P,2011ApJ...743...48W}. Similarly, the second indicator, H$\alpha$ emission, also saturates at values of $L_{\rm H\alpha}$/$L_{bol}$ $\approx$ 10$^{-3.8}$ \citep{2014ApJ...795..161D,2017ApJ...834...85N}. Here, $L_{\rm X}$, $L_{H\alpha}$, and $L_{bol}$ represent the X-ray, H$\alpha$, and bolometric luminosity, respectively. Notably, this saturation phenomenon is observed irrespective of the stellar spectral type. Saturation is observed to set in at rotation periods of 1-10 d for solar-type stars and early M dwarfs, corresponding to Rossby numbers ($R_{\rm o}$ = $P_{\rm rot}$/$\tau$, where $\tau$ is the convective turnover time; \citealt{1984ApJ...279..763N}) of order $R_{o} \sim$0.13 \citep{2011ApJ...743...48W}. 

\cite{2016MNRAS.463.1844S} investigated the connection between rotation and chromospheric activity in a sample of 32 M dwarfs using GALEX NUV and K2 data. Among these stars, 13 exhibited slower rotation rates and had periods that were considered less reliable. To isolate the contribution of upper atmospheric activity, they subtracted the photospheric component from the observed NUV emission, focusing exclusively on the excess NUV flux. This excess NUV flux was then utilized to analyze the relationship between rotation and chromospheric NUV activity. Most of the stars in their sample, which had NUV detections, were slow rotators. \cite{2016MNRAS.463.1844S} observed two distinct regimes, saturation regime and an unsaturated (linear) regime, in the plots of luminosity ($L^{'}_{\rm NUV}$) versus rotation period and $L^{'}_{\rm NUV}$/$L_{\rm bol}$ versus Rossby number. Here, $L^{'}_{\rm NUV}$ represents the luminosity associated with the excess NUV flux. Notably, the saturation in NUV luminosity persisted for rotation periods of up to $\sim$40 days, which extended beyond the critical period of $\sim$10 days observed in X-ray and H$\alpha$ activity indicators. 

\cite{2020AJ....160...12D} also used GALEX NUV data to study the rotation-activity relationship in a sample of 133 red giant stars. They found that the NUV excess is correlated with the rotation period and Rossby number, exhibiting similar trends to those found in M dwarfs. These trends include the saturation of activity among fast rotators. Likewise, \cite{2021ApJ...907...91L} studied the evolution of strong UV emission lines as a function of rotation and age for early M dwarfs (M0 - M2.5). They found that the surface fluxes of strong UV emission lines show saturation for rotation periods up to $\sim$10 days  and $R_{o}$ $\sim$0.1, and then a power-law decline in their values (see their Figures 4 and 5). 

Similar to the works done by \cite{2016MNRAS.463.1844S}, \cite{2020AJ....160...12D} and \cite{2021ApJ...907...91L}, we also attempt to investigate the excess NUV activity-rotation relationship for the sample of M dwarfs in this study. We used synthetic photospheric spectra generated using the PHOENIX atmosphere code \citep{Allard1995, Hauschildt1999, Allard2012, Husser2013}, computing a 64-layer photosphere model in radiative-convective equilibrium corresponding to the effective temperature, surface gravity, metallicity and mass of each star. We obtained these stellar properties from the NASA Exoplanet Archive\footnote{\url{https://exoplanetarchive.ipac.caltech.edu/}} and estimated metallicity using the V, J, K magnitude relation in \cite{2013AJ....145...52M}. These models yield fluxes consistent with that at the stellar surface, so we scaled the values by the radius and distance (R$^2$/d$^2$). We then estimated the photospheric contribution to the NUV flux through the \textit{Swift} UVM2 filter and subtracted it from the total NUV flux to estimate the excess NUV flux. 

\begin{figure*}
   \centering
   \includegraphics[width=0.45\textwidth,height=0.35\textwidth]{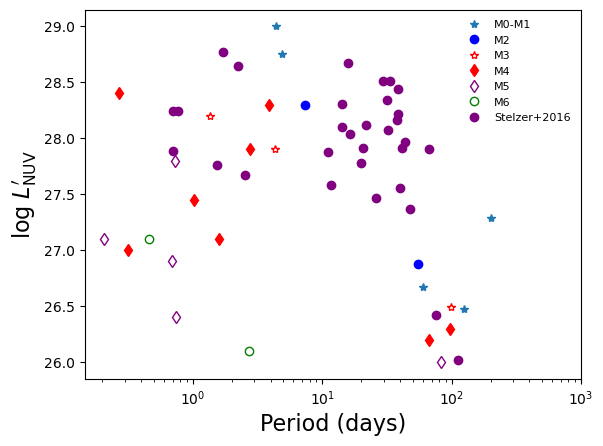}
   \vspace{0.00mm}
   \includegraphics[width=0.45\textwidth,height=0.35\textwidth]{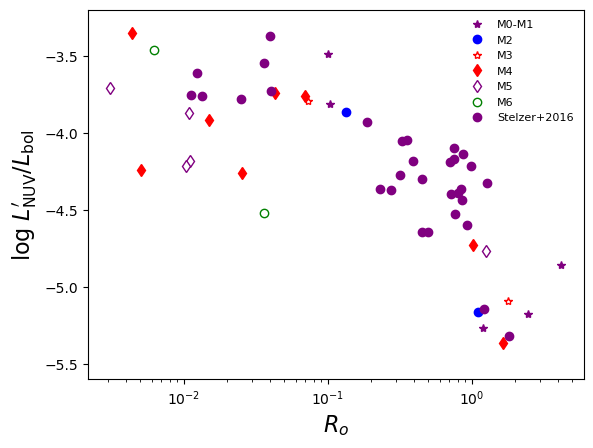}
    \caption{\textit{Left Panel}: Excess NUV luminosity plotted against rotation period for M dwarfs. The purple dots represent the results from \cite{2016MNRAS.463.1844S}. \textit{Left Panel}: Excess NUV to bolometric luminosity ratio ($L^{'}_{\rm Nuv}$/$L_{\rm bol}$) plotted against the Rossby number ($R_{o}$ = $P_{\rm rot}$/$\tau_{\rm conv}$). The purple dots represent the results from \cite{2016MNRAS.463.1844S}.}
     \label{fig:nuv rotation activity}
\end{figure*}

The left panel in Figure \ref{fig:nuv rotation activity} compares the excess NUV luminosity of our target stars to their rotation periods, with the \cite{2016MNRAS.463.1844S} data points shown for context. Different symbols are used for each spectral type. 
Interestingly, we observe no clear trend. 
In contrast to the trends observed with X-ray and H$\alpha$ activity indicators, it is noteworthy that some fast rotators like Wolf 359 and WX UMa exhibit relatively low levels of NUV emission, while some slow rotators like Lacaille 9352 and GJ 832 exhibit relatively high levels of NUV emission. Such a trend was also noticed by \cite{2016MNRAS.463.1844S}. This can be attributed to the differences in the magnetic dynamo operating in those stars. While Lacaille 9352 and GJ 832 are early M dwarfs, they have solar-like dynamo. On the other hand Wolf 359 and WX UMa are mid-M dwarfs, are fully convective and hence have a different magnetic dynamo. The timeline of the evolution of magnetic dyanamo and hence activity levels is a strong function of stellar mass \citep{2008AJ....135..785W,2021A&A...649A..96J}. 

In the right panel of Figure \ref{fig:nuv rotation activity}, we compare the fractional NUV luminosities ($L^{'}_{\rm Nuv}$/$L_{\rm bol}$) to the Rossby numbers ($R_{o}$). We also overplot the results of \cite{2016MNRAS.463.1844S} which are represented by red dots. For each star, we estimated the values of $\tau_{conv}$ based on its effective temperature ($T_{\rm eff}$) using equation 36 from \cite{2011ApJ...741...54C}, which includes an extrapolation for $T_{\rm eff}$ $<$ 3300 K. The determination of convective turnover times for M dwarfs beyond the fully convective boundary is still a subject of debate, primarily due to the likelihood of their dynamos being driven by fundamentally different processes compared to more massive stars \citep{2001ApJ...559..353M,2007ApJ...656.1121R}. However, \cite{2011ApJ...741...54C} note that the extrapolation of these turnover times to lower effective temperatures aligns reasonably well with the semi-empirical estimates conducted by \cite{2009ApJ...692..538R}. We compiled the values of $R_{o}$ for two stars Proxima Centauri and YZ Ceti from literature. For Proxima Cen., $R_{o}$ $\sim$0.65 \citep{2016Natur.535..526W} and for YZ Ceti $R_{o}$ $\sim$0.50 \citep{2023NatAs...7..569P}. The values of $R_{o}$ for those stars from our estimates are higher than the literature values by a factor of $\sim$2 which might be because of different theoritical models used to estimate them. 

Unlike in the left panel of Figure \ref{fig:nuv rotation activity}, we see some trend between $L^{'}_{\rm Nuv}$/$L_{\rm bol}$ and $R_{o}$ in the right panel. Though it is not very clear, it suggests that the saturation regime might extend beyond a Rossby number (Ro) value of $\sim$0.1. This aligns with the extension of saturation feature beyond the critical period of $\sim$10 days (for X-ray and H$\alpha$ activity indicators) observed in the left panel of Figure \ref{fig:nuv rotation activity} for the case of slow rotators. 
%
%
%

\section{Summary and Conclusions}\label{section:summary}
We obtained an unprecedented contemporaneous \textit{Swift} and TESS data set for two dozen nearby M dwarfs with supporting data from K2 and HST 
to examine the relationship between flare properties and activity at NUV and optical wavelengths.  
Our sample includes well-known nearby flaring M dwarfs such as AU Mic, Proxima Centauri, Wolf 359, YZ CMi, and Ross 614, as well as other M dwarfs with lower levels or even the absence of flare activity. 

We summarize the results from this study below:
\begin{itemize}
\item EV Lac, Ross 614, and YZ CMi exhibit the highest fractions of flaring time compared to the other stars in the sample. These stars flared for approximately 13\% of their total observation time. Notably, these stars are young, but youth does not seem to be the only factor at play as there are other young stars that did not flare. The average fraction of flaring time for M0-M3 dwarfs is estimated to be 2.1\%, while for M4-M6 dwarfs, it is 5\%.

\item We do not find any discernible relationship between the rotation period and either NUV or optical flare rates.

\item We identify optical counterparts for $\sim$27\% of our NUV flares. We find that almost all optical flares observed during Swift/UVOT observing windows have NUV counterparts which suggests that the fraction of NUV flares having optical counterparts would increase if the flares were observed in NUV and optical passbands using similar cadences; the longer TESS cadences mean that some short duration optical flares were missed. 

\item NUV flares are much more frequent in comparison to optical flares. For a small sample of 15 flares sampled simultaneously at fast cadence by both TESS and \textit{Swift}, we find that the NUV flare durations are similar to or longer than their optical counterparts.

\item Most of this study's M dwarf flares have energies and durations comparable to solar white light flares. This suggests that solar flares and M dwarf flares share common scaling laws connecting reconnection timescales, released flare energies, and magnetic energies. 

\item The majority (68\%) of the optical flares have amplitudes less than 2\% relative to quiescence and 98\% of the NUV flares had amplitudes greater than 1.5$\times$ the quiescent level. This significant difference in amplitudes of optical and NUV flares is because M dwarfs have low levels of quiescent UV fluxes, enhancing the contrast between flaring and quiescent levels.

\item We find that there is a decreasing trend in the energy fractionation ($E_{\rm TESS}$/$E_{NUV}$) as $E_{NUV}$ increases, but note there is large scatter.

\item We present an empirical relationship between NUV ($E_{NUV}$) and optical ($E_{\rm TESS}$) flare energies of the form $\log E_{\rm TESS}$ = 0.823$\pm$0.059 $\log E_{\rm NUV}$ + 5.630$\pm$1.8. Note that our optical flare energy derivation relies on the assumption of a 10,000 K blackbody and then conversion to TESS band energy ($E_{\rm TESS}$) due to the absence of an absolution flux calibration. So, others using this relationship should derive their optical flare energies in a similar fashion.

\item We evaluate the performance of various RHD models and blackbody models in predicting NUV flare energies from optical flare energies. We find the RHD model, F13+12.5*5F11, can predict the NUV flare energies from TESS flare energies more accurately than others.

\item The FFDs for both NUV and optical flares exhibit comparable slopes across all spectral types: M0-M3, M4, M5, and M6. This alignment suggests the presence of a common physical mechanism responsible for generating both UV and optical flares, as discussed in the existing literature.


\item  The FFD slopes for the M0-M3, M4 and M5 dwarfs (but not the M6 dwarfs) are shallow enough that they cross into the abiogenesis zone.
\end{itemize}

Many current studies assessing the impacts of flares on exoplanets extrapolate their findings from white light flare studies to the UV spectrum (e.g., \citealt{2020AJ....160..219F}, \citealt{2020ApJ...902..115H}). The Kepler/K2 and TESS missions have played a pivotal role in advancing the field of stellar flare science in relation to exoplanets, thanks to their high-cadence observations and extended mission durations. The relationship between NUV and optical flare properties established in this study will prove invaluable for leveraging the extensive dataset of optical flares obtained by Kepler/K2 and TESS. It will enable the estimation of approximate NUV flare rates on M dwarfs, providing essential input for models aiming to predict the effects of strong flares on the atmospheres of planets orbiting flaring M dwarfs. Furthermore, understanding energy fractionation in the optical/NUV bands is crucial for exploring new aspects of flare physics and for the development of more accurate flare models.

Furthermore, the results of this paper will play a crucial role in the planning of observations of NUV flares on M dwarfs, particularly for upcoming UV missions such as Star-Planet Activity Research CubeSat (SPARCS; \citealt{2018arXiv180809954A}) and Ultraviolet Transient Astronomy Satellite (ULTRASAT; \citep{2023arXiv230414482S}).

\section*{Acknowledgements}
We are thankful to S. Webb for helping to compile literature data of the solar and stellar flare durations obtained by various missions. This research has made use of the SVO Filter Profile Service (http://svo2.cab.inta-csic.es/theory/fps/) supported from
the Spanish MINECO through grant AYA2017-84089. This work has made use of
NASA’s Astrophysics Data System, the Simbad and Vizier
databases (operated at Centre de Données astronomiques de Strasbourg, Strasbourg, France). The material in this paper is based upon work supported by NASA under award No. 80NSSC19K0104, 80NSSC19K0315, 80GSFC21M0002, 80NSSC21K0362, 80NSSC23K0155, and 80NSSC22K0126. A portion of the presented work was also supported by NASA’s Astrophysics Data Analysis Program through grant 20-ADAP20-0016. 
This paper includes data collected by the TESS and K2 mission, which are available from the Mikulski Archive for Space Telescopes (MAST). Data was taken from the TESS Light Curves - All Sectors archive \citep{10.17909/t9-nmc8-f686}, the TESS "Fast" Light Curves - All Sectors archive \citep{10.17909/t9-st5g-3177} and the K2 Light Curves Campaign 14 archive \citep{10.17909/t9-sygx-1w59}.
Funding for the TESS mission is provided by NASA's Science Mission directorate. This work made use of data supplied by the UK \textit{Swift} Science Data Centre at the University of Leicester. This research has made use of the XRT Data Analysis Software (XRTDAS) developed under the responsibility of the ASI Science Data Center (ASDC), Italy. 

This research was carried out in part at the Jet Propulsion Laboratory, California Institute of Technology, under a contract with the National Aeronautics and Space Administration (80NM0018D0004). 

JRAD acknowledges support from the DiRAC Institute in the Department of Astronomy at the University of Washington. The DiRAC Institute is supported through generous gifts from the Charles and Lisa Simonyi Fund for Arts and Sciences, and the Washington Research Foundation.
L.D.V. acknowledges funding support from the Heising-Simons Astrophysics Postdoctoral Launch Program.
This research has made use of data obtained through the High Energy Astrophysics Science Archive Research Center Online Service, provided by the NASA/Goddard Space Flight Center. 

We thank the Space Telescope Science Institute for support through grant HST-GO-15463. This research is based on observations made with the NASA/ESA HST obtained from the Mikulski Archive for Space Telescopes (MAST) at the Space Telescope Science Institute, which is operated by the Association of Universities for Research in
Astronomy, Inc., under NASA contract NAS 526555. The HST observations are associated with program HST-GO-15463 (PI: A. Youngblood), obtained through the Director's Discretionary Time allocation. The specific observations analyzed can be accessed via https://doi.org/10.17909/kbv1-1244.

This project was supported in part by an appointment to the NRC Research Associateship Program at the U.S. Naval Research Laboratory in Washington, D.C., administered by the Fellowships Office of the National Academies of Sciences, Engineering, and Medicine. Basic Research in Astronomy at the U.S. Naval Research Laboratory is supported by 6.1 Base funding. 


\software{
Astropy \citep{astropy:2013, astropy:2018, astropy:2022}, Matplotlib \citep{Hunter:2007}, Numpy \citep{2020Natur.585..357H}, Lightkurve \citep*{lightkurve}, Python, IPython \cite{2007CSE.....9c..21P}, Jupyter \citep{Kluyver:2016aa}}
\facilities{TESS, \textit{Swift} (XRT \& UVOT), K2, HST} \\
\bibliographystyle{aasjournal}
\bibliography{astrobib}

\appendix
\section{Summary of NUV flare properties}
\label{appendix:NUV flares}




\begin{longtable}{|l|l|l|l|l|l|l|l|l|l|}
\caption{Summary of NUV flare properties. Here, $N$ is the total number of NUV flares observed on a given star in a given TESS sector, $N_{\rm U}$ is NUV flare ID, $T_{start}$ and $T_{stop}$ are flare start and stop times in units of BJD - 2457000 [d]. Likewise, `duration' is the total time in units of minutes during which the flare occurred and is estimated as the difference between $T_{start}$ and $T_{stop}$, `Peak Flux' is the flare flux during the peak of the flare, `amplitude' is the total height of the flare peak relative to the quiescent flux, `energy' is the NUV flare energy, and $L_{\rm q}$ is the quiescent NUV luminosity of a given star. A lower limit on the flare energy is given for each of the flares which were not observed for their full duration.} \label{tab:long} \\

\hline \multicolumn{1}{|c|}{\textbf{Name}} & \multicolumn{1}{c|}{\textbf{$N$}} & \multicolumn{1}{c|}{\textbf{$N_{\rm U}$}} & \multicolumn{1}{c|}{\textbf{$T_{\rm start}$}} & 
\multicolumn{1}{c|}{\textbf{$T_{\rm stop}$}} & \multicolumn{1}{c|}{\textbf{Duration}} & 
\multicolumn{1}{c|}{\textbf{Peak Flux}}  & \multicolumn{1}{c|}{\textbf{Amplitude}} & \multicolumn{1}{c|}{\textbf{$E_{\rm NUV}$}}
& \multicolumn{1}{c|}{\textbf{log $L_{q}$}}\\
\hline  \multicolumn{1}{|c|}{\textbf{}} & \multicolumn{1}{c|}{\textbf{}}  & \multicolumn{1}{c|}{\textbf{}} & \multicolumn{1}{c|}{\scriptsize{BJD - 2457000 [d]}} & 
\multicolumn{1}{c|}{\scriptsize{BJD - 2457000 [d]}} & \multicolumn{1}{c|}{[min]} & 
\multicolumn{1}{c|}{[cnts/s]}  & \multicolumn{1}{c|}{\textbf{}}  & \multicolumn{1}{c|}{[10$^{30}$ erg]}
& \multicolumn{1}{c|}{[erg s$^{-1}$]}\\
\hline 
\endfirsthead

\multicolumn{10}{c}%
{{\bfseries \tablename\ \thetable{} -- continued from previous page}} \\
\hline \multicolumn{1}{|c|}{\textbf{Name}} & \multicolumn{1}{c|}{\textbf{$N$}} & \multicolumn{1}{c|}{\textbf{$N_{\rm U}$}} & \multicolumn{1}{c|}{\textbf{$T_{\rm start}$}} & 
\multicolumn{1}{c|}{\textbf{$T_{\rm stop}$}} & \multicolumn{1}{c|}{\textbf{Duration}} & 
\multicolumn{1}{c|}{\textbf{Peak Flux}}  & \multicolumn{1}{c|}{\textbf{Amplitude}} & \multicolumn{1}{c|}{\textbf{$E_{\rm NUV}$}}
& \multicolumn{1}{c|}{\textbf{log $L_{q}$}}\\ \hline 
\multicolumn{1}{|c|}{} & \multicolumn{1}{c|}{}  & \multicolumn{1}{c|}{} & \multicolumn{1}{c|}{\scriptsize{BJD - 2457000 [d]}} & 
\multicolumn{1}{c|}{\scriptsize{BJD - 2457000 [d]}} & \multicolumn{1}{c|}{[min]} & 
\multicolumn{1}{c|}{[cnts/s]}  & \multicolumn{1}{c|}{}  & \multicolumn{1}{c|}{[10$^{30}$ erg]}
& \multicolumn{1}{c|}{[erg s$^{-1}$]}\\
\hline
\endhead

\hline \multicolumn{10}{|r|}{{Continued on next page}} \\ \hline
\endfoot

\hline \hline
\endlastfoot

AP Col & 16 & 1$^{\dag}$ & 1473.0841 & 1473.0869 & 4.2 & 25.4 & 39.1 & 3.8 & 27.4 \\
        &    & 2 & 1473.2044 & 1473.2063 & 2.9 & 2.4 & 3.7 & 0.33 & \\
        &   & 3$^{\dag}$ & 1482.0536 & 1482.0556 & 3.1 & 10.3 & 14.5 & 1.6 & \\
        &   & 4 & 1482.8307 & 1482.8344 & 5.5 & 8.6 & 9.8 & 1.3 & \\
        &   & 5 & 1482.9663 & 1482.9673 & 1.7 & 4.8 & 5.5 & 0.46 & \\
        &   & 6$^{\dag}$ & 1484.9602 & 1484.9648 & 7.2 & 2.0 & 2.4 & 0.59 & \\
        &   & 7 & 1485.8273 & 1485.8388 & 17.1 & 4.2 & 5.2 & 5.7 & \\
        &   & 8 & 1485.8874 & 1485.8886 & 2.2 & 1.1 & 1.7 & 0.12 & \\
        &   & 9 & 1485.9476 & 1485.9487 & 2.2 & 1.8 & 2.0 & 0.27 & \\
        &   & 10 & 1486.9588 & 1486.9619 & 5.0 & 1.9 & 2.8 & 0.39 & \\
        \hline
        & & & Cycle 3  & & & & & & \\
        \hline
        &  & 11$^{\dag}$ & 2190.0381 & 2190.0547 & 24.1 & 20.0 & 25.5 & 26.7 & 27.5 \\
        & & 12 & 2190.4401 & 2190.4409 & 1.7 & 1.9 & 2.3 & 0.27 & \\
        & & 13 & 2190.5027 & 2190.5038 & 1.7 & 3.5 & 3.7 & 0.37 &  \\
        & & 14 & 2190.5099 & 2190.5103 & 1.1 & 2.8 & 2.2 & 0.21 & \\
        & & 15 & 2190.9098 & 2190.9125 & 4.1 & 3.2 & 3.9 & 1.5 & \\
        & & 16 & 2191.1105 & 2191.1108 & 0.55 & 2.7 & 2.6 & 0.12 &  \\
        \hline
        AU Mic & 6 & 1 & 1340.7517 & 1340.7540 & 3.5 & 18.4 & 1.6 & 1.7 & 28.7 \\
             & & 2 & 1341.8232 & 1341.8249 & 2.8 & 23.5 & 2.2 & 3.5 & \\
             & & 3 & 1343.1545 & 1343.1556 & 2.2 & 13.0 & 1.5 & 2.4 & \\
             & & 4 & 1347.1380 & 1347.1421 & 6.1 & 19.7 & 1.6 & 6.8 & \\
             & & 5 & 1347.2673 & 1347.2767 & 13.6 & 35.5 & 2.9 & 45.5 &  \\
             \hline
             & & & Cycle 3 &  & & & & & \\
             \hline
             & & 6$^{\ddag}$ & 2058.2364 &  & $>$9.6 & 99.2 & 7.0 & $>$96 & 28.8 \\
     \hline
      DG CVn & 3 &    1 & 1953.3826 & 1953.3835 & 1.5 & 11.2 & 6.9 & 3.9 & 28.4 \\
             & &    2 & 1953.4511 & 1953.4536 & 3.9 & 6.4 & 3.9 & 8.7 & \\
             & &    3 & 1953.7063 & 1953.7071 & 1.3 & 4.2 & 2.7 & 1.3 & \\
    \hline
      DX Cnc & 0 &    & & & & & & & 27.1 \\
      \hline
      EV Lac & 9 &  1  &  1748.0401 &  1748.0408 & 1.1 & 14.0 & 2.5 &  0.40 & 27.9 \\
             & & 2$^{\ddag}$ &  1748.2379 & & $>$9.8 & 41.3 & 7.3 & $>$8.4 & \\
             & & 3$^{\dag}$ & 1749.0260 &  1749.0314 & 7.9 & 31.1 & 4.9 & 3.4 & \\
             & & 4&  1749.1059 & 1749.1075 & 2.4 & 15.0 & 2.4 & 0.57 & \\
             & & 5 &  1749.1663 & 1749.1669 & 0.92 & 60.2 & 9.6 & 1.4 & \\
             & & 6 & 1749.2232 &  1749.2238 & 1.1 & 19.1 & 2.8 & 0.48 & \\
             & & 7$^{\ddag}$ &  1749.2952 & & $>$1.5 & 339.3 & 41.8 & $>$13.3 & \\
             & & 8$^{*}$ & & 1750.2256 & $>$11.4 & 20.9 & 3.1 & $>$3.4  & \\
             & & 9$^{*}$ & 1750.2913 &  1750.2917 & 0.74 & 14.2 & 2.1 & 0.22 & \\
      \hline
      Fomalhaut C & 8 & 1 & 1361.8200 & 1361.8216 & 2.8 & 1.1 & 2.1 & 0.08 & 27.0 \\
                  & & 2 & 1361.8223 & 1361.8239 & 2.8 & 1.1 & 3.1 & 0.14 & \\
                  & &  3 & 1362.5573 & 1362.5584 & 1.8 & 2.1 & 7.2 & 0.24 &  \\
                  & & 4 & 1363.5597 & 1363.5603 & 1.1 & 3.1 & 9.5 & 0.21 & \\
                  & & 5 & 1364.7419 & 1364.7442 & 3.9 & 1.0 & 3.3 & 0.31 & \\
                  & & 6 & 1366.5982 & 1366.6007 & 3.9 & 5.7 & 14.5 & 0.83 & \\
                  & & 7 & 1366.6013 & 1366.6066 & 7.9 & 3.4 & 3.7 & 0.51 & \\
                  & & 8 & 1366.7375 & 1366.7387 & 1.8 & 2.1 & 5.3 & 0.16 & \\
    \hline
      GJ 1284 & 1 &1 & 1381.4844 & 1381.4870 & 3.9 & 6.8 & 4.5 & 4.3 & 28.3 \\
      \hline
      \hline
      GJ 832 & 0 & & & & & & & & 26.9 \\
      \hline
      GJ 1 & 0 & & & & & & & & 26.7 \\
      \hline 
      GJ 876 & 0 & & & & & & & & 26.3 \\
      \hline
      HIP 17695 & 2 &1 &1458.7890 & 1458.7916 & 3.9 & 3.9 & 3.4 & 3.7 & 28.3 \\
                & & 2 & 1461.1762 & 1461.1765 & 0.55 & 3.1 & 3.1 & 0.68  & \\
\hline
      Kapteyn's Star & 0 & & & & & & & & 26.5 \\
      \hline
      Lacaille 9352 & 1 & 1 & 1365.2023 & 1365.2056 & 5.0 & 15.3 & 4.2 & 0.43 & 27.3 \\
      \hline
      LP 776-25 & 3 & 1$^{\dag}$ & 1456.9436 & 1456.9449 & 2.0 & 8.4 & 7.3 & 2.3 & 28.2 \\
                & & 2$^{\dag}$ & 1461.8401 & 1461.8435 & 5.2 & 13.2 & 7.2 & 8.4 & \\
                & & 3 & & & $>$5.7 & $>$13.6 & $>$7.4 & $>$23.5 & \\
      \hline
      Luyten's Star & 1 & 1 & 1513.7822 & 1513.7868 & 7.2 & 0.9 & 4.1 & 0.07 & 26.5 \\
      \hline
      Prox. Cen. & 20 & & UVW1 Filter & & & & & & \\ 
      \hline
                & & 1$^{*}$ & 1584.2590 & 1584.2617 & 4.4 & 13.7 & 1.5 & 0.028 & 26.6 \\
                & & 2$^{*}$ & 1584.2660 & 1584.2669 & 1.5 & 17.4 & 2.2 & 0.016 & \\
                & & 3$^{*}$ & 1598.3523 & 1598.3537 & 2.2 & 16.3 & 2.2 & 0.022 & \\
                 & & 4 & 1604.3186 & 1604.3196 & 1.7 & 18.8 & 2.3 & 0.18 & \\
                 & & 5 & 1620.2601 & 1620.2613 & 2.2 & 20.8 & 1.5 & 0.015 & \\
                 & & 6 & 1620.3258 & 1620.3259 & 0.4 & 14.5 & 1.6 & 0.005 & \\
                 & & 7 & 1620.3287 & 1620.3306 & 3.3 & 16.3 & 1.8 & 0.025 & \\
                 \hline
                 & & & UVM2 filter& & & & & & \\
                 \hline
                 & & 8 & 1601.7311 & 1601.7314 & 0.55 & 4.8 & 5.3 & 0.006 & 26.0 \\
                 & & 9 & 1601.8003 & 1601.8007 & 0.74 & 3.1 & 3.6 & 0.007 & \\
                 & & 10 & 1602.6023 & 1602.6029 & 1.1 & 2.6 & 2.7 & 0.005 & \\
                 & & 11$^{\dag}$ & 1602.7232 & 1602.7239 & 1.3 & 59.7 & 49.7 & 0.061 & \\
                 & & 12 & 1602.7909 & 1602.7919 & 1.7 & 12.3 & 9.0 & 0.013 & \\
                 & & 13 & 1605.5754 & 1605.5760 & 1.1 & 5.5 & 4.4 & 0.008 & \\
                 & & 14 & 1605.6447 & 1605.6450 & 0.55 & 4.3 & 2.2 & 0.004 & \\
                 & & 15$^{*}$ & 1610.8212 & 1610.8225 & 2.0 & 4.6 & 3.3 & 0.009 & \\
                 & & 16 & 1652.8675 & 1652.8719 & 6.4 & 6.8 & 5.3 & 0.040 & \\
                 & & 17$^{*}$ & 1652.9252 & 1652.9282 & 4.4 & 13.2 & 9.4 & 0.061 & \\
                 & & 18$^{*}$ & 1659.6941 & 1659.6946 & 0.9 & 6.1 & 4.0 & 0.005 & \\
                 & & 19$^{*}$ & 1659.6957 & 1659.7026 & 10.1 & 11.1 & 9.4 & 0.12 & \\
                 & & 20$^{*}$ &  & 1665.5344 & $>$2.4 & 12.7 & 8.5 & $>$0.040 & \\
            \hline
      Ross 614 & 22 & 1$^{\dag}$ & 1472.9496 & 1472.9536 & 5.9 & 30.3 & 16.9 & 1.1 & 27.1 \\
                & & 2  & 1473.0121 & 1473.0126 & 0.92& 8.4 & 4.7& 0.085 & \\
                & & 3$^{*}$ & 1477.0694 &1477.0726 & 4.8 & 6.4  & 4.7 & 0.34 & \\
                & & 4$^{*}$ & 1477.0771 & 1477.0777 & 1.1 & 2.8 & 2.1 & 0.05 & \\
                & & 5$^{*}$ & 1477.2653 & 1477.2681 & 4.2 & 17.0 & 12.4& 0.92 & \\
                & & 6$^{*}$ & 1477.9891 & 1477.9915 & 3.7 & 5.4 & 3.8 & 0.31 & \\
                & & 7 & 1479.0510 & 1479.0581 & 10.3 & 35.6 & 20.7 & 1.2 & \\
                & & 8 & 1479.1241 & 1479.1243 & 0.55 & 5.2 & 3.0 & 0.05 & \\
                & & 9$^{\dag}$ & 1479.9115 & 1479.9195 & 11.8 & 11.4 & 7.1 & 1.2 & \\
                & & 10 & 1479.9762 & 1479.9765 & 0.55 & 3.9 & 2.4 & 0.03 & \\
                & & 11$^{\dag}$  & 1479.9832 & 1479.9908 & 11.0 & 30.7 & 19.2 & 2.5 & \\
                & & 12 & 1482.3105 & 1482.3111 & 1.1 & 3.7 & 2.6 & 0.07 & \\
                & & 13 &1483.0383  & 1483.0386 & 0.55 & 14.4 & 2.4 & 0.05 & \\
                & & 14  & 1483.0392 & 1483.0403 & 1.7 & 14.4 & 8.4 & 0.25 & \\
                & & 15 & 1483.0456 & 1483.0465 & 1.5 & 6.1 & 3.6 & 0.10 & \\
                & & 16$^{\dag}$ & 1483.0953 & 1483.0970 & 2.6 & 5.7 & 2.1 & 0.11  & \\
                & & 17$^{\dag}$ & 1483.0984 & 1483.0991 & 1.3 & 21.6 & 7.9 & 0.23 & \\
                & & 18$^{\dag}$ & 1483.1002 & 1483.1022 & 3.1 & 6.3 & 2.3 & 0.18 & \\
                & & 19 & 1483.9038 & 1483.9052 & 2.2 & 3.9 & 2.6 & 0.15 & \\
                & & 20$^{\dag}$ & 1483.9083 & 1483.9089 & 1.1 & 7.1 & 4.7& 0.14 & \\
                & & 21$^{\dag}$ & 1483.9587 & 1483.9611 & 3.7 & 3.8 & 2.5 & 0.18 & \\
                & & 22 & 1484.0268 & 1484.0281 & 2.0 & 3.5 & 2.4 & 0.07 & \\
      \hline
      TWa 22 & 6 & 1 &1551.9795 & 1551.9837 & 6.6 & 3.5 & 2.8 & 2.4 & 27.8 \\
            & & 2 & 1552.1094 & 1552.1117 & 3.9 & 1.5 & 2.0 & 0.87 & \\
            & & 3 & 1553.0385 & 1553.0402 & 2.6 & 2.7 & 6.6 & 1.7 & \\
            & & 4 & 1554.9663 & 1554.9686 & 3.9 & 1.1 & 2.6 & 1.3 & \\
            & & 5 & 1557.7611 & 1557.7631 & 3.3 & 1.1 & 2.9 & 1.6 & \\
            & & 6 & 1558.8177 & 1558.8183 & 1.1 & 1.3 & 3.7 & 0.59 & \\
      \hline
      V1005 Ori & 4 & 1$^{\dag}$ & 1448.8960 & 1448.8999 & 5.9 & 17.3 & 5.4 & 38.2 & 29.0 \\
                & & 2$^{*}$ & 1450.5578 & 1450.5609 & 4.6 & 11.3 & 4.0 & 25.6 & \\
                & & 3 & 1455.0022 & 1455.0024 & 0.55 & 8.4 & 2.5 & 2.0 & \\
                & & 4 & 1455.0027 & 1455.0036 & 1.5 & 6.6 & 2.0 & 4.6 & \\
    \hline
      Wolf 359          & 33 & 1$^{*}$ & 2566.6048 & 2566.6056 & 1.3 & 1.7& 8.8 & 0.023 & 26.10 \\
    (with TESS) & & 2$^{*}$ & 2566.6061 & 2566.6075 & 2.2 & 20.1 & 51.8 & 0.13 & \\
                & & 3 & 2567.0573 & 2567.0579 & 1.1 & 1.6 & 5.0 & 0.014 & \\
                & & 4 & 2567.1946 & 2567.1948 & 0.55 & 1.2 & 3.1 & 0.006 & \\
                & & 5$^{\dag}$ & 2567.4039 & 2567.4063 & 3.7 & 40.9 & 70.2 & 0.21 & \\
                & & 6 & 2569.3817 & 2569.3823 & 1.1 & 2.6 & 6.4 & 0.017 & \\ 
                & & 7 & 2569.0498 & 2569.0514 & 2.8 & 1.6 & 3.7 & 0.030 & \\
                & & 8 & 2568.9305 & 2568.9313 & 1.3 & 4.7 & 10.6 & 0.025 & \\
                & & 9 & 2568.7330 & 2568.7336 & 0.92 & 4.9 & 6.5 & 0.015 & \\
                & & 10 & 2568.4603 & 2568.4605 & 0.55 & 1.7 & 5.3 & 0.010 & \\
                & & 11$^{\dag}$ & 2568.3906 & 2568.3913 & 1.3 & 4.1 & 10.7 & 0.044 & \\
                & & 12 & & 2568.1949 & $>$2.6 & 7.0 & 8.2 & $>$0.032 & \\
                & & 13 & 2568.1967 & 2568.1969 & 0.55 & 3.0 & 3.5 & 0.005 & \\
                & & 14$^{\ddag}$ &  & 2569.0006 & $>$17.1 & 9.9 & 24.6 & $>$1.5 & \\
                & & 15$^{\dag}$ & 2570.3258 & 2570.3279 & 3.1 & 25.5 & 53.0 & 0.22 & \\
                & & 16 & 2570.2368 & 2570.2373 & 0.92 & 2.4 & 2.9 & 0.007 & \\
                & & 17 & 2570.2457 & 2570.2461 & 0.74 & 6.8 & 8.0 & 0.014 & \\
                & & 18 & 2570.1765 & 2570.1773 & 1.5 & 2.2 & 6.4 & 0.021 & \\
                & & 19$^{\dag}$ & 2570.1072 & 2570.1082 & 1.7 & 5.1 & 11.8 & 0.038 & \\
                & & 20 & 2569.9221 & 2569.9230 & 1.5 & 4.1 & 12.3 & 0.037 & \\
                & & 21 & 2569.8556 & 2569.8564 & 1.3 & 2.7 & 7.0 & 0.021 & \\
                & & 22 & 2569.7939 & 2569.7943 & 0.74 & 1.4 & 5.1 & 0.013 & \\
                & & 23 & 2569.5880 & 2569.5882 & 0.55 & 2.5 & 6.0 & 0.012 & \\
                & & 24 & 2569.5886 & 2569.5909 & 3.5 & 6.4 & 15.8 & 0.060 & \\
                & & 25$^{\dag}$ & 2570.5064 & 2570.5070 & 1.1 & 2.5 & 6.3 & 0.027 & \\
                & & 26 & 2570.5758 & 2570.5772 & 2.2 & 2.6 & 6.7 & 0.033 & \\
                & & 27 & 2570.9173 & 2570.9177 & 0.74 & 3.0 & 7.7 & 0.013 & \\
                & & 28 & 2570.9198 & 2570.9203 & 0.92 & 1.9 & 4.8 & 0.014 & \\
                & & 29 & 2571.0525 & 2571.0529 & 0.74 & 1.2 & 4.2 & 0.012 & \\
                & & 30$^{\dag}$ & 2571.1166 & 2571.1209 & 6.4 & 26.2 & 31.3 & 0.17 & \\
                & & 31 & 2571.2359 & 2571.2363 & 0.74 & 3.0 & 2.9 & 0.004 & \\
                & & 32 & 2571.8443 & 2571.8454 & 1.8 & 1.6 & 4.6 & 0.014 & \\
                & & 33 & 2572.0482 & 2572.0497 & 2.4 & 3.4 & 8.9 & 0.057 & \\
                
    \hline
     Wolf 359 & 15 & 34$^{\dag}$ & 3075.1950 & 3075.1953 & 0.74 & 5.8 & 3.3 & 0.006 & 26.10 \\
    (with $K2$) & & 35$^{\dag}$ & 3075.1981 & 3075.1987 & 0.92 & 5.4 & 3.7 & 0.008 & \\
                & & 36$^{\dag}$ & 3095.6563 & 3095.6573 & 1.7 & 5.1 & 7.8 & 0.04 & \\
                &  & 37 & 3095.7215 & 3095.7227 & 2.0 & 3.0 & 9.2 & 0.03 & \\
                & & 38$^{\dag}$ & 3096.6433 & 3096.6446 & 2.0 & 22.3 & 53.2 & 0.16 & \\
                & & 39 &  3096.7180 & 3096.7184 & 0.74 & 1.8 & 6.4 & 0.02 & \\
                & & 40 & 3096.7695 & 3096.7731 & 5.3 & 4.6 & 3.8 & 0.04 & \\
                & & 41$^{\dag}$ & 3116.5682 & 3116.5702 & 3.3 & 1.4 & 2.9 & 0.03 & \\
                & & 42$^{\dag}$ & 3116.8306 & 3116.8373 & 9.8 & 13.7 & 25.7 & 0.41 & \\
                & & 43 & 3117.6946 & 3117.6949 & 0.55 & 2.0 & 4.4 & 0.009 & \\
                & & 44$^{\dag}$ & 3117.6963 & 3117.6969 & 1.10 & 2.0 & 6.5 & 0.02 & \\
                &  & 45$^{\dag}$ & 3118.0430 & 3118.0435 & 0.92 & 2.8 & 5.8 & 0.02 & \\
                & & 46$^{\dag}$ & 3118.0461 & 3118.0470 & 1.5 & 3.8 & 7.8 & 0.03 & \\
                & & 47$^{\dag}$ & 3118.1099 & 3118.1106 & 1.10 & 1.7 & 3.4 & 0.007 & \\
                & & 48$^{\dag}$ & 3118.8221 & 3118.8229 & 1.3 & 5.1 & 11.9 & 0.03 & \\
    \hline
     
 Wolf 424 & 5 & 1 & 1948.2044 & 1948.2053 & 1.5 & 4.5 & 3.3 & 0.082 & 27.1 \\
                & & 2 & 1948.7302 &  & $>$1.3 & 5.8 & 6.1 & $>$0.23 & \\
                & & 3$^{\dag}$ & 1949.3953 & 1949.3990 & 5.5 & 34.0 & 35.1 & 1.5 & \\
                & & 4 & 1949.5879 & 1949.5882 & 0.55 & 3.4 & 3.2 & 0.060 & \\
                & & 5 & 1949.5959 & 1949.5964 & 0.92 & 3.8 & 3.5 & 0.087 & \\
      \hline
      YZ Ceti & 11 & 1 & 1401.6173 & 1401.6194 & 3.1 & 7.5 & 21.3 & 0.13 & 26.2 \\
             & & 2 & & 1402.2084 & $>$6.1 & 3.0 & 8.1 & $>$0.16 &  \\
             & & 3 & 1402.5384 & 1402.5407 & 3.5 & 2.3 & 10.6 & 0.11 & \\
             & & 4 & 1403.9915 & 1403.9924 & 1.5 & 1.1 & 4.8 & 0.019 & \\
             & & 5 & 1404.0674 & 1404.0682 & 1.3 & 1.1 & 4.8 & 0.024 & \\
             & & 6 & 1404.1342 & 1404.1349 & 1.3 & 2.0 & 8.2 & 0.023 & \\
             & & 7 & 1404.1360 &  & $>$0.74 & 2.0 & 8.4 & $>$0.032 & \\
             & & 8 & 1404.6538 & 1404.6559 & 3.1 & 7.1 & 21.9 & 0.11 & \\
             &  & 9 & 1404.6662 & 1404.6675 & 2.0 & 6.1 & 18.6 & 0.17 & \\
             & & 10 & 1404.7819 & 1404.7824 & 0.92 & 1.7 & 5.1 & 0.016 & \\
             &  & 11$^{*}$ & 1407.6387 & 1407.6393 & 1.1 & 4.7 & 22.3 & 0.064 & \\
     \hline
     YZ CMi & 21 & 1 & 1508.5441 & 1508.5475 & 5.2 & 17.6 & 3.9 & 2.3 & 27.9 \\
             & & 2$^{\ddag}$ & & 1508.6844 & $>$24.3 & 60.4 & 13.5 & $>$62.8 & \\
             & & 3 & 1509.5400 & 1509.5479 & 11.6 & 23.3 & 4.7 & 4.2 & \\
             & & 4 & 1509.5502 & 1509.5506 & 0.74 & 10.4 & 2.1 & 0.15 & \\
             & & 5 & 1509.5534 & 1509.5543 & 1.5 & 10.4 & 2.0 & 0.42 & \\
             & & 6 & 1509.5946 & 1509.5987 & 6.1 & 24.7 & 5.0 & 3.3 & \\
             & & 7 & 1509.6074 & 1509.6082 & 1.3 & 9.4 & 1.9 & 0.19 & \\
             & & 8 & 1509.6636 & 1509.6645 & 1.5 & 14.5 & 2.9 & 0.26 & \\
             & & 9 & 1510.5425 & 1510.5432 & 1.3 & 9.1 & 2.3 & 0.25 & \\
             & & 10 & 1511.7955 & 1511.7974 & 3.3 & 6.5 & 1.7 & 0.66 & \\
             & & 11 & 1511.8031 & 1511.8039 & 1.7 & 5.6 & 1.5 & 0.19 & \\
             & & 12$^{\dag}$ & 1513.1850 & 1513.1927 & 11.2 & 74.5 & 19.2 & 10.7 & \\
             & & 13 & 1513.3260 & 1513.3267 & 1.10 & 8.2 & 2.1 & 0.26 & \\
             & & 14 & 1514.3858 & 1514.3871 & 2.0 & 9.8 & 2.3 & 0.64 & \\
             & & 15 & 1514.9865 & 1514.9881 & 2.6 & 20.8 & 5.8 & 2.0 & \\
             & & 16$^{\dag}$ & 1514.9883 & 1514.9911 & 4.2 & 20.8 & 3.0 & 1.7 & \\
             & & 17 & 1515.0552 & 1515.0564 & 2.0 & 22.6 & 6.4 & 1.1 & \\
             & & 18 & 1515.0599 & 1515.0628 & 4.4 & 12.3 & 2.2 & 1.0 & \\
             & & 19$^{*}$ &  1516.1752 & 1516.1756 & 0.74 & 12.6 & 2.3 & 0.19 & \\
             & & 20$^{*}$ & 1516.3020 & 1516.3067 & 7.0 & 15.4 & 2.9 & 2.3 & \\
             & & 21$^{*}$ & 1516.3145 & 1516.3167 & 3.3 & 15.4 & 2.1 & 0.76 & \\
             \hline
             & & & Cycle 3 & & & & & & \\
             \hline
             & 10 & 1$^{\dag}$ & 2236.9036 & 2236.9162 & 18.4 & 83.5 & 26.6 & 29.1 & 27.8 \\
             & & 2 & 2236.9712 & 2236.9725 & 2.0 & 7.1 & 2.3 & 0.41 & \\
             &  & 3$^{\dag}$ & 2237.0412 & 2237.0467 & 8.1 & 28.8 & 9.2 & 4.1 & \\
             & & 4 & 2237.0542 & & $>$0.55 & 7.7 & 2.5 & $>$0.18 & \\
             & & 5$^{\dag}$ & 2237.7000 & 2237.7007 & 1.3 & 47.7 & 4.8 & 0.60 & \\
             & & 6$^{\ddag}$ & 2237.7061 &  & $>$9.2 & 41.6 & 4.2 & $>$6.4 & \\
             & & 7 & 2237.7658 & 2237.7662 & 0.74 & 15.8 & 2.6 & 0.28 & \\
             & & 8 & 2237.7671 & 2237.7675 & 0.74 & 21.1 & 3.3 & 0.32 & \\
             & & 9$^{\dag}$ & 2237.7801 & 2237.7815 & 2.2 & 41.0 & 6.5 & 2.0 & \\
             & & 10 & 2238.0325 & 2238.0339 & 2.2 & 10.5 & 5.8 & 1.5 & \\
     \hline
GJ 3631 & 16 & 1$^{*}$ & 2552.2628 & 2552.2643 & 2.8 & 1.1 & 4.0 & 0.21 & 26.9 \\ 
              & & 2$^{*}$ & 2552.4681 & 2552.4685 & 0.74 & 1.2 & 10.9 & 0.15 & \\
              & & 3 & 2556.2342 & 2556.2352 & 1.7 & 3.6 & 14.3 & 0.18 & \\
              & & 4 & 2559.3030 & 2559.3035 & 0.92 & 1.5 & 8.2 & 0.11 & \\
              & & 5$^{\dag}$ & 2561.0782 & 2561.0785 & 0.55 & 1.9 & 4.8 & 0.062 &  \\
              & & 6$^{\dag}$ & 2563.2749 & 2563.2908 & 23.0 & 6.9 & 63.1 & 16.9 & \\
              & & 7$^{\dag}$ & 2563.8788 & 2563.8798 & 1.7 & 2.2 & 6.9 & 0.17 &  \\
              & & 8 & 2564.4070 & 2564.4079 & 1.5 & 2.5 & 14.4 & 0.40 & \\
              & & 9 & 2564.4197 & 2564.4200 & 0.55 & 1.1 & 8.3 & 0.12 & \\
              & & 10 & 2568.7818 & 2568.7823 & 0.92 & 5.4 & 71.5 & 1.1 & \\
              & & 11$^{\dag}$ & 2568.1261 & 2568.1278 & 2.6 & 3.3 & 11.6 & 0.31 & \\
              & & 12 & 2570.7771 & 2570.7772 & 0.37 & 2.2 & 23.0 & 0.25 & \\
              & & 13 & 2575.6136 & 2575.6157 & 3.1 & 5.4 & 17.9 & 0.40 & \\
              & & 14 & 2575.6255 & 2575.6272 & 2.6 & 5.3 & 16.9 & 0.37 & \\
              & & 15 & 2576.4823 & 2576.4826 & 0.55 & 1.1 & 14.7 & 0.19 & \\
              & & 16 & 2576.4831 & 2576.4834 & 0.74 & 0.8 & 10.2 & 0.18 \\
    \label{table:UV flare properties}
\end{longtable}
\textbf{Note:}i)  A dagger ($\dag$) sign in the third column indicates that a given NUV flare has optical counterpart in TESS light curve and it was observed for full duration by Swift/UVOT. Likewise, a double dagger ($\ddag$) sign indicates that a given NUV flare has optical counterpart in TESS light curve and it was not observed for full duration by Swift/UVOT. An asterisk (*) sign indicates that simultaneous TESS data is not available during the flare. ii) Optical counterpart for three NUV flares with IDs $N_{U}$ = 16,17,18 appears to be a single complex flare ($T_{\rm start}$ = 1483.0925)  in TESS light curve because of longer cadence size, and hence are not resolved as separate flares. \\ 

\section{Properties of TESS/K2 flares which have NUV counterparts}
\label{appendix:TESS flares}

\begin{longtable}{|l|l|l|l|l|l|l|l|}
\caption{Properties of TESS/K2 flares which have NUV counterparts. The column $N_{\rm U}$ gives the ID of NUV flare in Appendix \ref{appendix:NUV flares} 
for which a given TESS/$K2$ flare in this table is its optical counterpart. The flare energies ($E_{\rm bol}$) are the 10,000 K BB bolometric energies and can be converted to TESS band energies ($E_{\rm TESS}$) using the ratio $E_{\rm TESS}$/$E_{\rm bol}$ = 0.18. Similarly, the ratio of $E_{\rm K2}$ to $E_{\rm bol}$ is 0.32 for the $K2$ band energy.} 
\\

\hline \multicolumn{1}{|c|}{\textbf{Name}} & \multicolumn{1}{c|}{\textbf{$N_{U}$}}  & \multicolumn{1}{c|}{\textbf{$T_{\rm start}$}} & 
\multicolumn{1}{c|}{\textbf{$T_{\rm stop}$}} & \multicolumn{1}{c|}{$T_{\rm peak}$} & 
\multicolumn{1}{c|}{\textbf{ED}}  & \multicolumn{1}{c|}{\textbf{log $E_{\rm bol}$}} & \multicolumn{1}{c|}{\textbf{Amplitude}}
\\
\hline  \multicolumn{1}{|c|}{\textbf{}} & \multicolumn{1}{c|}{\textbf{}}   & \multicolumn{1}{c|}{\scriptsize{BJD - 2457000 [d]}} & 
\multicolumn{1}{c|}{\scriptsize{BJD - 2457000 [d]}} & \multicolumn{1}{c|}{BJD - 2457000 [d]} & 
\multicolumn{1}{c|}{[min]}  & \multicolumn{1}{c|}{\textbf{[erg]}}  & \multicolumn{1}{c|}{} \\
\hline 
\endfirsthead

\multicolumn{8}{c}%
{{\bfseries \tablename\ \thetable{} -- continued from previous page}} \\
\hline \multicolumn{1}{|c|}{\textbf{Name}} & \multicolumn{1}{c|}{\textbf{$N_{U}$}}  & \multicolumn{1}{c|}{\textbf{$T_{\rm start}$}} & 
\multicolumn{1}{c|}{\textbf{$T_{\rm stop}$}} & \multicolumn{1}{c|}{$T_{\rm peak}$} & 
\multicolumn{1}{c|}{\textbf{ED}}  & \multicolumn{1}{c|}{\textbf{log $E_{\rm bol}$}} & \multicolumn{1}{c|}{\textbf{Amplitude}}  \\ \hline 
\multicolumn{1}{|c|}{} & \multicolumn{1}{c|}{}   & \multicolumn{1}{c|}{\scriptsize{BJD - 2457000 [d]}} & 
\multicolumn{1}{c|}{\scriptsize{BJD - 2457000 [d]}} & \multicolumn{1}{c|}{BJD - 2457000 [d]} & 
\multicolumn{1}{c|}{[min]}  & \multicolumn{1}{c|}{[erg]}  & \multicolumn{1}{c|}{}] \\
\hline
\endhead

\hline \multicolumn{8}{|r|}{{Continued on next page}} \\ \hline
\endfoot

\hline \hline
\endlastfoot
AP Col & 1 & 1473.08486&	1473.08763&	1473.08486&	1.8&	31.1 &	1.008\\
&3& 1482.05429&	1482.05429&	1482.05429&	0.51&	30.6&	1.004 \\
&6& 1484.96121&	1484.96538&	1484.96121&	1.3&	31.0&	1.003\\
&11&2190.03796 &	2190.05625&	2190.04074&	14.1 &	32.0 &	1.023\\
\hline
AU Mic & 6 &  2058.2366&	2058.29632&	2058.2366&	15.6&	33.4 &	1.009 \\
\hline
EV Lac & 2 & 1748.2384&	1748.2467&	1748.2384&	1.5&	31.5&	1.003\\
 &3 &  1749.0273&	1749.0328&	1749.0286&	1.2&	31.3&	1.003	\\
 & 7 &  1749.2953&	1749.3036&	1749.2967&	4.8&	32.0&	1.013\\
\hline
GJ 3631&5&2561.07581&	2561.07766&	2561.07581 &	2.0 &	30.9 &	1.02 \\
&6&2563.2758&	2563.2883&	2563.2758&	32.7 &	32.2 &	1.06\\
&7&2563.87887 &	2563.8791 &	2563.87887 &	0.74	&30.5 &	1.02\\
&11&2568.12672 &	2568.12857 &	2568.12672 &	2.8 &	31.1 &	1.02\\
\hline
LP 776-25 & 1 &  1456.94462&	1456.94601&	1456.94462&	0.60&	31.4 &	1.003 \\
&2& 1461.84041&	1461.84874&	1461.84041&	2.6 &	32.0 &	1.006 \\
\hline
Prox. Cen. &11 &  1602.72307&	1602.72307&	1602.72307&	0.16&	29.5 &	1.001 \\
\hline
Ross 614 & 1& 1472.94935&	1472.95213&	1472.95074&	0.88&	30.9 &	1.004\\
&9&1479.91469&	1479.91885&	1479.91469&	1.09&	31.0 &	1.003 \\
&11&1479.98413&	1479.99524&	1479.98413&	3.0&	31.4 &	1.006 \\
&20&1483.90914&	1483.91331&	1483.90914&	0.72&	30.8 &	1.002\\
&21&1483.95775&	1483.95914&	1483.95775&	0.49&	30.7 &	1.002\\
\hline
V1005 Ori & 1& 1448.89678&	1448.896787&	1448.89678&	0.31&	31.6&	1.003\\
\hline
Wolf 359 & 5 &  2567.40392 & 2567.40647	& 2567.40392 &	1.4 &	30.2 &	1.02 \\
(TESS)&11 & 2568.39059&	2568.39082&	2568.39059&	0.25&	29.5 &	1.007 \\
&14&2568.98741&	2569.0057&	2568.98741 &	11.5&	31.1 &	1.012\\
&19&2570.10743&	2570.10789&	2570.10743&	0.36&	29.6 &	1.008 \\
&15  &2570.32597&	2570.32805&	2570.32597&	1.3&	30.2 &	1.018 \\
&25 &2570.50654&	2570.50723&	2570.50654&	0.43&	29.7 &	1.008 \\
&30  &2571.11678 &	2571.11864&	2571.11678&	1.2&	30.1 &	1.013 \\
\hline
Wolf 359 &34&3075.19455&	3075.19592&	3075.19455&	0.47&	29.6&	1.006 \\
($K$2)&35 &3075.19728&	3075.20068&	3075.19728&	1.12&	29.95&	1.004 \\
& 36&3095.656&	3095.66491&	3095.656&	1.31&	30.03&	1.005 \\
& 38&3096.64361&	3096.64905&	3096.64361&	2.25&	30.26&	1.0239	\\
& 41&3116.56805&	3116.57895&	3116.56805&	0.97&	29.89&	1.0023 \\
&42 &3116.83094&	3116.84252&	3116.83094&	8.05&	30.81&	1.0200 \\
&44 &3117.69657&	3117.69861&	3117.69657&	0.125&	29.00&	1.0012\\
& 45&3118.04322&	3118.04322&	3118.04322&	0.06&	28.695&	1.0010\\
& 46&3118.04595&	3118.04799&	3118.04595&	0.18&	29.16&	1.002 \\
&47 &3118.10929&	3118.11269&	3118.10929&	0.21&	29.22&	1.001 \\
& 48&3118.82235&	3118.82371&	3118.82235&	0.36&	29.46&	1.003 \\
\hline
Wolf 424 &3  & 1949.3954&	1949.3968&	1949.3954&	0.52&	30.0 &	1.002 \\
\hline
YZ CMi & 2 & 1508.65524&	1508.71219&	1508.65524&	122.6&	33.3 &	1.132\\
&12&1513.18573&	1513.19406&	1513.18573&	2.8&	31.6 &	1.009\\
&16&1514.98985&	1514.99263&	1514.98985&	0.81&	31.1 &	1.003\\
 &22 & 2236.9053&	2236.90993&	2236.9053&	4.5	&31.8 &	1.021\\
&24&2237.04165&	2237.0435&	2237.04165&	0.60&	31.0 &	1.006\\
&26&2237.7002& 2237.70113 &	2237.7002&	0.20 &	30.5 &	1.005\\
&27&2237.70622&	2237.71826&	2237.70622&	6.26&	32.0&	1.013\\
&30&2237.78053&	2237.78238&	2237.78053&	0.79&	31.1 &	1.009
\end{longtable}

\end{document}